\documentclass[11pt,a4paper]{article} 
\pdfoutput=1
\usepackage[utf8x]{inputenc}     
\usepackage{jheppub}
\usepackage{enumerate}
\usepackage{epsfig} 
\usepackage{float} 
\usepackage[caption = false]{subfig}
\usepackage{wasysym} 
\usepackage{mathrsfs} 
\usepackage{amsfonts} 
\usepackage{amsbsy} 
\usepackage{amscd} 
\usepackage{pstricks} 
\usepackage{multirow} 
\usepackage{tikz}
\usepackage{color}
\usepackage{slashed}
\usepackage{array}
\usepackage{tablefootnote}

\usetikzlibrary{arrows,positioning,shapes.geometric} 
\usepackage[compat=1.1.0]{tikz-feynman}          
\usepackage[font=small,labelfont=bf]{caption}
\usepackage{slashed}
\usepackage{multirow}
\usepackage{tabularx}
\usepackage{soul}  
\usepackage{listings} 
\usepackage{color}

\def\l{\lambda}
\usepackage{amsthm}

\title{Top-philic Dark Matter in a Hybrid KSVZ axion framework}

\author[a,b]{Anupam Ghosh,}
\author[a]{Partha Konar,}
\author[c]{Rishav Roshan}

\affiliation[a]{Physical Research Laboratory, Ahmedabad, 380009, Gujarat, India}
\affiliation[b]{Indian Institute of Technology, Gandhinagar, 382424, Gujarat, India}
\affiliation[c]{Department of Physics, Kyungpook National University, Daegu, 41566, Korea}

\emailAdd{anupam@prl.res.in}
\emailAdd{konar@prl.res.in}
\emailAdd{rishav.roshan@gmail.com}

\abstract{ We explore a two-component dark matter scenario in an extended Kim-Shifman-Vainshtein-Zakharov (KSVZ) axion framework. This hybrid setup incorporates an extra $SU(2)_L$ complex singlet scalar whose lightest component plays the role of one of the dark matter, while the QCD axion of the KSVZ model acts as a second dark matter candidate. In this work, we focus on accentuating the role of vector-like quark that naturally emerges in the KSVZ extension on the dark matter and collider phenomenology. Here, we demonstrate that the presence of this colored particle can significantly affect the allowed dark matter parameter space of the scalar dark matter by opening up additional co-annihilation as well as the direct detection channels. Moreover, the interaction between the color particle with the top quark and scalar dark matter provides a unique topology to generate a boosted-top pair with considerable missing transverse momentum at the LHC. Using jet substructure variables and multivariate analysis, here we show that one can already exclude a vast region of parameter space with 139 $\text{fb}^{-1}$ integrated luminosity at 14 TeV LHC.
}
\preprint{\today}


\begin{document}
\maketitle

\section{Introduction}
\label{Intro}

  The Standard Model (SM) of fundamental particles is one of the outstanding achievements of modern-day physics, which has been experimentally verified at many frontiers. Ever since the discovery of the Higgs boson \cite{CMS:2012qbp,ATLAS:2012yve}, the last missing piece of SM, the particle physics community is eagerly scrutinizing collider data to witness if nature can offer a glimmer of hope in the ongoing hunt for new physics. 
Physics beyond the SM (BSM) is envisaged as inevitable since SM still fails to account for various issues to give us a fully coherent description of nature. Some of the principal concerns are related to the Strong CP problem \cite{Peccei:1977hh,Peccei:1977ur}, the existence of the dark matter (DM) \cite{Sofue:2000jx,Clowe:2006eq} in the Universe, non-zero but minuscule neutrino masses \cite{Super-Kamiokande:1998kpq,SNO:2002tuh,K2K:2002icj}, matter-antimatter asymmetry of the Universe \cite{Riotto:1999yt,Dine:2003ax} etc. The incapability of the SM in explaining these issues motivates us to look at possible extensions accommodating these aspects.

Different celestial observational evidence \cite{Sofue:2000jx,Clowe:2006eq} at diverse length scales suggests the existence of a non-baryonic, non-luminous, gravitationally interacting form of matter in the Universe, which is popularly known as the \emph{dark matter}. Precision measurement of anisotropies in cosmic microwave background radiation (CMBR) in Wilkinson Microwave Anisotropy Probe (WMAP) \cite{Hinshaw:2012aka} and PLANCK \cite{Akrami:2018odb} played a crucial role in providing us with a valuable estimate of the present abundance of DM relic density. Despite these shreds of evidence in different formats, the particle nature of the DM is still obscured from us. Different DM paradigms exist in the literature attempting to explain the DM's particle nature. Some of the most popular ones are weakly interacting massive particles (WIMP) 
\cite{Jungman:1995df,Bertone:2004pz,Konar:2009ae,Feng:2010gw,Arcadi:2017kky,PhysRevD.105.115038,Konar:2020wvl,Konar:2020vuu}, feebly interacting massive particle (FIMP) \cite{Hall:2009bx,Bernal:2017kxu,Barman:2020plp,Barman:2021tgt,Datta:2021elq,Barman:2021qds,Konar:2021oye,Chakrabarty:2022bcn}, asymmetric dark matter models \cite{DuttaBanik:2020vfr,Barman:2021ost}, models with axion or axion-like particle (ALP) DM \cite{Preskill:1982cy,Abbott:1982af,Dine:1982ah,Marsh:2015xka} etc. Although these models can explain the DM relic density of the Universe, their parameter spaces are restricted from different experimental searches. For instance, WIMP dark matter is severely constrained from null detection of the DM in the collider searches \cite{Kahlhoefer:2017dnp,CMS:2017dcx} as well as the direct~\cite{Akerib:2016vxi,Tan:2016zwf,Cui:2017nnn,Aprile:2018dbl} and indirect~\cite{Kohri:2009yn,Ahnen:2016qkx,Eiteneuer:2017hoh} search experiments.

Interestingly such a null outcome in experimental searches can instead be an indication of a much richer dark sector, and the dark sector can possibly be composed of more than one component of DM. In recent times, there have been several proposals for a multicomponent dark matter \cite{Borah:2019aeq,Bhattacharya:2019fgs,Bhattacharya:2019tqq,DuttaBanik:2020jrj,Chakrabarty:2021kmr,DuttaBanik:2016jzv,Borah:2019epq,Bhattacharya:2021rwh,Abdallah:2019svm,Pandey:2020hoq,Dasgupta:2013cwa,Chatterjee:2018mac,Lazarides:2022spe,Konar:2009qr} where the dark sector is made up of different DM components. For example, one can think of a scenario with WIMP-WIMP \cite{Borah:2019aeq,Bhattacharya:2019fgs,Bhattacharya:2019tqq,DuttaBanik:2020jrj,Chakrabarty:2021kmr} dark matter candidates, WIMP-FIMP \cite{DuttaBanik:2016jzv,Borah:2019epq,Bhattacharya:2021rwh} dark matter candidates, FIMP-FIMP \cite{Abdallah:2019svm,Pandey:2020hoq} dark matter candidates, WIMP-axion \cite{Dasgupta:2013cwa,Chatterjee:2018mac} dark matter candidates, etc. In the present work, we aim to explore one such scenario where the dark sector constitutes a WIMP-type dark matter, and the QCD axion plays the role of a second dark matter.

The minimal extension of the SM that can accommodate a WIMP dark matter is its extension by a scalar singlet field with a Higgs portal interaction \cite{Silveira:1985rk,McDonald:1993ex,Burgess:2000yq}. Here, the DM is assumed to have an odd charge under a discrete unbroken $Z_2$ symmetry that guarantees its stability. This particular DM scenario is in tension with direct search experiments. The recent direct detection data from the XENON1T experiment \cite{Aprile:2018dbl} has nearly ruled out the DM with mass below 1 TeV \cite{Bhattacharya:2019fgs,DuttaBanik:2020jrj,Borah:2020nsz} apart from the near Higgs resonance region. Reviving this sub-TeV parameter space of the scalar singlet DM could be an exciting undertaking since it can open the possibility of its testing at different frontiers like direct, indirect, and collider searches. One possible way to revive this sub-TeV parameter space is to embed this scalar DM in a two-component framework \cite{Bhattacharya:2019fgs,DuttaBanik:2020jrj}. With various extensions of two-component DM with a scalar singlet already in place, an interesting exercise could explore the insertion of a scalar singlet in a QCD axion DM framework, as it can enlarge the scalar singlet DM parameter space and can also provide rich collider phenomenology.

Extension of the SM with a global Peccei-Quinn (PQ) symmetry \cite{Peccei:1977hh,Peccei:1977ur} provides solutions for two of the critical issues discussed above in one go, $i.e.$, the \emph{Strong CP problem} and the existence of \emph{dark matter}. This global symmetry is expected to be broken at a scale much larger than the Electroweak (EW) scale. The breaking of $U(1)_{PQ}$ predicts a pseudo-Goldstone particle, popularly known as the \emph{QCD axion}, that is not absolutely stable but can have a lifetime much greater than the age of the Universe \cite{Preskill:1982cy,Abbott:1982af,Dine:1982ah,Marsh:2015xka} to play the role of DM. There are primarily three different QCD axion models that can simultaneously explain the presence of the DM in the Universe and solve the Strong CP problem. The (i) Peccei-Quinn-Weinberg-Wilczek (PQWW) \cite{Peccei:1977hh,Wilczek:1977pj,Weinberg:1977ma} model introduces an additional singlet scalar that also obtains a non-zero vacuum expectation value (vev) at the time of EW phase transition. This setup is already ruled out from the experiments. The (ii) Kim-Shifman-Vainshtein-Zakharov (KSVZ) \cite{Kim:1979if,Shifman:1979if} model introduces an extra colored particle together with a complex scalar that breaks the PQ symmetry. Anomaly-free condition is ensured by the introduction of vector-like quarks (VLQ). Finally, the (iii) Dine-Fischler-Srednicki-Zhitnitsky (DFSZ) \cite{Dine:1981rt,Zhitnitsky:1980tq} model incorporates an additional Higgs field apart from the PQ breaking scalar. It is also interesting to point out that the breaking of PQ symmetry in these models also leaves a remnant $Z_2$ symmetry that remains intact. If such a setup is extended with an extra particle that also carries a non-trivial $Z_2$, then this unbroken symmetry can naturally ensure its stability. This motivates us to study two-component DM scenarios in these models. 

In the present work, we aim to explore a hybrid KSVZ scenario, where an extra complex scalar singlet extends the particle spectrum of the KSVZ setup in addition to the usual complex scalar that breaks the PQ symmetry. As mentioned above, the breaking of $U(1)_{PQ}$ symmetry in the KSVZ construction leaves remnant $Z_2$ symmetry which remains unbroken throughout, and the VLQ present in the setup carries an odd charge under it. If the model is extended by a complex scalar that also holds a non-trivial charge under the same $Z_2$, the lightest component of the scalar can play the role of the second dark matter, making the dark sector two-component.

Note that our setup is similar to the one considered in \cite{Dasgupta:2013cwa,Chatterjee:2018mac} but with some crucial differences. For example, the present framework considers an up-type VLQ rather than a down-type VLQ, and it opens up a non-trivial possibility in collider analysis. Current construction also aims to explore the possibility of DM effective annihilations (including the DM co-annihilation with the vector-like quark as well as the annihilation of the vector-like quark to the SM particles) to the SM particles in obtaining the correct relic density of the scalar (WIMP) DM. Next, both \cite{Dasgupta:2013cwa,Chatterjee:2018mac} did not consider the VLQ's contribution to the DM-nucleon spin-independent scattering, which can play a crucial role in the DM phenomenology of WIMP dark matter. This feature of DM effective annihilation to the SM particles, as well as the role of the VLQ in the tree level spin-independent DM-nucleon scattering, was also discussed in \cite{Borah:2020nsz}. Ref.\cite{Borah:2020nsz} featured a VLQ doublet and VLQ singlet fermions, unlike ours with a singlet VLQ which naturally occurs in a KSVZ scenario. We also want to emphasize that, although \cite{Dasgupta:2013cwa,Chatterjee:2018mac,Borah:2020nsz} featured some discussions on the collider aspect of this kind of setup, none of them explored it in detail. In the present work, we also perform an exhaustive colliders analysis and aim to find out the relevant parameter space, which would not only validate the dark matter phenomenology but also be consistent with the collider searches.

The paper is organized as follows. We introduce our model in section \ref{model} where the particle spectrum together with their charges under different symmetry groups have been discussed. Various theoretical and experimental constraints in our model are presented
in section \ref{constraints}. In section \ref{DMphenomenology} we discuss the dark matter phenomenology of the model. The collider analysis and the result based on multivariate analysis are presented in section \ref{collider}. Finally, we summarize our findings in section \ref{conclusions}.

\section{The Model}
\label{model}

As stated in the introduction, the present work aims to study dark matter and collider phenomenology in a hybrid KSVZ framework of QCD axion. As is well known, the vanilla KSVZ model requires a complex scalar singlet $\eta$ that breaks a global symmetry, popularly known as $U(1)_{\text{PQ}}$. In addition, this model also demands a $SU(2)_L$ singlet colored fermion $\Psi$ with a $+1$ unit of $U(1)_{\text{PQ}}$ charge. This extra quark is vector-like and hence does not introduce any chiral anomaly. In addition, the hybrid KSVZ model also introduces an additional complex singlet scalar $S$ charged under the $U(1)_{\text{PQ}}$. The BSM fermion and scalar content of the model and their respective charges are listed in Tab. \ref{tab:particle}. The most general renormalizable and gauge-invariant Lagrangian for the present setup can be written as,
\begin{eqnarray}
-\mathcal{L}^{\text{VLQ}}&=& f_{i}S \overline{\Psi}_L {u_i}_R + f_\Psi \eta \overline{\Psi}_L \Psi_R + h.c.,
\label{Lag}
\end{eqnarray}
where, $u_R$ represents right-handed up-type quarks in the SM with $i=u,c,t$. Here, $L$ and $R$ denote left- and right-handed projections. Note that the hypercharge of the newly introduced VLQ depends on its interaction with SM quarks. The relevance of introducing an up-type VLQ in this setup will be clear once we discuss the DM and collider phenomenologies in sections \ref{DMphenomenology} and \ref{collider} respectively.

\begin{table}[tb!]
\begin{center}
 \begin{tabular}{|c|c|c|c|c|}
\hline
 & $\eta$ & $S$ & $\Psi$   \\
\hline\hline
 $SU(3)_C$ & 1 & 1& 3  \\  
\hline
 $SU(2)_L$ & 1 & 1& 1 \\  
\hline
 $U(1)_Y$ & 0 & 0& 2/3 \\  
\hline
 $U(1)_{PQ}$ & 2 & 1& 1  \\  
\hline
 \end{tabular} 
\caption{Particle contents and their respective charge assignments under different symmetry groups.}
\label{tab:particle}
\end{center}
\end{table}

Moving on to the scalar part of the Lagrangian, the most general renormalizable scalar potential of our model,
$V(H,\eta, S)$ can be written as,
\begin{eqnarray}
V(H,\eta,S)&=& \lambda_H (|H|^2 - v_H^2/2)^2 +\lambda_\eta (|\eta|^2-F_a^2/2)^2+\lambda_{\eta H} (|H|^2 - v_H^2/2) (|\eta|^2-F_a^2/2) \nonumber \\
&+&\mu_S^2|S|^2+\lambda_S |S|^4 +\lambda_{SH} |H|^2|S|^2 +\lambda_{S \eta } |\eta|^2 |S|^2 +[\epsilon_S \eta^* S^2 +h.c].
\label{potential}
\end{eqnarray}
After the breaking of both $U(1)_{\text{PQ}}$ and the SM gauge symmetry, the different scalars involved in the present setup take the following form,
\begin{eqnarray}                
H = \left( \begin{array}{c}
                         0  \\
        \frac{1}{\sqrt{2}}(v_H+h_0)  
                 \end{array}  \right) \, , \, \, \,                     
 \eta =e^{\frac{ia}{F_a}} \frac{(F_a+\sigma_0)}{\sqrt{2}}\, , \, \, \, 
 S=\frac{S_1+iS_2}{\sqrt{2}}\, ,
\label{e2a}
\end{eqnarray}
where  $v_H$  denotes the vacuum expectation value (vev) of $H$ obtained after the electroweak symmetry breaking (EWSB) and $F_a$ represents the $U(1)_{\text{PQ}}$ breaking scale. It is to be noted that, after the breaking of both symmetries, a non-zero $h_0-\sigma_0$ mixing leads to the following mass terms:
\begin{equation}
M^2 \equiv
\begin{pmatrix} 
2v_H^2 \lambda_H & F_a v_H \lambda_{\eta H} \\ 
F_a v_H \lambda_{\eta H} & 2 F_a^2 \lambda_\eta 
\end{pmatrix}.
\end{equation}
The mass matrix can be diagonalised using
\begin{equation}
\begin{pmatrix} h_0  \\ \sigma_0 \end{pmatrix}
 =
  \begin{pmatrix}
   \cos \theta_m & \sin \theta_m\\
   -\sin \theta_m & \cos \theta_m
   \end{pmatrix} 
   \begin{pmatrix} h  \\ \sigma \end{pmatrix}
\end{equation}
where the mixing angle is given by, 
\begin{equation}
\tan(2\theta_m) =\dfrac{F_a v \lambda_{\eta H}}{F_a^2 \lambda_\eta - v^2 \lambda_H} \,\,.
\end{equation}
Finally, after diagonalization, the physical masses of the $h$ and $\sigma$ are given as,
\begin{eqnarray}
M_{h,\sigma}^2 = (\lambda_H v^2 +\lambda_\eta F_a^2 ) \pm \sqrt{(\lambda_H v^2 -\lambda_\eta F_a^2 )^2+ F_a^2 v^2 \lambda_{\eta H}^2} \,\,.
\label{phys_mass}
\end{eqnarray}
Next, as an artifact of two different symmetry breakings, the masses of the different components of the $S$ can be expressed as,
\begin{eqnarray}
M_{S_{1,2}}^2&=&\frac{1}{2}(2\mu_S^2+v_H^2\l_{SH}+F_a^2\l_{S\eta}\mp 2\sqrt{2}\epsilon_sF_a).
\label{DM_mass}
\end{eqnarray}
Notice that the presence of the term proportional to $\epsilon_S$ in Eq. \ref{potential} plays a crucial role in generating the mass splitting among the components of $S$. Subsequently, the mass of the VLQ is given as,
 \begin{eqnarray}
 M_{\Psi}&=& f_{\Psi}\frac{F_a}{\sqrt{2}}.
 \label{mass_vlq}
 \end{eqnarray}
At this stage, it is interesting to point out that, even after the breaking of both the symmetries, there still exists a remnant $Z_2$ symmetry under which both the Lagrangian as well as the scalar potential remains invariant. This remnant $Z_2$ can remain intact if $S$ does not acquire a non-zero vev. Under such a scenario, the lightest neutral component of $S$ can provide a vital DM candidate.

Finally, the setup also contains a pseudo-Nambu Goldstone boson $a$, associated with scalar $\eta$, popularly known as \emph{axion}. The axion obtains a mass as a result of non-perturbative QCD effects given as~\cite{Preskill:1982cy,Dine:1982ah},
\begin{eqnarray}
m_a\simeq 0.6~\text{meV}\times \bigg(\frac{10^{10}~\text{GeV}}{F_a}\bigg).
\label{axion_mass}
\end{eqnarray}
Note that a suitable choice of decay constant $F_a$ can adjust the fraction of which QCD axion can contribute toward the relic density of the dark matter. That makes the preset setup a tunable two-component dark matter scenario. The role of QCD axion as a DM candidate and its constraints are elaborated in section \ref{DMphenomenology}. Now with the knowledge of all the particles and their interactions in this hybrid KSVZ setup, we are in a position to list the set of independent parameters important for the dark matter and collider phenomenology:
$$\{M_{\Psi},M_{S_1},M_{S_2},M_{\sigma},F_a,\lambda_{SH},\lambda_{S \eta },f_{i}\}.$$

\section{Experimental and Theoretical Constraints}
\label{constraints}

The extended KSVZ model under consideration is subjected to various theoretical as well as experimental constraints. In this section, we summarize all the relevant ones. 

\begin{itemize}
\item
{\bf{Stability and Perturbativity:}} The scalar sector is extended over the vanilla model. Hence, different scalars in the present setup can help stabilize the electroweak vacuum. The stability of the electroweak vacuum also demands that the scalar potential should be bounded from below in all the field directions of the field space. On the other hand, a perturbative theory demands that the model parameters should obey: 
\begin{eqnarray}
|\lambda_{i}|< 4\pi~ {\rm{and}}~ |g_i|, |y|, |f_{i}|<\sqrt{4\pi}  .
\label{pert} 
\end{eqnarray}
where ${g}_i$ and ${y}$ are the SM gauge and Yukawa couplings, whereas $f_{i}$ are Yukawa couplings involving different BSM fields, respectively. 

\item
{\bf{Relic density, Direct and Indirect detection of DM:}} For any dark matter model, it is essential to satisfy the observed abundance of DM relics from the precision measurement in the Planck experiment \cite{Aghanim:2018eyx},
\begin{eqnarray}
\Omega_{\text{DM}}h^2&=&0.120\pm 0.001.
\end{eqnarray} 

Apart from DM relic density, the DM-nucleon scattering cross-section is also constrained by various direct search experiments like LUX \cite{Akerib:2016vxi}, PandaX-II \cite{Tan:2016zwf,Cui:2017nnn}, and XEXON1T \cite{Aprile:2018dbl}. Finally, the DM annihilation to the SM particles are also subjected to the constraints coming from the indirect search experiments like PAMELA~\cite{Kohri:2009yn}, Fermi-LAT~\cite{Eiteneuer:2017hoh}, MAGIC~\cite{Ahnen:2016qkx} etc. Nonetheless, in all these cases, one also needs to take care of the multi-component nature of DM in our extended scenario, which is further discussed in section \ref{DMphenomenology}.

\item
{\bf{Flavor constraints:}} The Yukawa interactions of the complex singlet scalar $S$ with VLQ and the SM right-handed quarks like $u$ and $c$ in the present setup can contribute towards the $D^0-\bar{D}^0$ mixing~\cite{Garny:2014waa}. The measured value of the $D-$meson mass splitting significantly constrained this mixing. The Feynmann diagrams that contribute to this mixing are shown in Fig. \ref{DDbar}; each diagram has four possible configurations with a total of sixteen diagrams. Effective operator contributing to this mixing in the present setup can be expressed as
\begin{eqnarray}
\mathcal{L_\text{eff}}=\frac{\tilde{z}}{M_{\Psi}^2}\bar{u}_R^\alpha\gamma^\mu c_R^\alpha \bar{u}_R^\beta \gamma_\mu c_R^\beta.
\end{eqnarray}
where
\begin{eqnarray}
\tilde{z}=-\frac{f_u^2f_c^2}{96\pi^2}[g_{\psi}(M_{S_1}^2/M_{\Psi}^2)+g_{\psi}(M_{S_2}^2/M_{\Psi}^2)-2g_{\psi}(M_{S_1}M_{S_2}/M_{\Psi}^2)].
\end{eqnarray} 
Here $g_{\Psi}(x)=24xf_6(x)+12\tilde{f_6}(x)$ where the expressions of $f_6$ and $\tilde{f_6}$ can be found in~\cite{Gedalia:2009kh}. The measurement of the $D-$meson mass splitting demands $|\tilde{z}|\lesssim 5.7\times10^{-7}(M_{\Psi}/\text{TeV})^2$~\cite{Gedalia:2009kh,Garny:2014waa} 

\item
\textbf{LHC diphoton searches}: As a result of mixing between $h$ and $\sigma$, all the tree level interactions with the SM Higgs get modified. In such a case, the signal strength in the di-photon channel takes a form:
\begin{eqnarray}
\mu_{\gamma \gamma} = c^2_\theta 
\frac{BR_{h \to \gamma \gamma}}{BR_{h \to \gamma \gamma}^{\text{SM}}} \simeq 
c^2_\theta \frac{\Gamma_{h \to \gamma \gamma}}{\Gamma_{h \to \gamma \gamma}^{\text{SM}}}.
\end{eqnarray}
LHC sets a limit on this new mixing angle as $|\sin{\theta}|\leq 0.36$ \cite{Robens:2016xkb}.

\item
\textbf{Invisible Higgs decay}: Involvement of the new interactions of SM Higgs with various BSM particles in the present setup can lead to its new decay modes if kinematically allowed. These extra decays of Higgs can contribute toward invisible Higgs decay. In such a situation, we need to employ the bound on the invisible Higgs decay width as \cite{ATLAS-CONF-2020-052}: 
\begin{subequations}
\begin{eqnarray}
Br(h\rightarrow \rm{Invisible}) \equiv
\frac{\Gamma(h\rightarrow \rm{Invisible})}{\Gamma(h\rightarrow SM)+\Gamma(h\rightarrow \rm{Invisible})} < 0.11 .
\label{invi}   
\end{eqnarray} 
\end{subequations}
In the case of light DM, the Higgs can decay to a pair of it when kinematically allowed. However, in our present analysis, we primarily focus on the parameter space where $m_{i}>~\frac{m_h}{2}$ so the above constraint is not applicable.

\item
{\bf{Direct collider constraints:}}
Due to the presence of colored vector-like quarks, the present model is subjected to various collider constraints. Being non-trivially charged under the $U(1)_{PQ}$ allows the VLQ to couple with the complex scalar and the SM up type quarks. If kinematically allowed, the heavier states can always decay into the DM and an SM quark. Therefore a generic collider signature of this model contains a considerable amount of missing (transverse) energy from the escape of final DM particles from detection at the detector.

Vector-like fermion can be pair produced through electroweak interaction performed at CERN's Large Electron Positron Collider (LEP):
\begin{equation}
e^+ e^- \rightarrow \gamma^{*}, Z \rightarrow \Psi \bar{\Psi}
\end{equation}
The interaction between vector-like fermion, light SM quarks, and the DM can lead to the decay of $\Psi$ to a light quark associated with DM at LEP if kinematically allowed. The reinterpreted LEPII results of squark search \cite{Giacchino:2015hvk, OPAL:2002bdl} exclude the mass of $\Psi$ up to 100 GeV. Such constraint is incorporated in our final exclusion plots. Please follow the brown region in Fig. \ref{result_final}.
Similar searches were also carried out at the LHC. In a recent ATLAS search, the vector-like mediator is searched while it decays into an invisible particle and light quark up (charm) when the mass difference between the mediator and DM is less than the top quark mass. The green region in Fig. \ref{result_final} is excluded from the reinterpreted result \cite{Giacchino:2015hvk} of the ATLAS search \cite{Marjanovic:2014eca} for multijet (2-6 jets) plus missing transverse momentum at center-of-mass (CM) energy $\sqrt{s}=8$ TeV with an integrated luminosity of 20.3 $\text{fb}^{-1}$. 
Exploring a larger mass difference between the mediator and DM candidate, top-antitop plus missing transverse momentum signal has been extensively studied by both CMS and ATLAS collaborations, particularly superpartners searches of the top quark~\cite{ATLAS:2017eoo, ATLAS:2017drc, ATLAS:2017avc, ATLAS:2017www, ATLAS:2017tmw, CMS:2017okm, CMS:2017mbm, CMS:2017gbz, CMS:2017jrd} and some dedicated dark matter searches~\cite{CMS:2017dcx}. The vector-like mediator can be pair produced at the LHC mainly through strong interaction and then decay into an up-type quark and invisible particle. So, the search for a top pair along with the missing transverse momentum signature by ATLAS and CMS can be reinterpreted to exclude some of the parameter spaces of this model. The CMS analysis~\cite{CMS:2017jrd} is reinterpreted in Ref~\cite{Colucci:2018vxz} at 13 TeV LHC for an integrated luminosity of 35.9 $\text{fb}^{-1}$, assuming vector-like mediator decay with $100\%$ branching fraction into the top and invisible particle. In their analysis, the signal consists of two oppositely charged isolated leptons from leptonic decays of both top and anti-top. The signal also consists of at least two hard jets; one of them is b-tagged and a large missing transverse momentum. The olive region in Fig. \ref{result_final} is the exclusion region ($2\sigma$) obtained from this analysis.  
\end{itemize}

The existing LHC search relies on finding the top pair based on two hard leptons and a b-tagged jet. It is evident that the sensitivity of such detection deteriorates when these tops are boosted, especially while decaying from a heavy mother particle. We propose an alternative search strategy in this work by recognizing these boosted double top jets with a large missing transverse energy signature using jet substructure variables and multivariate analysis. We are examining the spectrum where the mass difference between vector-like mediator and DM is larger than the mass of the top quark such that on-shell decay into the top is possible. Our search strategy helps to explore the significant parameter space of this model that gives observed relic density of DM and also allowed from the direct-detection experiment with the current luminosity of the LHC.

\section{Dark Matter Phenomenology}
\label{DMphenomenology}

In this section, we aim to elaborate on the DM phenomenology of the model under consideration. As discussed earlier, the setup is a hybrid of the KSVZ model that includes an extra complex scalar ($S$) whose lightest component ($S_1$) plays the role of one of the DM while the part of the second DM is played by the QCD axion of the KSVZ setup. The involvement of the two DMs in this extended KSVZ scenario makes the layout a two-component DM system. Besides the QCD axion, the KSVZ setup naturally demands a presence of an extra colored fermionic $SU(2)_L$ singlet. This fermion plays a non-trivial role in the DM phenomenology and the collider searches of the DM as it talks directly to it through the Yukawa interaction given in Eq. \ref{Lag}. Next, we discuss the DM phenomenology of both the DM candidates of the present model.

\subsection{Relic density and DM detection}  

 Apart from providing a solution to the \emph{strong CP problem}, another interesting consequence of introducing a PQ symmetry is the emergence of the Nambu Goldstone boson, popularly known as \emph{axion}. If the breaking scale ($F_a$) of the PQ symmetry is chosen appropriately, the resulting axion can be light as well as stable. This QCD axion can be an excellent DM candidate in such a scenario. Axions can be produced non-thermally as a result of the misalignment mechanism. Here, the axion field begins to coherently oscillate around the minimum of the PQ vacuum when its mass becomes comparable to the Hubble parameter. This coherent oscillation of the axion field behaves like a cold matter in the Universe. The relic density of the axion in such a case is approximately given by~\cite{Dasgupta:2013cwa,Chatterjee:2018mac,Bae:2008ue},
\begin{equation}
\Omega _a h^2 \simeq 0.18 \hspace{1mm} \theta_a ^2 \hspace{1mm} \bigg(\frac{F_a}{10^{12} \text{GeV}}\bigg)^{1.19} \, .
\end{equation} 
Here, $\theta_a$ represents the initial misalignment angle of the axion.

For the case of the scalar DM, we consider the mass hierarchy $M_{S_2}>M_{S_1}$ such that the lightest scalar component represents the second DM candidate. Its interactions with the SM Higgs and the VLQ keep it in equilibrium with the thermal bath in the early Universe. As the temperature of the Universe drops below the DM mass, its production from the thermal bath stops while its annihilation of the SM particle continues. Once the Universe's expansion rate becomes larger than the interaction rate of the DM, its annihilation to the SM bath also stops, and its abundance freezes out. DM can annihilate to the SM particles through: (a) its contact interactions, (b) Higgs-mediated channels~\footnote{From Eq. \ref{phys_mass}, it is evident that until and unless $\l_{\eta}$ is very small, $M_{\sigma}$ will remain much heavier than $M_h$ and hence the $\sigma$ mediated annihilation channels be very much suppressed.} and (c) VLQ mediated channels (as a result of Yukawa interaction given in Eq. \ref{Lag}). The presence of the Yukawa interaction also allows the DM to co-annihilate if the mass-splitting between the DM and newly introduced quark is sufficiently small. Note that as the VLQ and $S_2$ share the same $Z_2$ charge similar to the DM, their annihilations would also be important for evaluating the effective annihilation cross-section. In appendix~\ref{appendixFR}, we present all the important annihilation and co-annihilation channels of the DM that are crucial in determining its final relic abundance~\footnote{Just for completeness we have also shown the DM annihilation to the axion final states. These annihilations are highly suppressed and do not contribute towards the relic density of scalar dark matter. This is because most of the vertices involved in these annihilation cross-sections are either proportional to $1/F_a$ or $\frac{\sin{\theta}}{F_a}$ or $\frac{\epsilon_S}{F_a}$ or $\epsilon_S$. }. Once all the important annihilation and co-annihilation channels are identified, one can use them to determine the final relic density of the DM, which can be expressed as~\cite{McDonald:1993ex},
\begin{eqnarray}
\Omega_{S_1} h^2&=&\frac{1.09\times10^9 ~\rm{GeV^{-1}}}{g_*^{1/2}~M_{Pl}}\frac{1}{J(x_f)},
\end{eqnarray}
\label{relic_expression}
where $J(x_f)$ is given by,
\begin{eqnarray}
J(x_f)&=& \int_{x_f}^{\infty}\frac{\langle \sigma |v|\rangle_{\rm eff}}{x^2}~\rm{dx}.
\label{th_int}
\end{eqnarray}
$\langle \sigma |v|\rangle_{\rm eff}$ in Eq. \eqref{th_int} is the effective thermal average DM annihilation cross-sections including contributions from the co-annihilations and is given by,
\begin{eqnarray}
\langle \sigma |v|\rangle_{\rm eff} &=& \frac{g_{s_1}^2}{g_{\rm eff}^2}\sigma(\overline{S_1}S_1)+2\frac{g_{s_1} g_{s_2}}{g_{\rm eff}^2}\sigma(\overline{S_1}S_2)(1+\Delta_{12})^{3/2}~\text{exp}[-x\Delta_{12}]+2\frac{g_{s_1} g_{\Psi}}{g_{\rm eff}^2}\sigma(\overline{S_1}\Psi)(1+\Delta_\Psi)^{3/2}\nonumber \\
&&~\text{exp}[-x\Delta_\Psi]+\frac{g_{S_2}^2}{g_{\rm eff}^2}\sigma(\overline{S_2}S_2)(1+\Delta_{12})^{3}~\text{exp}[-2x\Delta_{12}] +\frac{g_{\Psi}^2}{g_{\rm eff}^2}\sigma(\overline{\Psi}\Psi)(1+\Delta_\Psi)^{3}\nonumber \\
&&~\text{exp}[-2x\Delta_\Psi].  
\label{eff_cs} 
\end{eqnarray}
In the equation above, $g_{s_1}$, $g_{s_2}$ and $g_{\Psi}$ are the spin degrees of freedom for $S_1$, $S_2$ and $\Psi$. 
Here, $x=\frac{M_{S_1}}{T}$ representing dimensionless parameter with inverse of temperature, while $\Delta_\Psi$ and $\Delta_{12}$ are two dimensionless parameters qualifying mass splittings from dark scalar candidate:
\begin{eqnarray}
\Delta_\Psi = \frac{M_\Psi-M_{S_1}}{M_{S_1}}; \;
\Delta_{12}=\frac{M_{S_2}-M_{S_1}}{M_{S_1}}.
\end{eqnarray}  
The effective degrees of freedom in Eq. \eqref{eff_cs} is given by,
\begin{eqnarray}
g_{\text{eff}} &=& g_{s_1}+g_{s_2}(1+\Delta_{12})^{3/2}~\text{exp}[-x\Delta_{12}]+g_{\Psi}(1+\Delta_{\Psi})^{3/2}~\text{exp}[-x\Delta_\Psi].
\end{eqnarray} 

In the following analysis, we first generate the model using FeynRules~\cite{Alloul:2013bka} and then implement it in \texttt{micrOMEGAs} -v5 \cite{Belanger:2018ccd} to find the region of parameter space that corresponds 
to correct relic abundance for our scalar DM candidate in accordance with the relation,
\begin{eqnarray}
\Omega_{\text{T}}h^2&=&\Omega_{a}h^2+\Omega_{S_1}h^2,
\label{totalRelic}
\end{eqnarray}
where $\Omega_{\text{T}}h^2$ corresponds to the total relic density of the DM satisfying PLANCK constraints \cite{Aghanim:2018eyx}.

Next, the present model is subjected to the constraints coming from the direct search experiments for the dark matter. Experiments like LUX \cite{Akerib:2016vxi}, PandaX-II \cite{Tan:2016zwf,Cui:2017nnn} looks for the DM recoil in the DM-nucleon scattering and subsequently provides a bound on the DM-nucleon scattering cross-section. Being a two-component DM system, the direct detection cross-section of the scalar DM should be rescaled as,

\begin{eqnarray}
\sigma_{S_{1,\text{eff}}}^{\text{SI}}=\frac{\Omega_{S_1}}{\Omega_{\text{T}}}\sigma_{S_{1}}^{\text{SI}}
\label{singletdd}
\end{eqnarray} 
 
As mentioned earlier, due to the direct Yukawa interaction of the scalar DM with the up-quark, two other scattering processes contribute to the direct detection cross-section of the scalar apart from the usual SM Higgs-mediated scattering. In appendix~\ref{appendixFR} we listed all the scattering processes of the DM $S_1$ with the detector nucleon.

 Finally, the model is also subjected to the constraints coming from the indirect search experiments. Indirect search experiments looking for an excess of gamma rays can help in probing the WIMP dark matter. DM particles can annihilate and produce SM particles, out of which photons (and also neutrinos), being electromagnetically neutral, have better chances of reaching the detector from the source without getting deflected. Experiments like PAMELA~\cite{Kohri:2009yn}, Fermi-LAT~\cite{Eiteneuer:2017hoh}, MAGIC~\cite{Ahnen:2016qkx} etc. look for such excess in order to confirm the particle nature of the DM. The present set up being a two-component DM scenario, the indirect detection cross-section of the scalar DM should be rescaled as well,
\begin{eqnarray}
	\sigma_{S_{1,\text{eff}}}^{\text{ID}}=\bigg(\frac{\Omega_{S_1}}{\Omega_{\text{T}}}\bigg)^2\sigma_{S_{1}}^{\text{ID}}.
	\label{singletdd}
\end{eqnarray} 

At this stage, it is worth commenting on some of the detection possibilities of the axion as a DM candidate. Several ongoing and proposed experiments rely on axion being a DM. All these experiments lean on different detection techniques. For example, ADMX \cite{ADMX:2009iij} searches for DM-photon conversion in the presence of the magnetic field. CASPEr \cite{Budker:2013hfa} uses nuclear magnetic resonance to hunt for the axion DM; it is known that if the axion exists, it will modify Maxwell's equation. ABRACADABRA \cite{Kahn:2016aff} utilizes this by using a toroidal magnet to source an effective electric current, and finally, MADMAX \cite{Caldwell:2016dcw} is a proposed experiment that uses dielectrics haloscopes.
 
\subsection{Parameter Space of Hybrid KSVZ Axion Framework} 

\begin{figure}[htb!]
\centering
\includegraphics[scale=0.4]{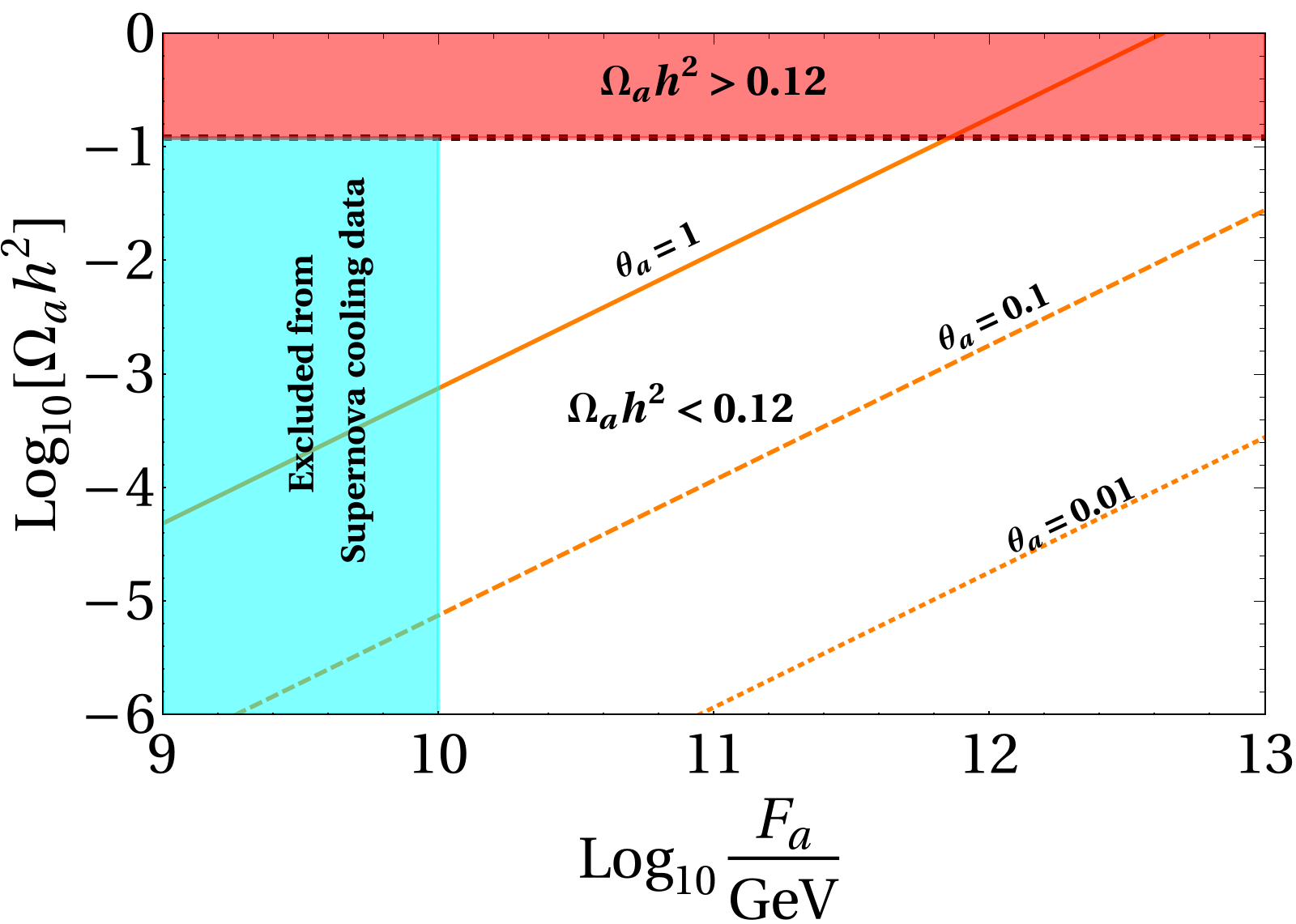}
\caption{Variation of QCD axion relic density with the decay constant $F_a$ for three different values of misalignment angles: $\theta_a=1.0$ (solid), $\theta_a=0.1$ (dashed), and $\theta_a=0.01$ (dotted). Black thick dashed line corresponds to observed relic $\Omega_{\text{DM}}h^2=0.12$. The cyan region is disallowed from the Supernova cooling data. The light pink region corresponds to the parameter space where the DM relic density remains overabundant.}
\label{axn_PS}
\end{figure}

It is well known that the KSVZ model provides a DM in the form of QCD axion. For this axion to play the role of the DM or contributes sufficiently towards the relic density of the DM, the decay constant $F_a$ should lie in the range,
\begin{eqnarray}
10^{10}~\text{GeV}\leq F_a \leq 10^{12}~\text{GeV}.
\end{eqnarray} 
The lower bound on $F_a$ comes from the supernova cooling data \cite{Raffelt:1987yt} whereas the upper bound results from the overproduction of the axion or, in other words, the relic density of the axion become overabundant. To understand this, in Fig. \ref{axn_PS} we study the variation of the axion relic density ($\Omega_a h^2$) with the decay constant for three different values of the misalignment angles $i.e.$  $\theta_a=1.0$ (solid), $\theta_a=0.1$ (dashed), and $\theta_a=0.01$ (dotted). The region in cyan is ruled out from the supernova cooling data, whereas the light pink region corresponds to the overproduced DM relic density. As can be seen from Fig. \ref{axn_PS}, for $\theta_a=1.0$ and $F_{a}\simeq 10^{12}~\text{GeV}$, the axion alone can contribute $100\%$ towards the relic density of the dark matter. Finally, the white region corresponds to the parameter space where QCD axion as a DM remains underabundant.

The present setup is an extended version of the KSVZ scenario, which contains an additional DM candidate as a singlet scalar. The presence of this extra DM here demands us to choose a parameter space for axion from Fig. \ref{axn_PS} where the relic density of the axion remains underabundant so that the relic density of the axion together with the scalar can satisfy the Planck limit. For a demonstrative purpose, we fix $F_a = 10^{11}$ GeV and choose the misalignment angle as $\theta_a=1$ for the rest of the analysis. This choice of $F_a$ and $\theta_a$ corresponds to $\Omega_ah^2 =0.012$. Without losing generality in our analysis, we set a heavier  $M_{\sigma}$ at 50 TeV (as the setup requires it to be quite heavy). 
At this stage, we would like to point out that the DM matter couples to $\sigma$ through $S_1-S_1-\sigma$ interaction. This interaction can also help DM to annihilate into the SM particles through scalar mixing. Although these annihilations will have suppression coming from the mass of $\sigma$, they might still not be that small as these annihilations are also proportional to the $F_a\lambda_{S\eta}$. With $F_a=10^{11}~\text{GeV}$ and not so small value of $\lambda_{S\eta}$, the DM can still have significant annihilation cross-sections and such cross-sections might violate perturbative unitarity \cite{Profumo:2019ujg}. This demands $\lambda_{S\eta}$ to be extremely tiny. For simplicity, we set $\lambda_{S\eta}=0$ throughout our analysis. Next, for the analysis purpose, we also define a mass-splitting, $\Delta M = M_{S_2}-M_{S_1}$ and consider it to be a free parameter rather than $M_{S_2}$. It is interesting to point out that once $\Delta M$ and $F_a$ are fixed, the parameter $\epsilon_S$ automatically gets fixed, as can be seen from Eq. \ref{DM_mass}.  
Before diving into the detailed analysis of the second DM candidate, we will like to mention the set of parameters that are relevant for the analysis of the DM phenomenology of the second DM candidate:
$$\{M_{\Psi},M_{S_1},\Delta M,F_a,\lambda_{SH},f_{i}\}.$$

\begin{figure}[H]
\centering
\includegraphics[scale=0.35]{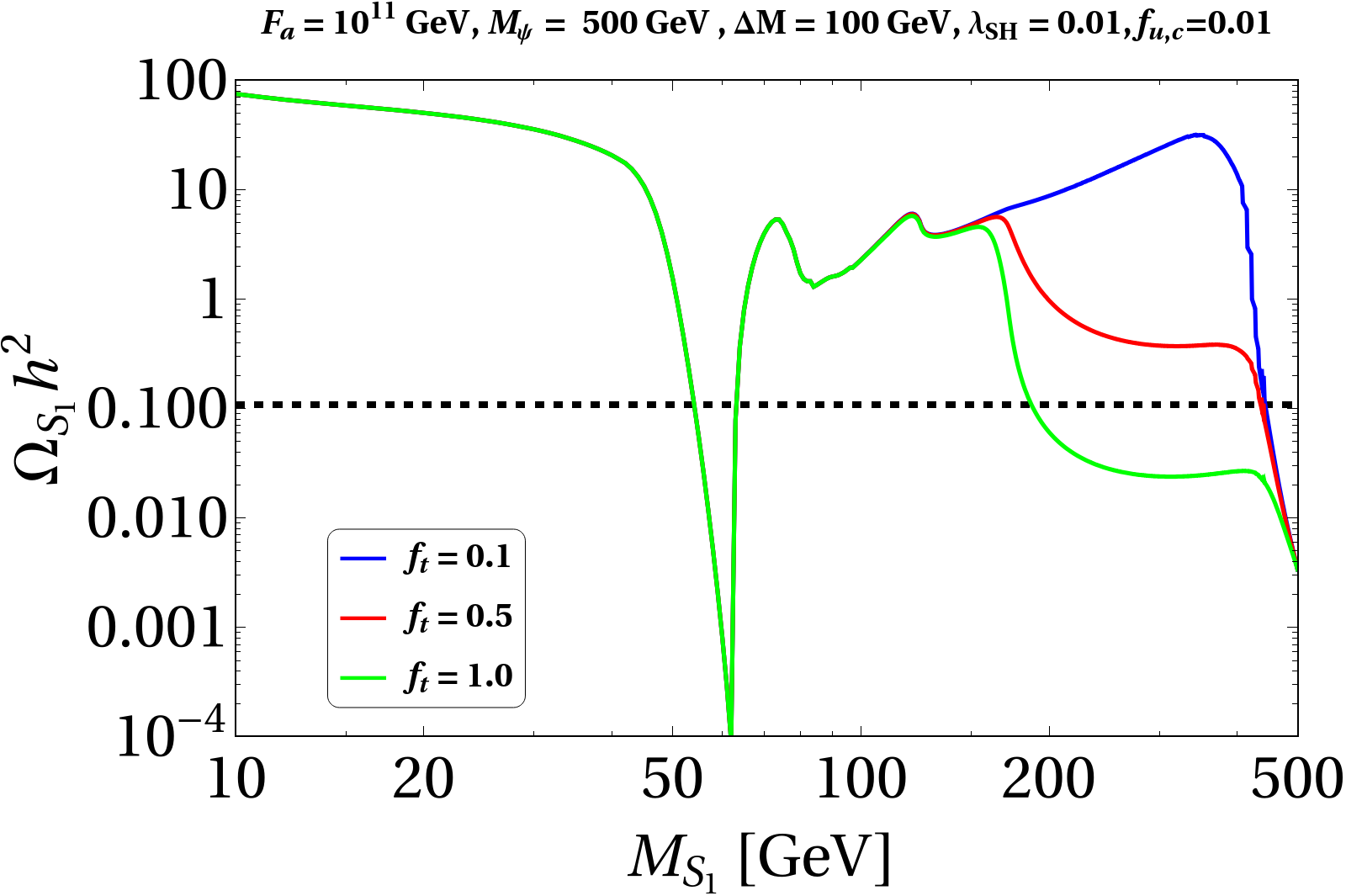}
\includegraphics[scale=0.35]{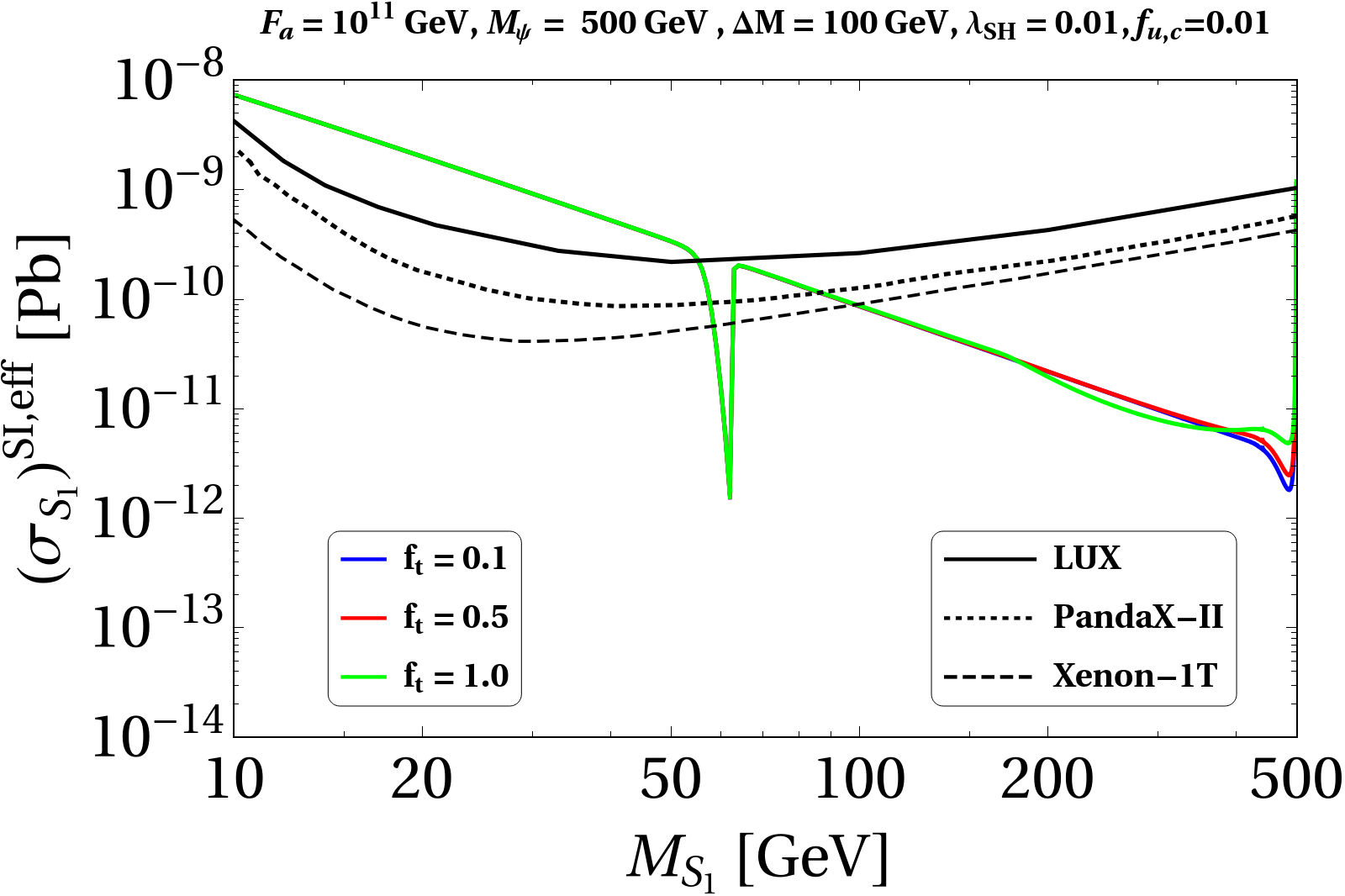}
\includegraphics[scale=0.35]{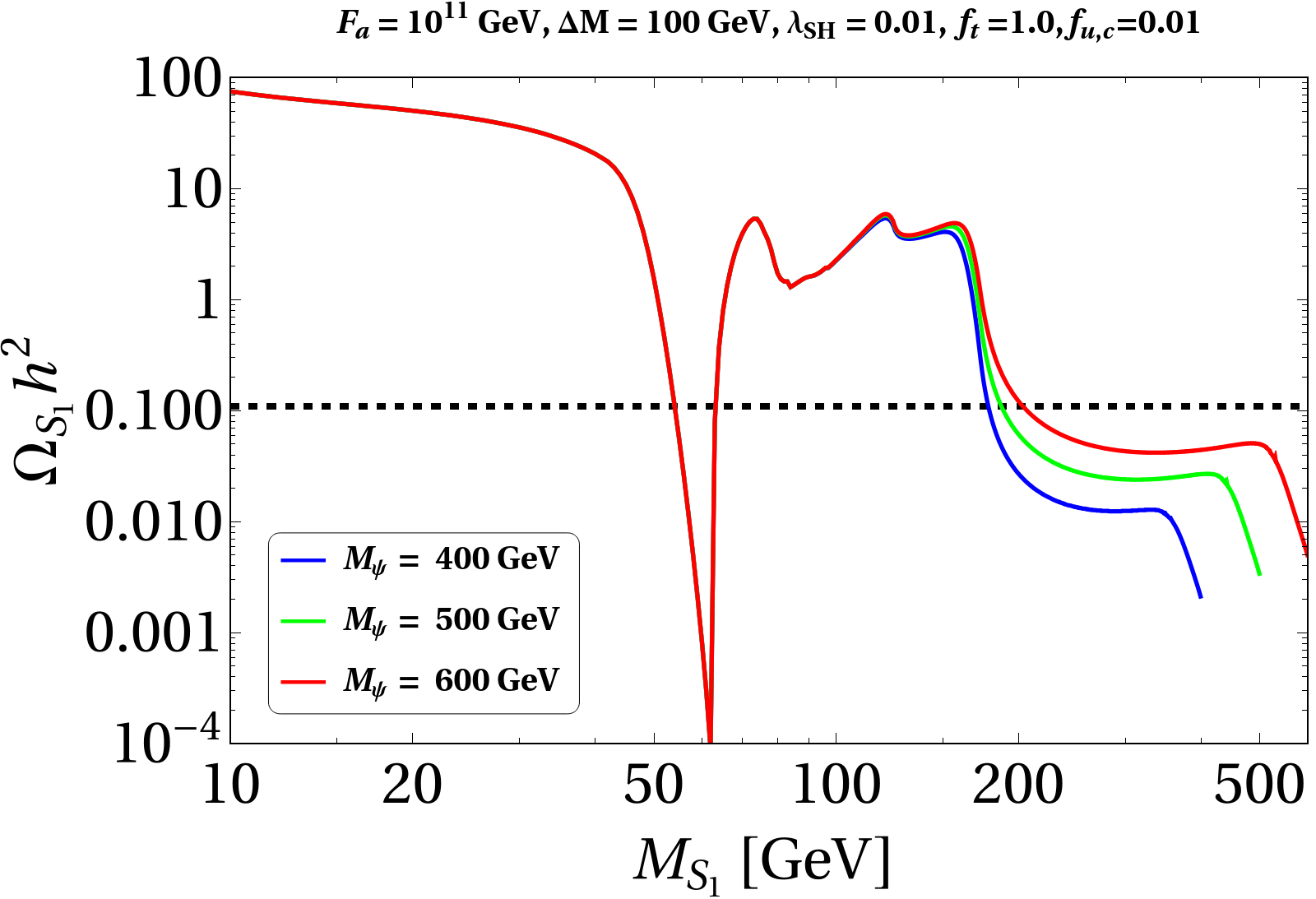}
\includegraphics[scale=0.35]{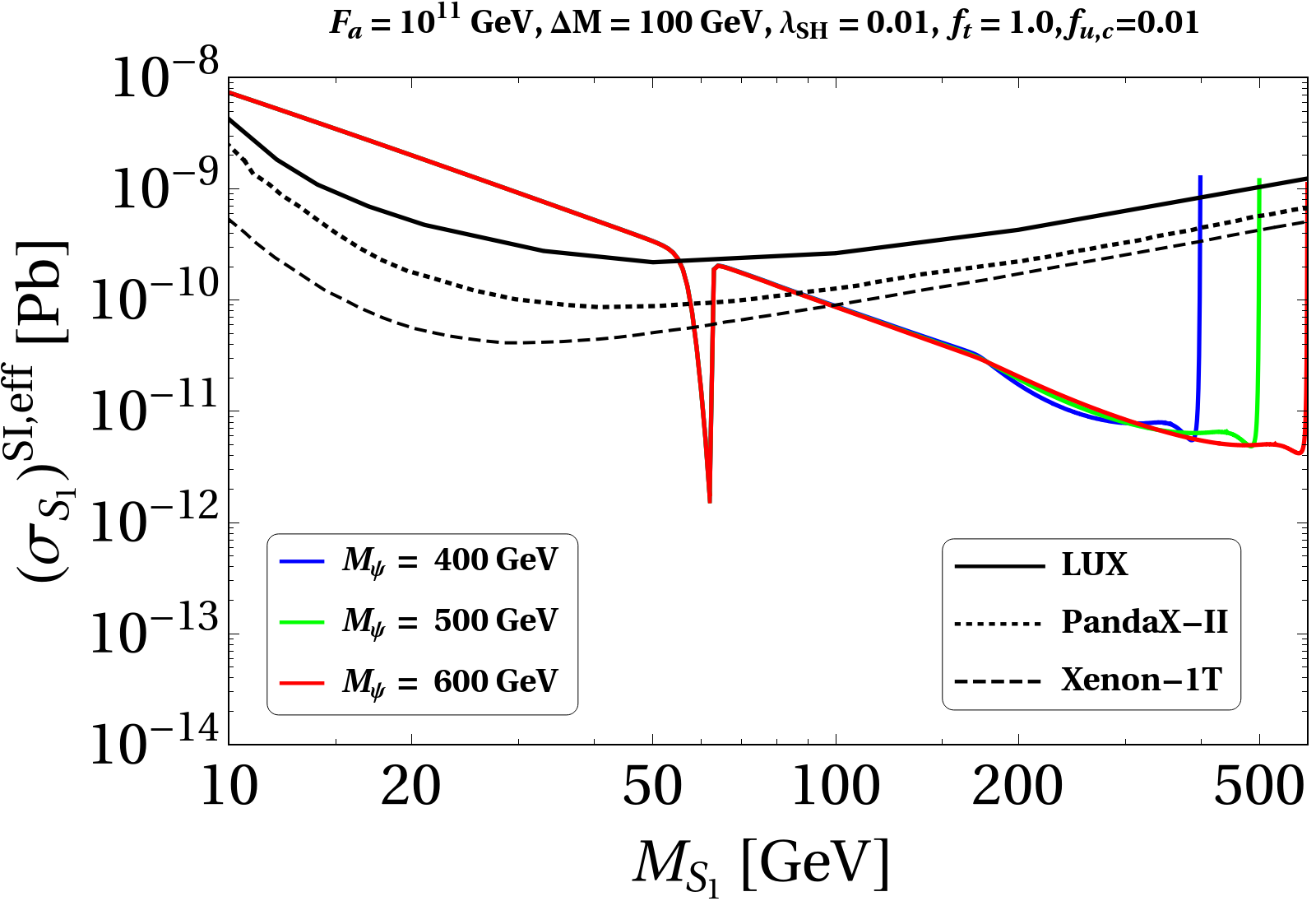}
\includegraphics[scale=0.35]{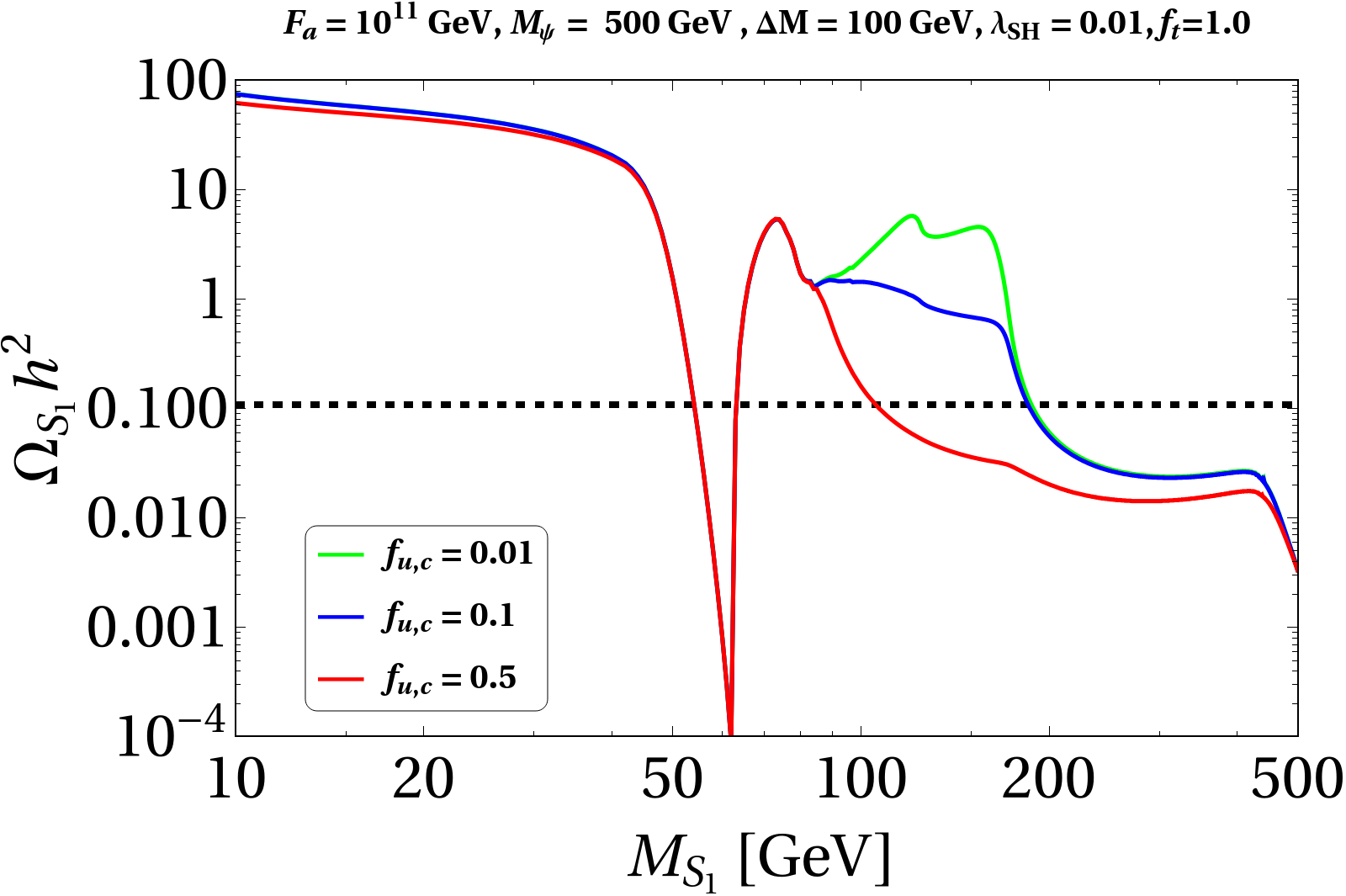}
\includegraphics[scale=0.35]{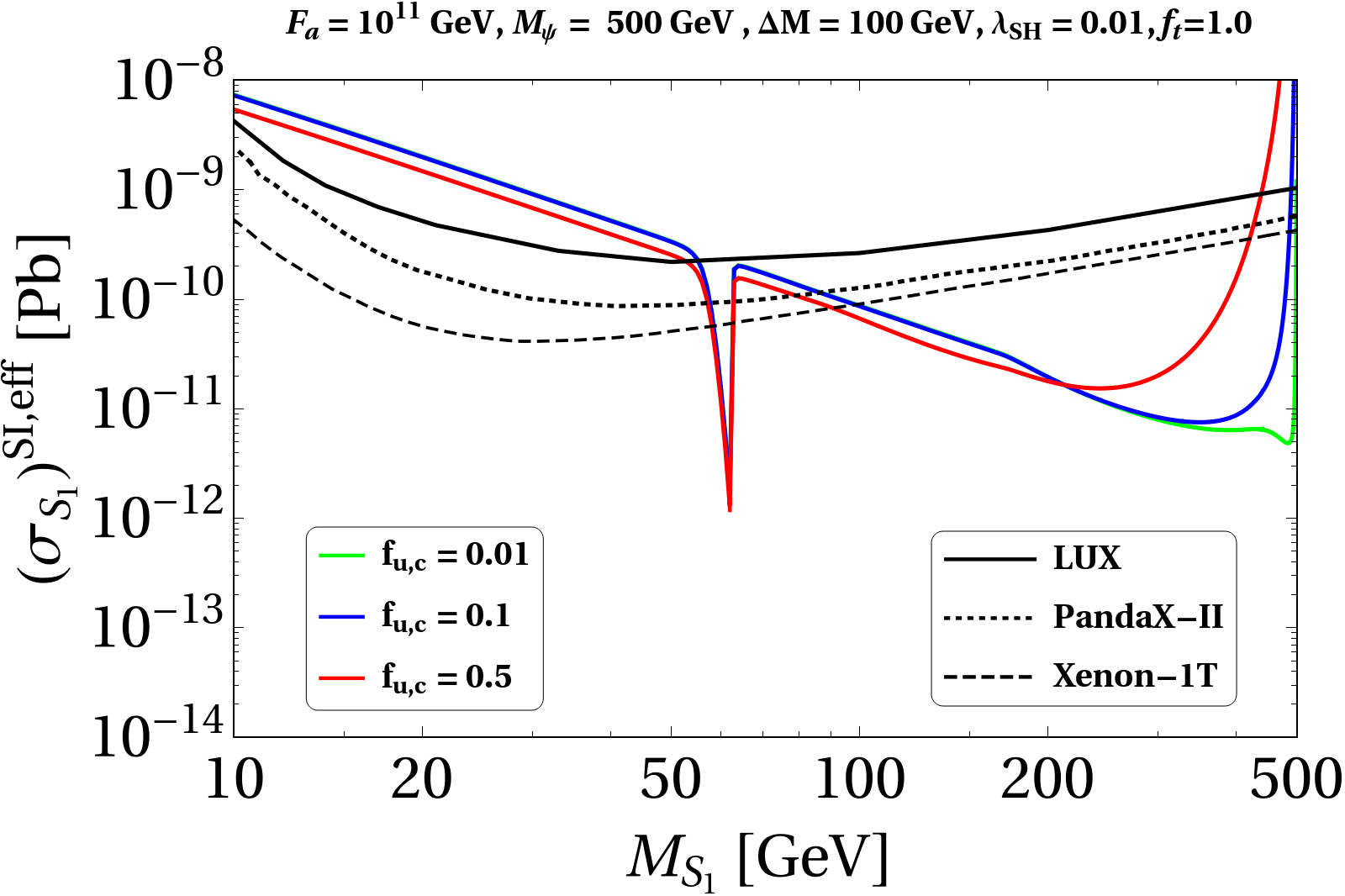}
\includegraphics[scale=0.35]{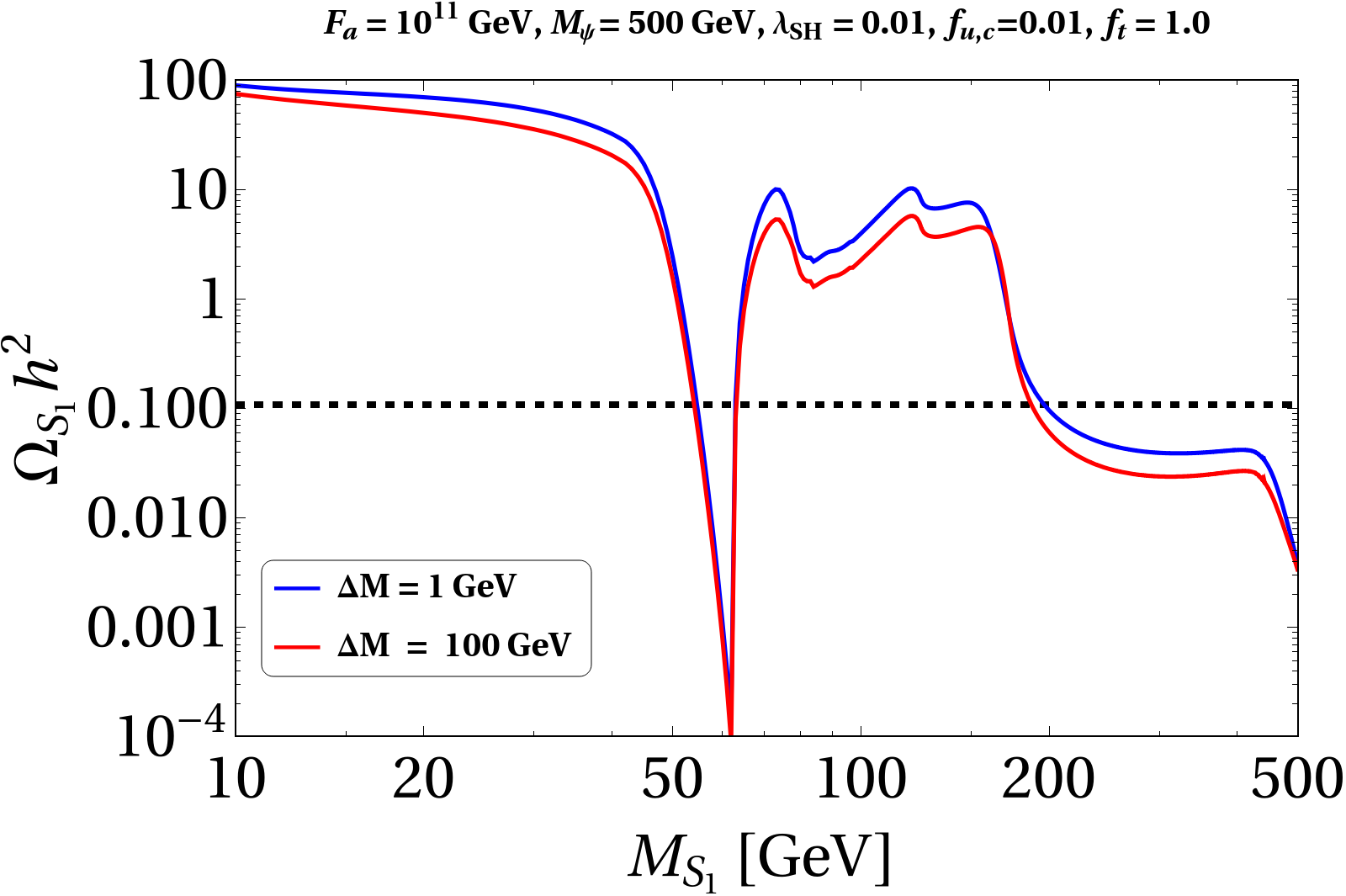}
\includegraphics[scale=0.35]{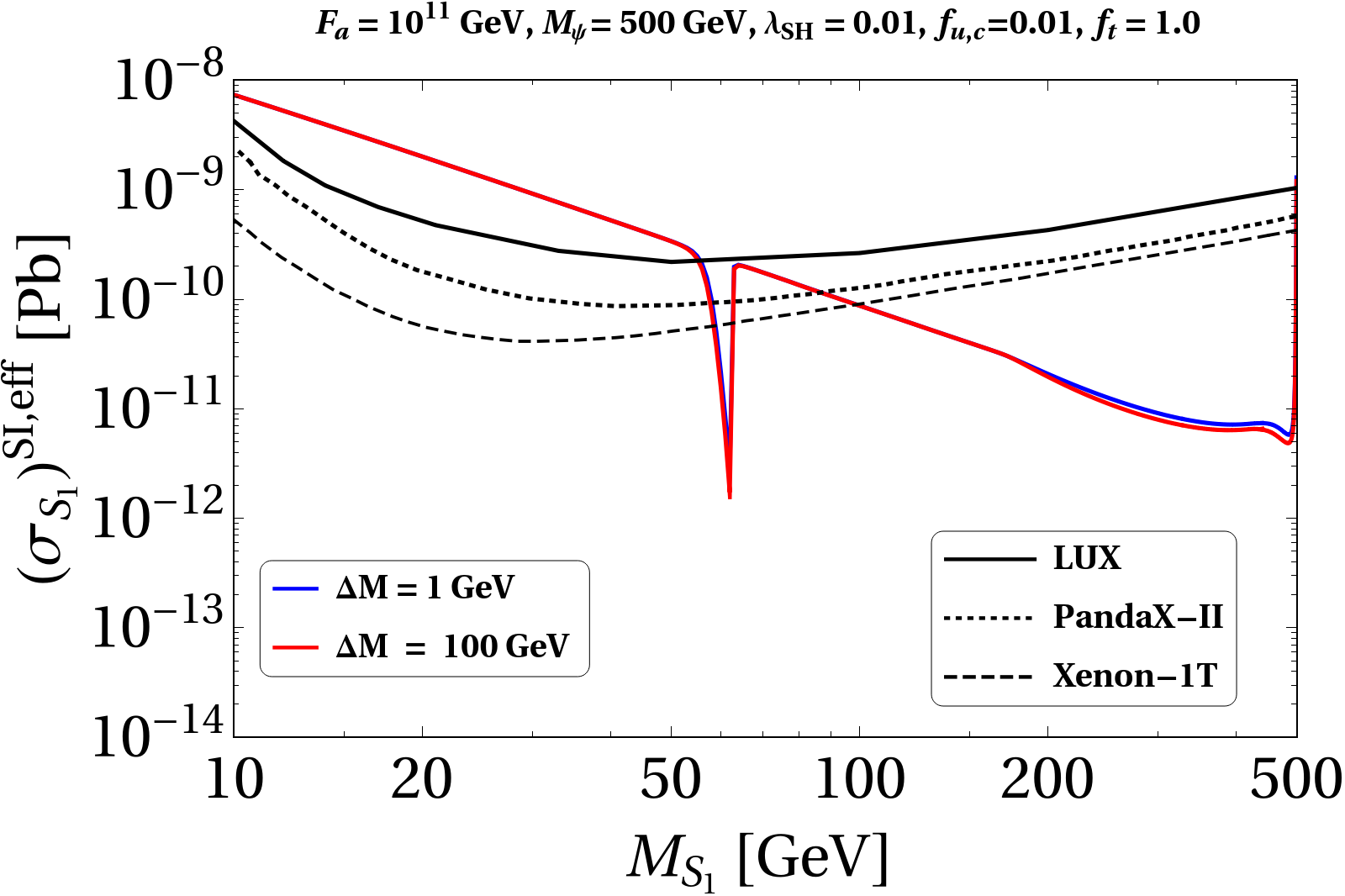}
\caption{Variation of $\Omega_{S_1}h^2$ (left panel) and $\sigma_{S_{1,\text{eff}}}^{\text{SI}}$ (right panel) versus dark  matter mass $M_{S_1}$. In all the plots we fix $F_a=10^{11}~\text{GeV},~\l_{SH}=0.01$. The Black dashed line in all the left plots corresponds to $0.120-\Omega_ah^2$.
}
\label{relic_S_up}
\end{figure}

\begin{figure}[H]
\centering
\includegraphics[scale=0.35]{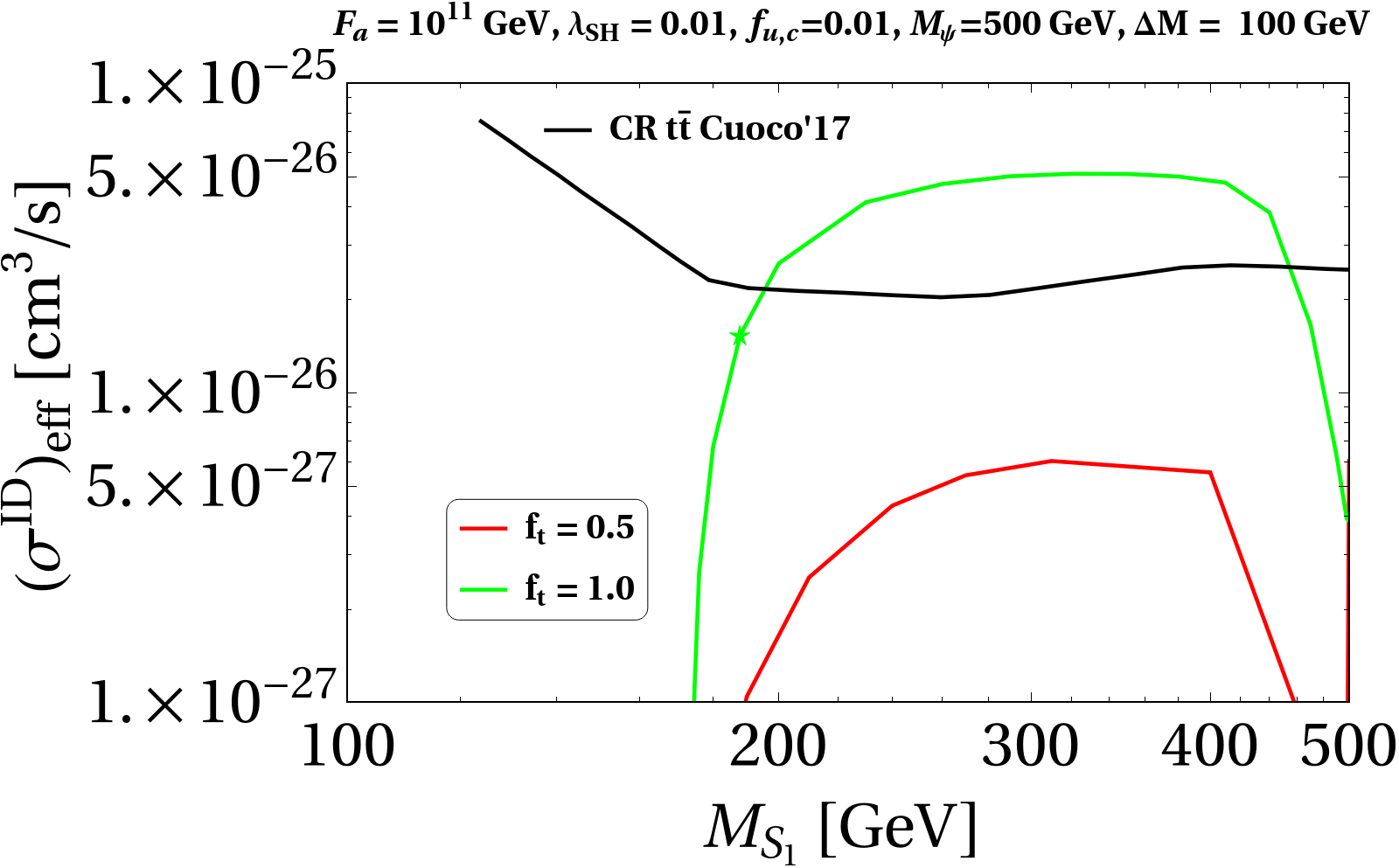}
\includegraphics[scale=0.35]{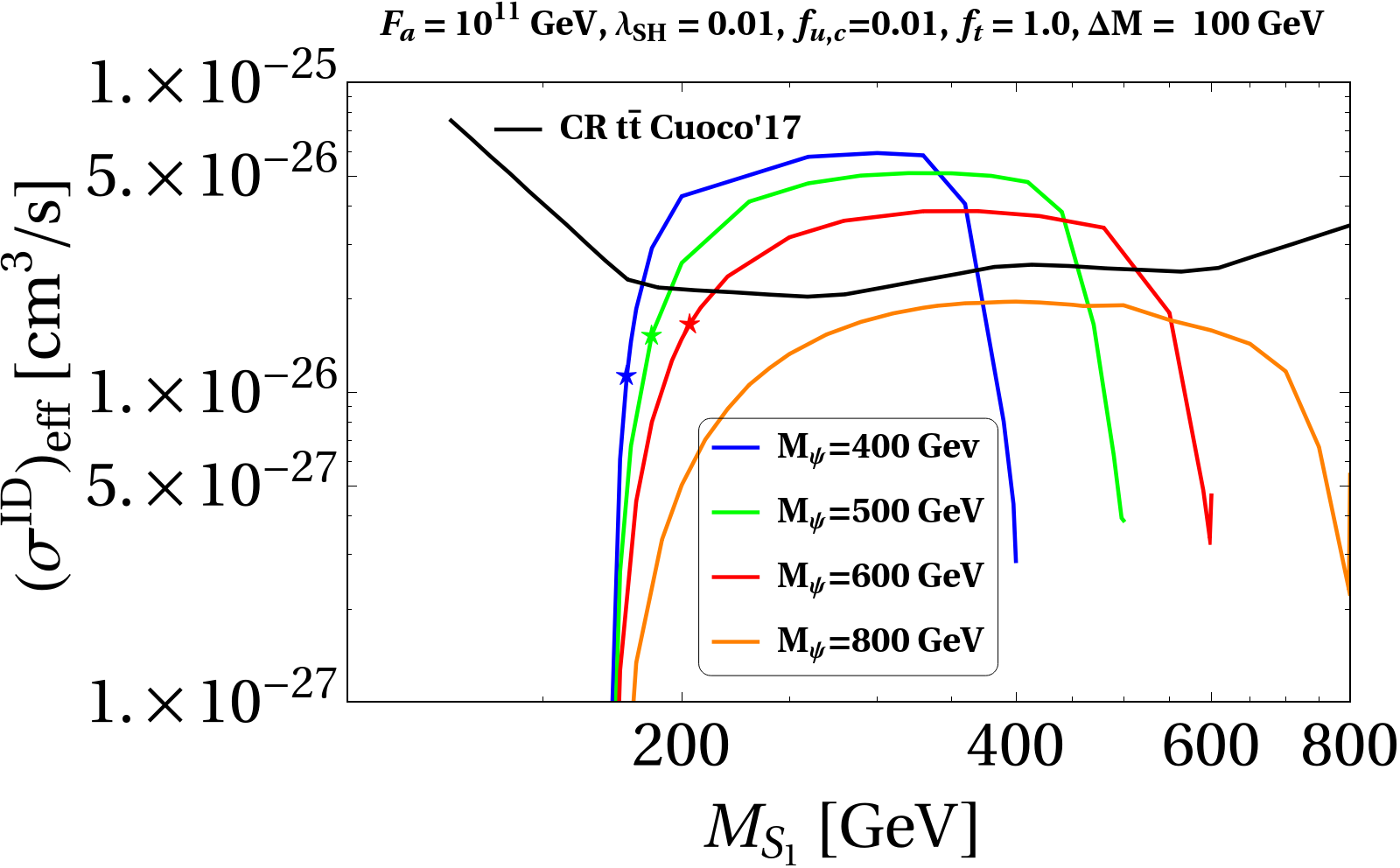}
\includegraphics[scale=0.35]{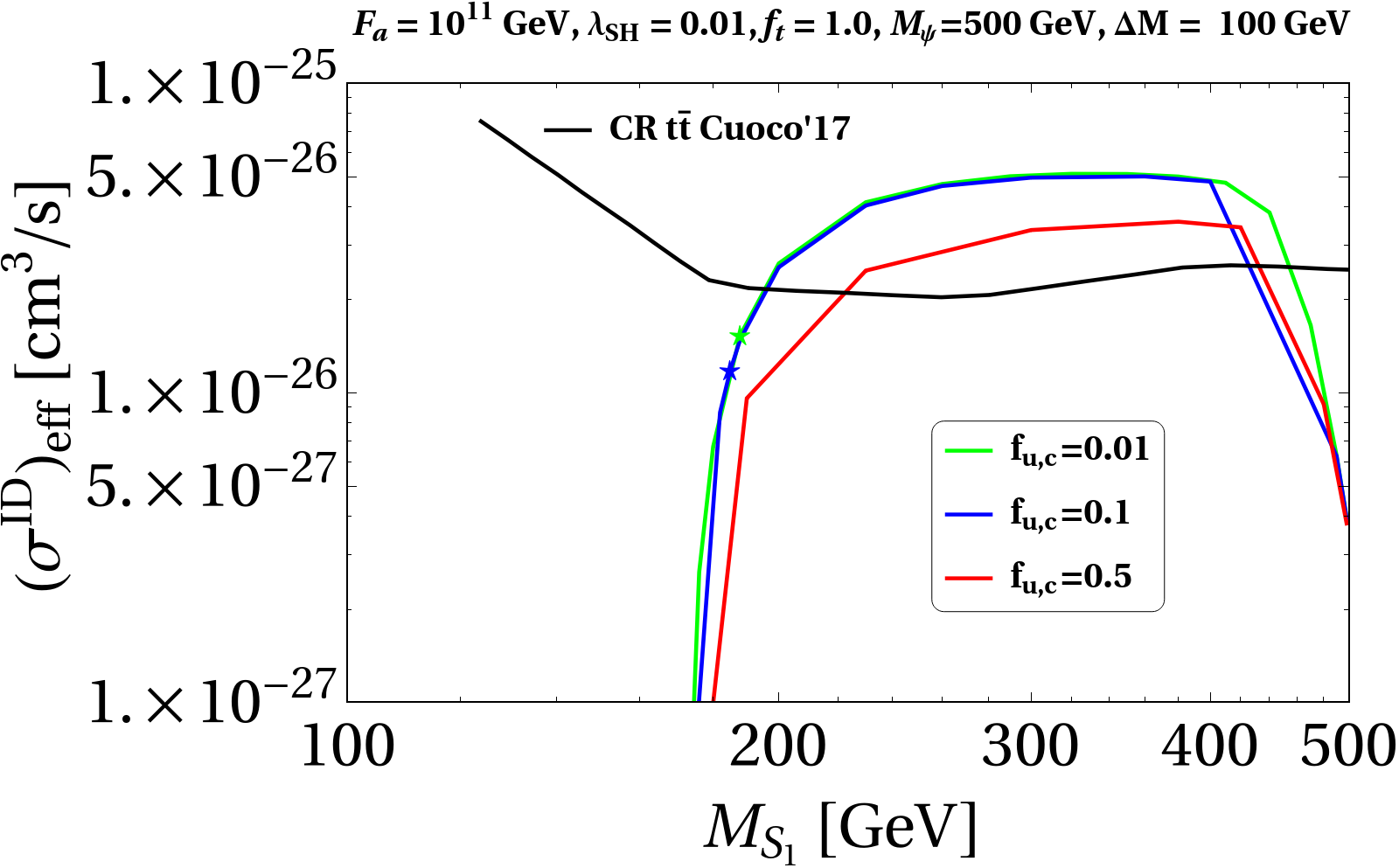}
\includegraphics[scale=0.35]{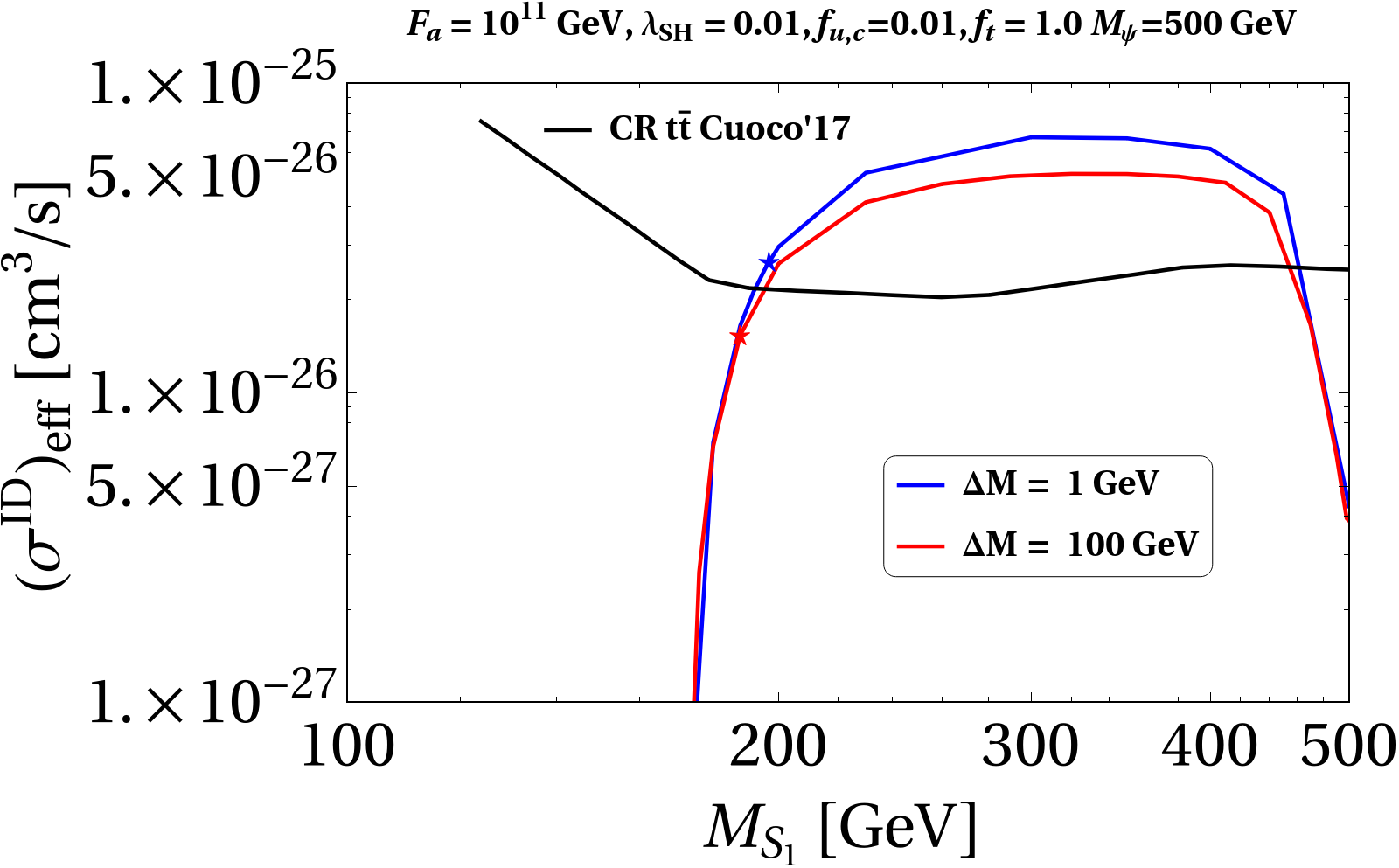}
\caption{Variation of effective indirect detection cross-section versus DM mass. Variation for different values of $f_t$, the mass of VLQ, $f_{u,c}$, and $\Delta M$ are shown in top left, top right, bottom left and bottom right panels respectively. In all the plots we fix $F_a=10^{11}~\text{GeV},~\l_{SH}=0.01$. The solid black line shows the experimental upper bound in $t\bar{t}$ final state.
}
\label{ID}
\end{figure}

To demonstrate the above discussions, we display the variation of the relic density of $S_1$ with its mass in all the left panel plots of Fig. \ref{relic_S_up}. In the right panel, we also exhibit the variation of the effective direct detection cross-section with $M_{S_1}$ for different choices of parameters. In the top left panel of Fig. \ref{relic_S_up}, we project the importance of the Yukawa coupling $f_t$ while choosing fixed values of  $\Delta M =100$ GeV, $M_{\Psi}=500$ GeV and $\l_{SH}=0.01$. It is interesting to point out that for $\l_{SH}=0.01$, the DM does not satisfy the correct relic density in a pure scalar singlet DM scenario. In these plots, we also set both Yukawa couplings $f_u=f_c=0.01$ to highlight the importance of the top Yukawa coupling $f_t$ for three values of $f_t$: 0.1 (blue), 0.5 (red) and 1.0 (green). Notice that for $f_t=0.1$, with the increase in the DM mass, we first observe a resonance dip at $M_{S_1}=M_h/2$ \footnote{In a lower DM mass regime, DM annihilating to the three body final states $q \bar{q} g$ can also contribute significantly towards the relic density, where chirality suppression in the lower order process is lifted by final state radiation. We do not consider this three-body final state in our analysis as the entire low mass regime of the DM is already ruled out from the DD searches, as can be seen from the right panels of Fig.~\ref{relic_S_up}.  }, next, a fall is observed at $M_{S_1}=80$ GeV where the annihilation of the DM to the $W^{\pm}$ boson opens up after which the relic density increases with the increase in the DM mass ($\langle \sigma v \rangle \propto 1/M_{S_1}$) and again drops at $M_{S_1}=125$ GeV when the DM starts annihilating into the Higgs boson \footnote{We would also like to point out that for DM mass $M_{S_1}< M_{t}$, DM annihilations to $gg$ (via a box diagram)~\cite{Deppisch:2018eth,Baek:2016lnv} or three-body final states like $tWb$~\cite{Deppisch:2018eth} can also contribute towards the relic density of the DM for a significantly large Yukawa coupling $f_t$. We do not consider these processes in our analysis as we found that these processes remain suppressed for the choice of $f_t$ we are interested in for our analysis. }. Finally, at a larger value of $M_{S_1}$ ($M_{S_1}= 345 $ GeV), when the mass difference between $M_{\Psi}$ and $M_{S_1}$ becomes relatively small, and the effect of DM co-annihilation with the VLQ comes into the picture, and a sharp drop in DM relic density is observed. In this region, although the DM co-annihilates with $M_{\Psi}$, the annihilation of $\Psi$ with $\bar{\Psi}$ to the gluons dominates the effective annihilation cross-section (see Eq. \ref{eff_cs}) of the DM. Due to these co-annihilations, the relic density dark matter finally satisfies the condition $0.120-\Omega_ah^2$ (denoted by the black dashed line) at $M_{S_1}=444$ GeV. Further increasing $f_t$ to the higher values like 0.5 and 1.0, one notices that the DM annihilations to top quarks mediated by $M_{\Psi}$ (see appendix~\ref{appendixFR}) start to dominate once the threshold of $M_{S_1}=M_{top}$ is crossed and the relic density can also satisfy the condition $0.120-\Omega_ah^2$ near about $M_{S_1}\simeq 200$ GeV for $f_t=1.0$. This figure illustrates the importance of VLQ in the DM phenomenology of the present setup as it makes huge parameter space allowed from the relic density, which was originally disallowed in the pure scalar singlet DM scenario. 

In the top-right panel, we plot the spin-independent effective direct detection cross-section of $S_1$ with the DM mass and compare it with the experimental results. Notice that only the coupling $f_u$ enters the direct detection cross-section of DM apart from the Higgs-portal coupling $\l_{SH}$. Note that the reduced values in the effective direct detection cross-section are the result of the rescaling in the two-component DM system (see Eq.~\ref{singletdd}). Additionally, Higgs resonance dip at $M_{S_1}=M_h/2$ and rise at $M_{S_1}\simeq 500$ GeV because of the interference among the direct detection diagrams (see appendix~\ref{appendixFR}). Here, one notices that the near resonance region remains discarded from the experimental bounds. Still, the other regions where relic density can be satisfied remain allowed from the direct detection searches.

In the left panel of the second row in Fig. \ref{relic_S_up} we fix $f_t=1.0$ and then study the effect of varying $M_{\Psi}$ in $\Omega_{S_1}h^2-M_{S_1}$ plane. As expected, the final fall in the relic density pattern happens at three different positions corresponding to the three different values of $M_{\Psi}$. With the heavier propagator mass $i.e.~M_{\Psi}=600$ GeV, the effective annihilation cross-section of the DM to the top quark remains smaller in comparison to what is observed for $M_{\Psi}=400$ GeV and hence relatively larger relic density is observed for $M_{\Psi}=600$ GeV than for $M_{\Psi}=400$ GeV. Similarly, in the left panel of the third row, we depict the effect of varying $f_{u}~\text{and}~f_{c}$. For simplicity, here we also assume $f_{u}~=~f_{c}$. As expected, for a large value of $f_{u,c}$ the annihilation of DM to top and up (charm) quark final state also becomes dominant the moment the threshold $2M_{S_1}=M_{top}+M_{u(c)}$ is achieved. This leads to an increase in the DM annihilation cross-section, and consequently, a decrease in the relic density is observed. Next, in the left panel of the fourth row we show the effect of varying $\Delta M$ on the relic density of the DM. As expected, a smaller $\Delta M$ results in a larger effective DM annihilation cross-section and hence a smaller relic, so in order to satisfy the correct relic density a heavier DM mass is required. On the other hand, a larger $\Delta M $ requires a smaller DM mass to satisfy the observed relic density and hence the plot shifts towards the lower DM mass. Finally, the middle and bottom right panel of Fig. \ref{relic_S_up} can be followed from the one observed in the top right panel.

As can be seen from Fig.~\ref{relic_S_up}, the correct relic density is mostly satisfied in the parameter space where $M_{s_1}>M_t$. Hence, one needs to check the prospects of indirect detection of DM in our model specifically focusing on $t\bar{t}$ final states from the DM annihilations. We display our findings in Fig.~\ref{ID} where we plot the effective indirect detection cross-section with DM mass and show the experimental upper bound (black solid line) of DM annihilating to $t\bar{t}$ final states that can be obtained from antiproton cosmic ray data~\cite{Colucci:2018vxz}. In the top left panel of Fig.~\ref{ID}, we find that the DM mass for which the observed relic density satisfied (shown by green $\star$) in the top left panel of Fig.~\ref{relic_S_up} is also allowed from the constraints coming from the indirect search bound. A similar situation is also observed in the top right and bottom left panels. On the other hand, in the bottom right panel where the variation with $\Delta M$ is studied, it found that a larger $\Delta M\sim100$ GeV is prefered if one also considers the constraints coming from indirect search experiments. For this reason, we fix $\Delta M=100$ GeV throughout our analysis.

\begin{figure}[tb!]
\centering
\includegraphics[scale=0.46]{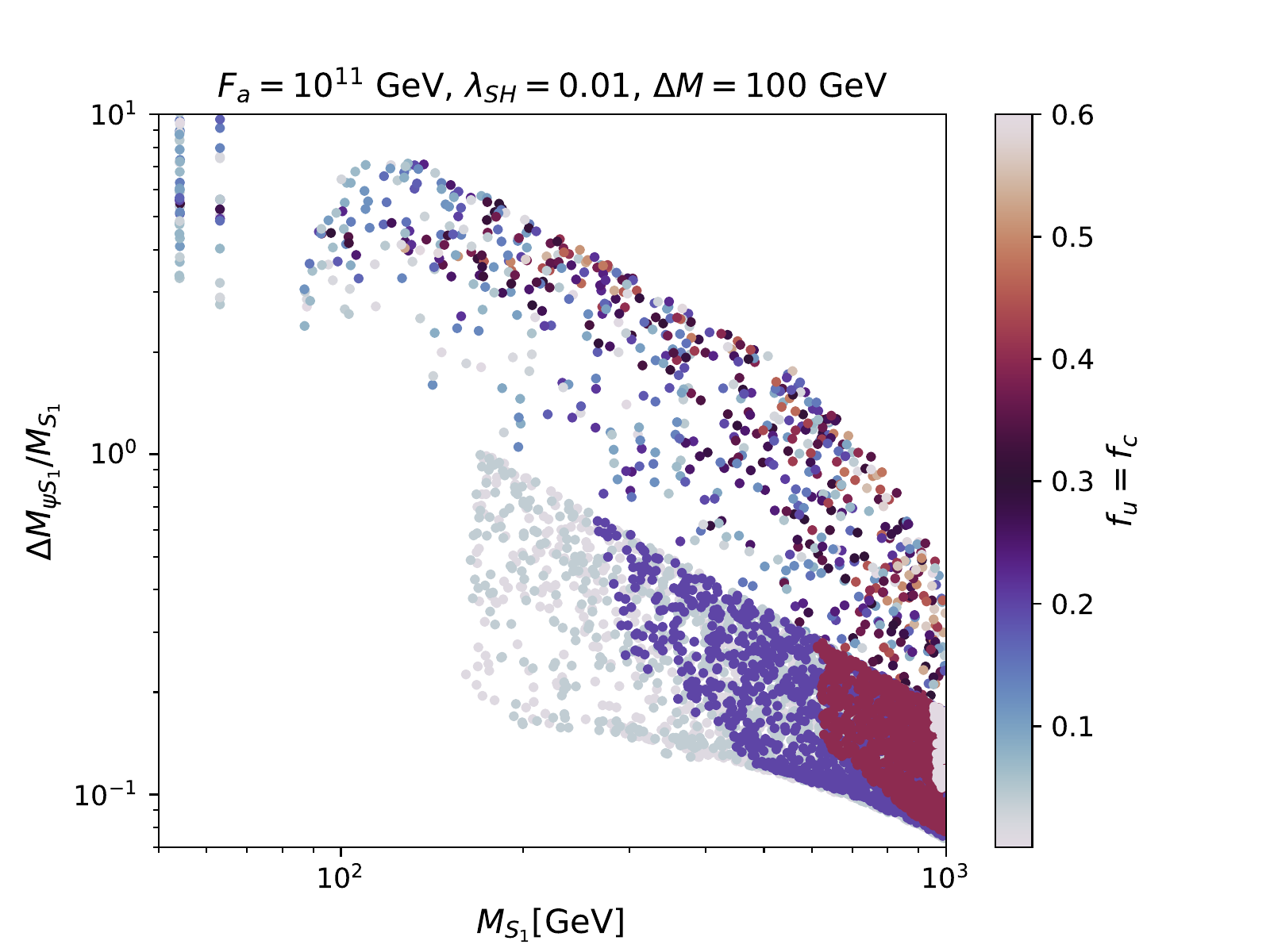}
\includegraphics[scale=0.46]{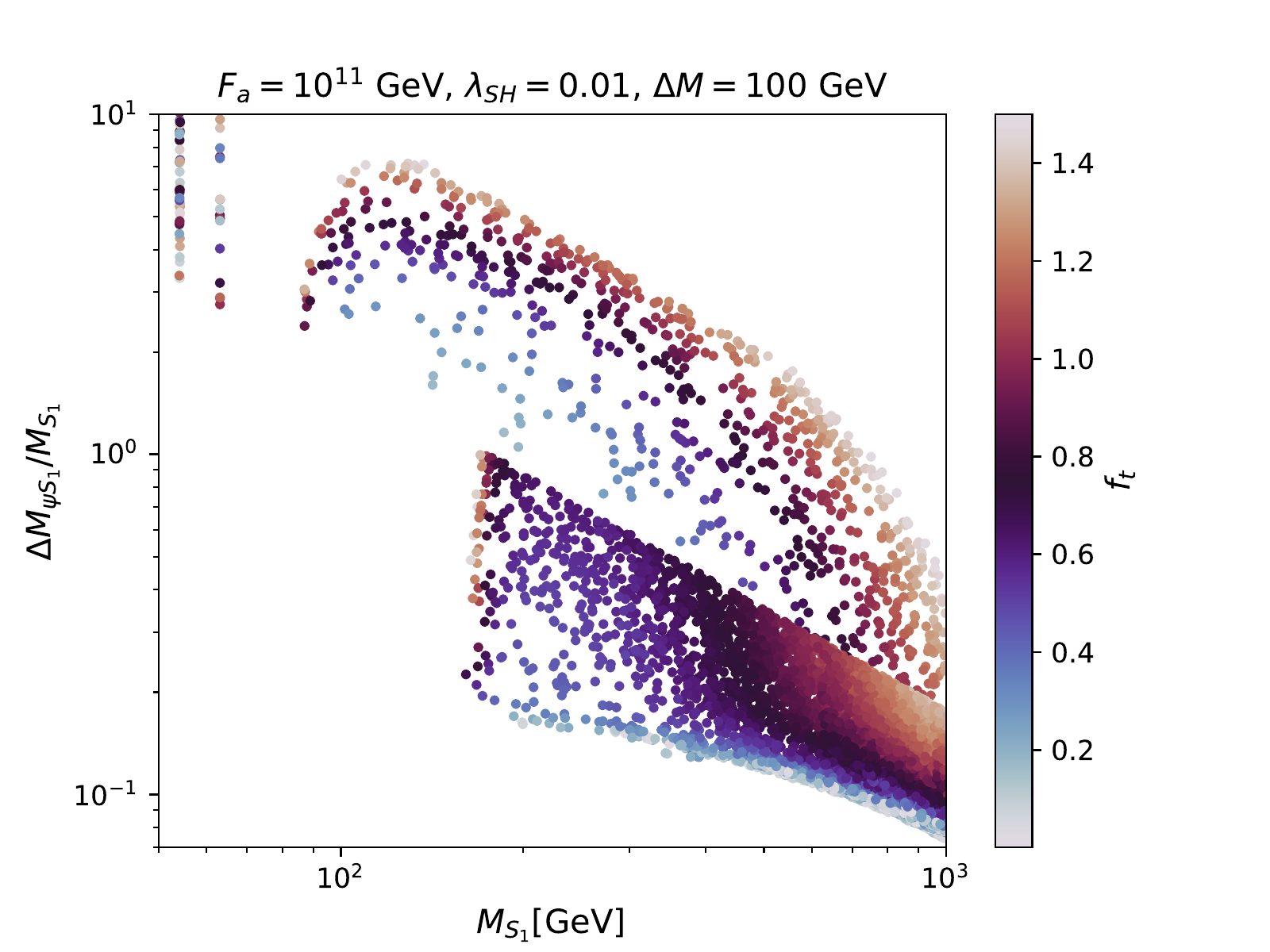}
\caption{Parameter space satisfying observed DM abundance and also allowed by the direct search experiments in the bi-dimensional plane of $\frac{\Delta M_{\Psi S_1}}{M_{S_1}}-M_{S_1}$, where the colour coding is done with respect to: Left Panel:  the Yukawa couplings $f_{u}=f_c$ and Right Panel:  the Yukawa coupling $f_{t}$. In both the plots we fix $F_a=10^{11}~\text{GeV},~\l_{SH}=0.01,~ \Delta M=100$ GeV while we vary $f_{t}$  in the range 0 - 1.5 and $f_u=f_c$ in the range 0 - 1.5.}
\label{scan}
\end{figure}

In Fig. \ref{scan}, we show the parameter space that remains consistent with the DM constraints and is also allowed by the constraints that come from the flavor observable like $D^0-\bar{D}^0$ mixings  in the bi-dimensional plane of $\Delta M_{\Psi S_1}/M_{S_1}~Vs~M_{S_1}$, where $\Delta M_{\Psi S_1}$ is the mass difference between VLQ and DM, $M_{\Psi}-M_{S_1}$. Here, the dependence upon different Yukawa couplings ($f_u=f_c$ in the left panel and $f_t$ in the right panel of the figure) is spotted with a continuous color map. Two discrete narrow slices at the top-left corner due to Higgs resonance. We are primarily interested in the non-resonant continuous region extended over vast parameter space. At the lower $M_{S_1}$ end, this continuous region opens up when the DM pair annihilate into a top quark and up (charm) quark (see appendix~\ref{appendixFR}). Eventually, for a choice of heavier mass, the DM pair starts annihilating into the top pair.

This allowed region can be categorized into two distinct parts as upper and lower regions separated by a line where the mass difference between VLQ and DM equates to the top mass. Hence the upper region can be probed at the collider with on-shell production of top quark from VLQ decay, while the lower region is sensitive to a probe with light quark search. We will further demonstrate in the next section how top quark searches from the boosted top jet can improve the search strategy in this region.

As a  consequence of the narrow mass gap between scalar DM and VLQ, co-annihilation takes a leading role in most parts of the lower region. Precisely because of the same reason, this region is also susceptible to the direct detection probe. Variations of color contours for different $f_{u}$ values are evident in the lower region of the left plot in Fig. \ref{scan}. This reflects the gradually larger parameter space excluded due to direct detection constrain for a choice of larger $f_{u}$ values. The lower value of mediator mass increases the direct detection cross-section. In order to keep this cross-section below the current direct detection bounds, a smaller $f_u$ is required.
On the contrary, the direct search experiments allow the upper region irrespective of the choice of $f_u$, and hence a uniform distribution of the colors is observed. This is because even for a large $f_u$, the DM-nucleon scattering cross-section still remains small due to the presence of a heavier mediator $i.e~M_\Psi$. 
 
At the right part of the same plot, one finds that with an increasing DM mass, a relatively smaller $\Delta M_{\Psi S_1}/M_{S_1}$ is required in order to satisfy the correct relic density, while the interplay between the DM mass, mediator's mass and the $f_u$ makes these points allowed from the direct detection constraints. In this region, the DM dominantly annihilates into the top-quark pair and sub-dominantly into the top quark and up (charm) quark final states. Next, in the right panel of Fig. \ref{scan}, we show the color coding with respect to $f_t$ in order to highlight its significance. One observes the correct relic density in the top-left region of the plot due to the involvement of a large $f_t$ as is also evident from the top left panel of Fig. \ref{relic_S_up}. As expected for a lighter mediator mass, a relatively smaller $f_t$ is required to satisfy the correct relic density, as is also observed while moving downward in the plot. Finally, the role of $f_t$ becomes more prominent once the $M_{S_1}=M_{top}$ threshold is opened, as can also be seen from the right side of the plot.

\section{Collider Analysis and Results}
\label{collider}
The involvement of VLQ ($\Psi$) in the present setup opens up interesting collider prospects as they can be produced either in pair or associated with a scalar in the proton-proton collision at LHC. Among these production channels,  $p p \rightarrow \Psi \bar{\Psi}$ and 
$p p \rightarrow \overset{ \scalebox{0.4}{$(\mkern-1mu-\mkern-1mu)$}}{\Psi}S_{1,2}$,
the cross-section of the second process strongly depends on the Yukawa coupling $f_{u,c}$, while the pair of VLQs is produced primarily by the strong interaction and hence model-independent.
Once produced, the VLQ can decay preferably into scalar DM candidate or its heavier pair along with one of the up-type quarks as allowed kinematically. Hence primary LHC searches rely on identifying such quark jets along with missing transverse energy (MET or $\slashed{E}_T$) from DM production, as discussed in section \ref{constraints}.

It is noteworthy that a substantial parameter space exists in this model where the mass difference between VLQ and DM is significantly larger than the top quark mass while providing correct dark matter relic density and also allowed from the direct detection experiments. Here produced top quarks are expected to be fairly boosted by production from the decay of heavy mother (VLQ) particles. Such a prospect motivates us to look at this signal possessing a unique topology where hadronic decay of the top retains its collimated structure as a boosted fatjet~\footnote{We encountered a similar feature in the succession of different BSM scenarios \cite{Das:2017gke, Bhardwaj:2018lma, Bhardwaj:2019mts, Bhardwaj:2020llc}, where boosted fatjet is probed in association with MET. Fatjets, in these searches, still harbor the intrinsic footprint of their root and manifest such features inside the jet substructure. Exploring this can provide additional tools to deal with a significant background involving QCD jets.} 
and is identified as a top-like-fatjet ($J_t$).

\begin{figure}[tb!]
\centering
\includegraphics[scale=0.45]{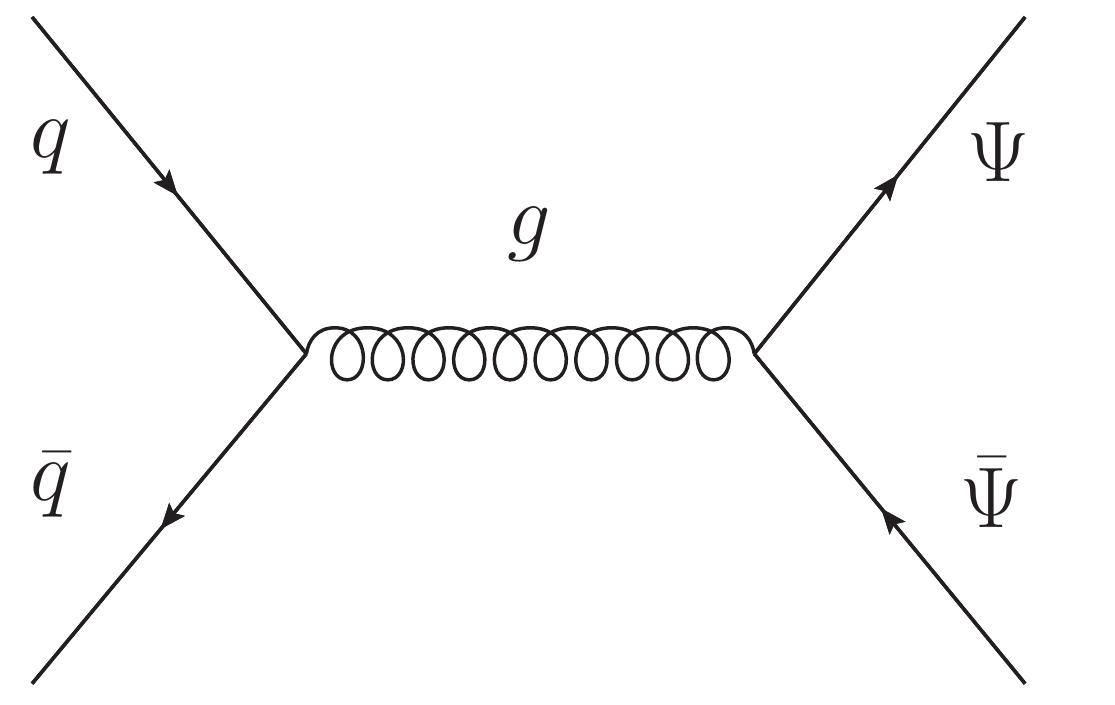}
\includegraphics[scale=0.45]{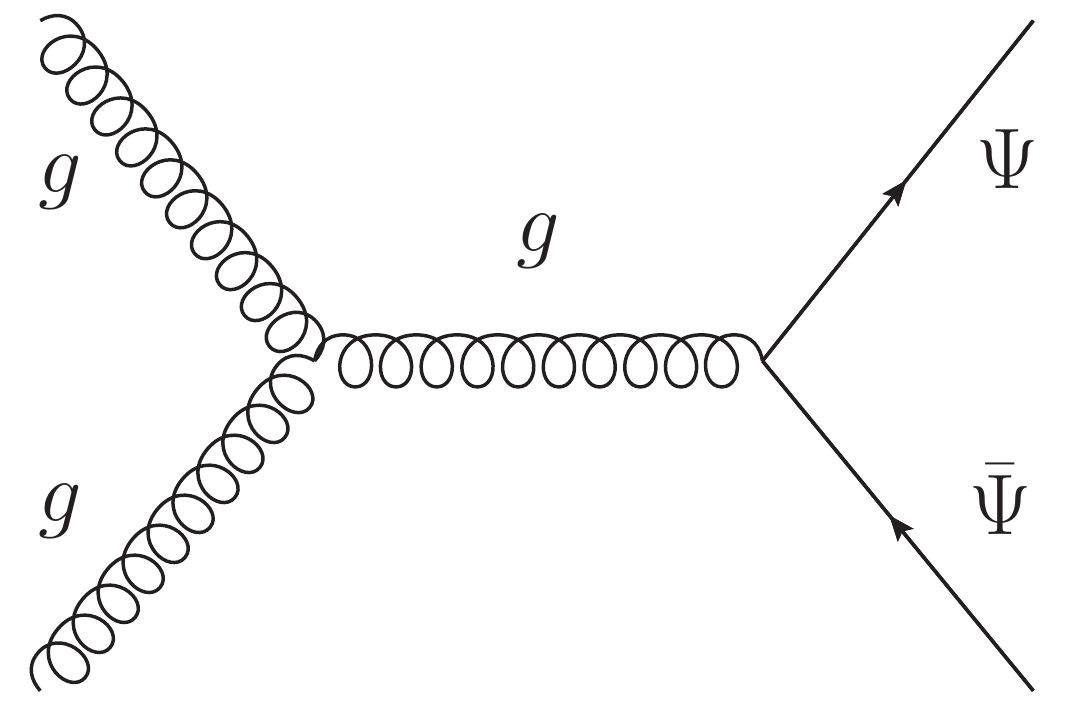}
\includegraphics[scale=0.45]{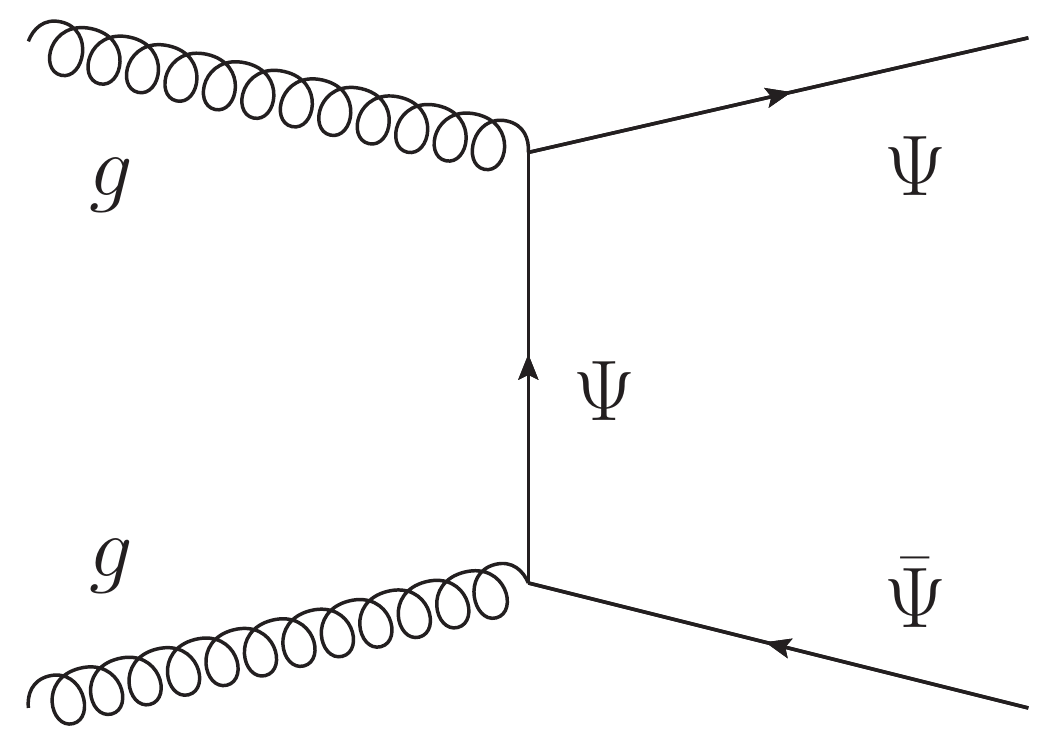}\\
\includegraphics[scale=0.45]{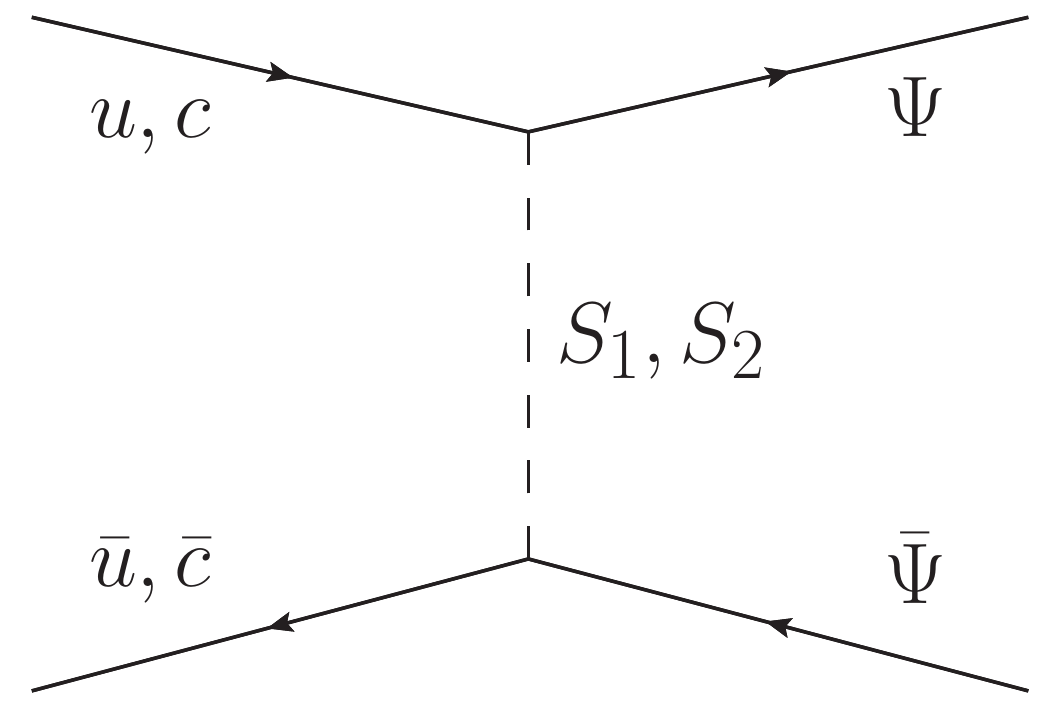}
\includegraphics[scale=0.45]{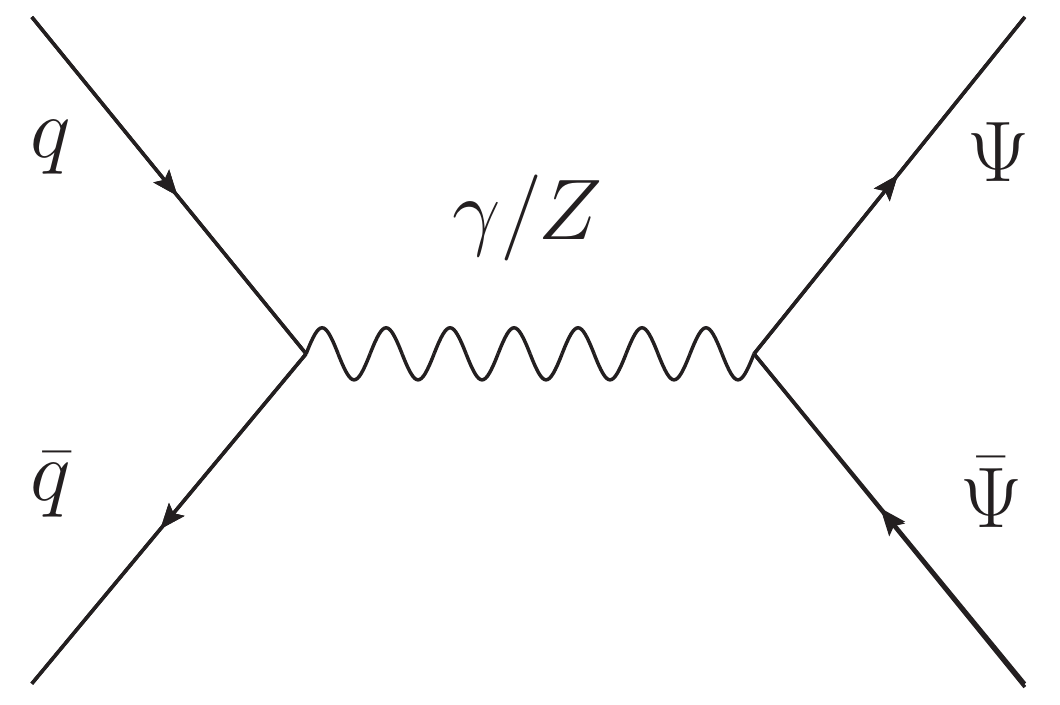}
\caption{Representative Feynman diagrams for leading order partonic processes contributing VLQ pair production $p p \rightarrow \Psi \bar{\Psi}$ at LHC.}
\label{FD_production}
\end{figure}

To probe these regions at the LHC, we consider pair production of VLQs and each of those further decay into the top quark associated with the scalar (DM or $S_2$). Here we adopt a significantly smaller Yukawas $f_u=f_c$ ($=0.01$) so that the primary branching fraction of the decay of $\Psi$ into the top quark is close to $100\%$. The signal topology is below, where we identify two final top fatjets associated with significant missing transverse momentum from dark matter.
\begin{equation}
p p \rightarrow \Psi \bar{\Psi}\rightarrow (t,S_{1,2}), (\bar{t},S_{1,2}) \equiv 2J_t +\slashed{E}_T
\end{equation}
Note, $S_2$ can decay through two-body $(S_2\rightarrow S_1 a)$, three-body $(S_2\rightarrow S_1 \hspace{0.5mm} j \hspace{0.5mm} j)$, and four-body $(S_2\rightarrow S_1 \hspace{0.5mm} j \hspace{0.5mm} b \hspace{0.5mm} W)$ decay modes, where suppressed multibody decay occurs through off-shell VLQ. Partonic level Feynman diagrams of the production of VLQ pair are shown in Fig. \ref{FD_production}. Although the main contribution comes from the strong interaction, we keep all the diagrams for completeness. 
Few representative benchmark points (BPs) are listed in Tab. \ref{parameters}; those provide observed relic density of DM and allowed from the direct and indirect detection experiments along the constraints coming from the theoretical and LHC data as listed in section \ref{constraints}.
Also, the production cross-section of the partonic process $pp\rightarrow \Psi \bar{\Psi}$ at LO for different benchmark points before decaying into SM quark and scalar at 14 TeV LHC is shown at the last column Tab. \ref{cross-section}. For our analysis, we have used an NLO QCD  $K-$factor of 1.33 for the $pp\rightarrow \Psi \bar{\Psi}$ production~\footnote{
We estimate an approximate NLO (QCD) $K-$factor for the process $pp\rightarrow \Psi \bar{\Psi}$ by replacing $\Psi$ with the top quark of mass $m_\Psi$ at the {\sc MadGraph5\_aMC@NLO} and took the most conservative value over this mass range.}.

\begin{table}[tb!]
\begin{center}
 \begin{tabular}[b]{|c|c|c|c|c|c|c|c|}
\hline\hline
Benchmark& $M_{S_1}$& $\Delta M_{\Psi S_1}$ & $\Delta M$ & $f_t$ & $\Omega_{S_1} h^2$ & $\sigma^{SI}_{S_1, eff}$ & $\sigma(pp\rightarrow \Psi \bar{\Psi} )$ \\ 
 points   & (GeV)        &      (GeV)                      &         (GeV) &   &  &  (pb) & (fb) \\%
\hline\hline
BP1 & 301 & 305& 100 & 0.8 & 0.108 & $9.24 \times 10^{-12}$ & 966\\
\hline
BP2 & 302 & 475 & 100 & 1 & 0.104 & $9.96 \times 10^{-12}$ & 223\\
\hline
BP3 & 403 & 405 & 100 & 1 & 0.109  & $5.77 \times 10^{-12}$ & 175\\
\hline
BP4 & 358 & 448 & 100 & 1 & 0.109  & $6.48 \times 10^{-12}$ & 177\\
\hline
BP5 & 433 & 364 & 100 & 1 & 0.107  & $4.6 \times 10^{-12}$ & 188\\
\hline
BP6 & 459 & 326 & 100 & 1 & 0.109  & $4.2 \times 10^{-12}$ & 208\\
\hline
BP7 & 494 & 273 & 100 & 1 & 0.107  & $3.79 \times 10^{-12}$ & 239\\
\hline
BP8 & 510 & 238 & 100 & 1 & 0.099  & $3.69 \times 10^{-12}$ & 278\\
\hline
BP9 & 527 & 224 & 200 & 1 & 0.103  & $3.56 \times 10^{-12}$ & 272\\
\hline
BP10 & 542 & 188 & 100 & 0.98 & 0.106  & $3.57 \times 10^{-12}$ & 321\\
\hline
 BP11 & 678 & 349 & 100 & 1.3 & 0.109  & $2.14 \times 10^{-12}$ & 37 \\
\hline
 \end{tabular}
 \caption{Different benchmark points satisfy the observed relic density of DM, direct and indirect detection (not shown in the table) bounds, along with the constraints coming from the theoretical and LHC data, as listed in the text. $M_{S_1}$ is the mass of the DM, $S_1$. $f_t$ is the coupling strength of the interaction between top quark, VLQ, and the scalar $S$ (see: Eq. \ref{Lag}). $\Delta M_{\Psi S_1}=M_\Psi -M_{S_1}$ and $\Delta M=M_{S_2}-M_{S_1}$. $\Omega_{S_1} h^2$ (see: Eq. \ref{totalRelic}) and $\sigma^{SI}_{S_1, eff}$ (see: Eq. \ref{singletdd}) are the relic density of DM, $S_1$ and effective direct detection cross-section, respectively. Other parameters are $F_a=10^{11}~\text{GeV},~\l_{SH}=0.01,~\text{and}~ f_{u}=f_{c}=0.01$. The production cross-section of the partonic process $pp\rightarrow \Psi \bar{\Psi}$ at LO for different benchmark points before decaying into SM quark and scalar at 14 TeV LHC is given at the last column.
}\label{parameters}  \label{cross-section}
\end{center}
\end{table}

\subsection{Simulation Details with Signal and Backgrounds}
\label{setup}

In preparation for our investigation of this Hybrid KSVZ framework through VLQ pair production at the LHC, we require a realistic setup to simulate both the signal processes as well as a careful selection of background processes that can mimic the signal. 

We implement this Hybrid KSVZ framework in {\sc FeynRules} \cite{Alloul:2013bka} to generate the UFO model file required for matrix element generation for Monte-Carlo event generator. Parton level events are generated in the {\sc MadGraph5\_aMC@NLO} environment \cite{Alwall:2014hca} and further pass through {\sc Pythia8} \cite{Sjostrand:2001yu, Sjostrand:2014zea} for showering, fragmentation and hadronization.
Background events are generated along with two to four additional jets {\sc MLM} matching \cite{Mangano:2006rw, Hoeche:2005vzu} with virtually-ordered Pythia showers to avoid any double counting. We include higher-order corrections for different processes by multiplying the appropriate $K$ factor. An in-built NN23LO1 pdf set is adopted for the parton distribution functions (PDF), and a default dynamical factorization scale is used for events generation. 
The showered events are further passed through {\sc Delphes3} \cite{deFavereau:2013fsa} to include detector effects with the default  CMS card. Jets ($j$) of radius parameter 0.5 are constructed with the $\mbox{anti-k}_T$ \cite{Cacciari:2008gp} clustering algorithm, where we used the particle-flow towers and particle-flow tracks as input. We implement the Cambridge-Achen (CA) \cite{Dokshitzer:1997in} algorithm to construct large radius fatjets - $J$. {\sc Fastjet 3.2.2} \cite{Cacciari:2011ma} is used for clustering fatjets of radius parameter $R=1.5$. A boosted top gives a fatjet whose radius parameter is approximately govorned by $R \sim 2m_t/P_T$, where $m_t$ ($P_T$) is the mass (transverse momentum) of the top quark. Hence, the minimum transverse momentum required by each top to form such a fatjet is $P_T \gtrsim 200$ GeV. Finaly, we implement the adaptive Boosted Decision Tree (BDT) algorithm to perform the multivariate analysis (MVA) in the {\sc TMVA} \cite{Hocker:2007ht} framework.

Our analysis considers all the backgrounds that significantly contribute to the two boosted top fatjets with large missing transverse momentum, as listed below.\\

\noindent
\underline{$t \bar{t}+$ jets:} Top pair production with the semi-leptonic top decays is the most dominant background for our signal process. Although pure hadronic decay of tops can offer two boosted top jets, the requirement of a considerable amount of missing energy reduces this background by a significant factor of $~100$, where mismeasurement of hadronic activities acts as a source of MET. In the semileptonic decay, one top decay hadronically and is reconstructed as boosted top jet, and the other top decay leptonically gives a significant amount of missing energy when the lepton escapes detection. Other boosted jet comes from the QCD radiation. This background is matched with the MLM matching scheme up to two extra jets.\\

\noindent
\underline{QCD background:} QCD background is enormous at the LHC but can be reduced to a negligible contribution (see, for example,  Ref.\cite{Plehn:2010st}). Even after the requirements of two boosted fatjets, we are left with a remarkably large number of events from this background. We further require at least one b tag within the leading or sub-leading fatjet. Contribute negligibly after additional suppression of ~100 comes from fake MET from hardons and another ~50 from the requirement of $b$  tag fatjet. We do not include this background in our analysis.\\

\noindent
\underline{$tW+$ jets:} Single top production associated with W boson significantly contributes to the SM background. The top is reconstructed as the boosted top where the b quark is tagged within it, and the W boson decays leptonically to give rise to missing transverse momentum. In contrast, another boosted fatjet arises from QCD jets. MLM matching up to two extra jets is done for this process.\\

\noindent
\underline{$V+$ jets:} (Simi-)invisible decay of W/Z vector boson in addition to QCD radiation that emulates the fatjet can contribute sizably even with a requirement of sizeable reconstructed mass of the fatjet. We do MLM matching up to four extra jets for both processes. A generation level cut $\slashed{E}_T>100$ GeV is applied for both processes to obtain statistically significant background events.\\

\noindent
\underline{di-boson + jets:} Minor contribution can come from Di-boson + jets. We retain all the three di-boson background processes ($pp \rightarrow WZ,\hspace{1mm} WW,\hspace{1mm} ZZ$) in our analysis. Among the three, WZ + jets contribute the most. All three processes are matched up to two extra jets with an MLM matching scheme. In all the cases, one of the vector boson decay invisibly $(Z \rightarrow \nu \nu) $ or leptonically $(W \rightarrow l \nu)$ to give $\slashed{E}_T$. One of the boosted fatjet comes from QCD jets, and another fatjet comes from either hadronically decaying vector boson or the QCD jets.\\

\noindent
\underline{$t \bar{t} V$:} Such processes have three body phase spaces and have less cross-section than other background processes mentioned above. Both the tops can be reconstructed as boosted fatjets, while $\slashed{E}_T$ comes from the invisible or leptonically decay of the Z and W boson, respectively. Among these two, $t \bar{t} Z$ contributes the most because of the larger cross-section and more significant efficiency when applying $\slashed{E}_T$. 

We consider all contributions generating those events at leading order and normalize with the NLO (QCD) cross-section. Higher-order QCD corrected production cross-section at the 14 TeV LHC for different background processes accounted in this analysis are listed in Tab. \ref{cross-section:BG}.

\begin{table}[tb!]
\begin{center}
 \begin{tabular}{|l|l|r|}
\hline
 \multicolumn{2}{|c|}{ Background}   & $\sigma$ (pb)   \\ 
\hline
top pair \cite{Muselli:2015kba} & $t \bar{t}+$ jets & 988.57 [$N^3$LO] \\
\hline
single top \cite{Kidonakis:2015nna}  & $t W$  & 83.1 [$N^2$LO] \\
\hline
\multirow{2}{*}{mono-V boson \cite{Catani:2009sm, Balossini:2009sa} } & $Z+$ jets  & $6.33\times 10^4$ [$N^2$LO] \\
\cline{2-3}
 & $W+$ jets  & $1.95\times 10^5$ [NLO] \\
 \hline
 \multirow{3}{*}{di boson \cite{Campbell:2011bn} } & $ZZ+$ jets  & 17.72 [NLO] \\
\cline{2-3}
 & $WW+$ jets  & 124.31 [NLO] \\
\cline{2-3}
 & $WZ+$ jets  & 51.82 [NLO] \\
 \hline
 {mono-V + $t \bar{t}$} & $t \bar{t} Z$  & 0.911 [NLO] \\
 \cline{2-3}
 & $t \bar{t} W^\pm$ & 0.636 [NLO]\\
 \hline
 \end{tabular} 
\caption{Higher-order QCD corrected cross-section at the 14 TeV LHC of different background processes considered in our study. The order of QCD correction is given in brackets.
For the final process, higher-order QCD corrected cross-section in five massless quark flavors at 14 TeV LHC obtained from {\sc MG5\_aMC@NLO}. Default factorization and renormalization scales and an in-built NN23NLO pdf set are used.
}
\label{cross-section:BG}
\label{tab:cross-sec ttz}
\end{center}
\end{table}


\subsection{Construction of High-Level Variables and Cut-Based Analysis}
\label{variables}

Once we have generated our signal and background processes after the realistic detector-level simulation, the next task is constructing high-level event variables sensitive to kinematic configuration signal and background processes. For example, the unique point of this collider study counts on the fatjet characteristic and its different properties related to the mass-energy distribution within these fatjets. We categorize some of the useful variables for our analysis in the following bulleted points:\\

\noindent
\underline{N-subjettiness ratio:} 
In the case of a highly boosted top quark, one can capture all three hadronically decayed constituents of the top quark within a single large-radius jet (fatjet). The whole energy of a reconstructed top-fatjet is distributed around three subjet axes. Assuming N number of subjets belong to the fatjet, N-subjettiness is defined by the angular distance in the transverse plane of constituents of the fatjet from the nearest subjet axis and weighted by the transverse momentum of the constituents as below \citep{Thaler:2010tr, Thaler:2011gf}:
\begin{equation}
\tau_N = \dfrac{1}{\mathcal{N}_0}\, \sum_i \, P_{T,i}\, \text{min} \{ \Delta R_{i,1},\Delta R_{i,2},...,\Delta R_{i,N}\}.
\end{equation}
Here, the summation goes over all the particles inside the jet. The denominator is $\mathcal{N}_0=\sum_i \, P_{T,i} \, R$, where $P_{T,i}$ and $R$ are the transverse momentum of the i-th constituent and radius of the jet, respectively.
Since N-subjettiness determines the jet shape, the N-subjettiness ratios, such as $\tau_{31}$ and $\tau_{32}$ are good observables in signal background analysis. $\tau_{32}$ effectively distinguishes the top signal from two-prong fatjets arising from the boosted W or Z boson in the background. In contrast, $\tau_{31}$ is also effective for separating the top signal from the one-prong QCD fatjets that contribute significantly to the background.\\

\noindent
\underline{Pruned jet mass:} 
Jet-mass is a good variable for classifying a boosted top-fatjet from the two-prong fatjets from the boosted W/Z boson or one-prong QCD fatjets. The jet mass, $M_J=(\sum_{i\in J} P_i)^2$, where four-momentum of i-th energy hit in the calorimeter is $ P_i$. Since large radius jets pick additional soft contributions from underlying QCD radiations, we must remove these soft and wide-angle radiations for more realistic predictions. Different jet grooming techniques, pruning, filtering, and trimming \cite{Krohn:2009th, Butterworth:2008iy, Ellis:2009su, Ellis:2009me} are available to remove those softer and wider angle radiations while we consider pruning in our analysis. 
In the first step of pruning, we define fatjet using the CA algorithm, and in the second step, we pruned its constituents in each recombination step.
\begin{equation}
Z=\text{min}(P_{Ti},P_{Tj})/P_{T(i+j)} < Z_{\text{cut}} \hspace{9mm} \text{and} \hspace{9mm} \Delta R_{ij} > R_{\text{fact}} \hspace{1mm} .
\end{equation} 
The merging $i,j\rightarrow J$ is vetoed when both the conditions are satisfied. Pruning is parametrized by two parameters, the softness parameter, $Z$, and the angular distance of the constituents, $\Delta R_{ij}$. We chose $Z_{cut}=0.1$ \cite{Ellis:2009su} and $R_{\text{fact}}=0.86 \hspace{1mm} (\sim m_t/P_{T,\text{top}}) $ \cite{Ellis:2009me} in our analysis.

\begin{figure}[htbp!]
\centering
\subfloat[] {\label{fig:FJ_MJ0} \includegraphics[width=0.33\textwidth]{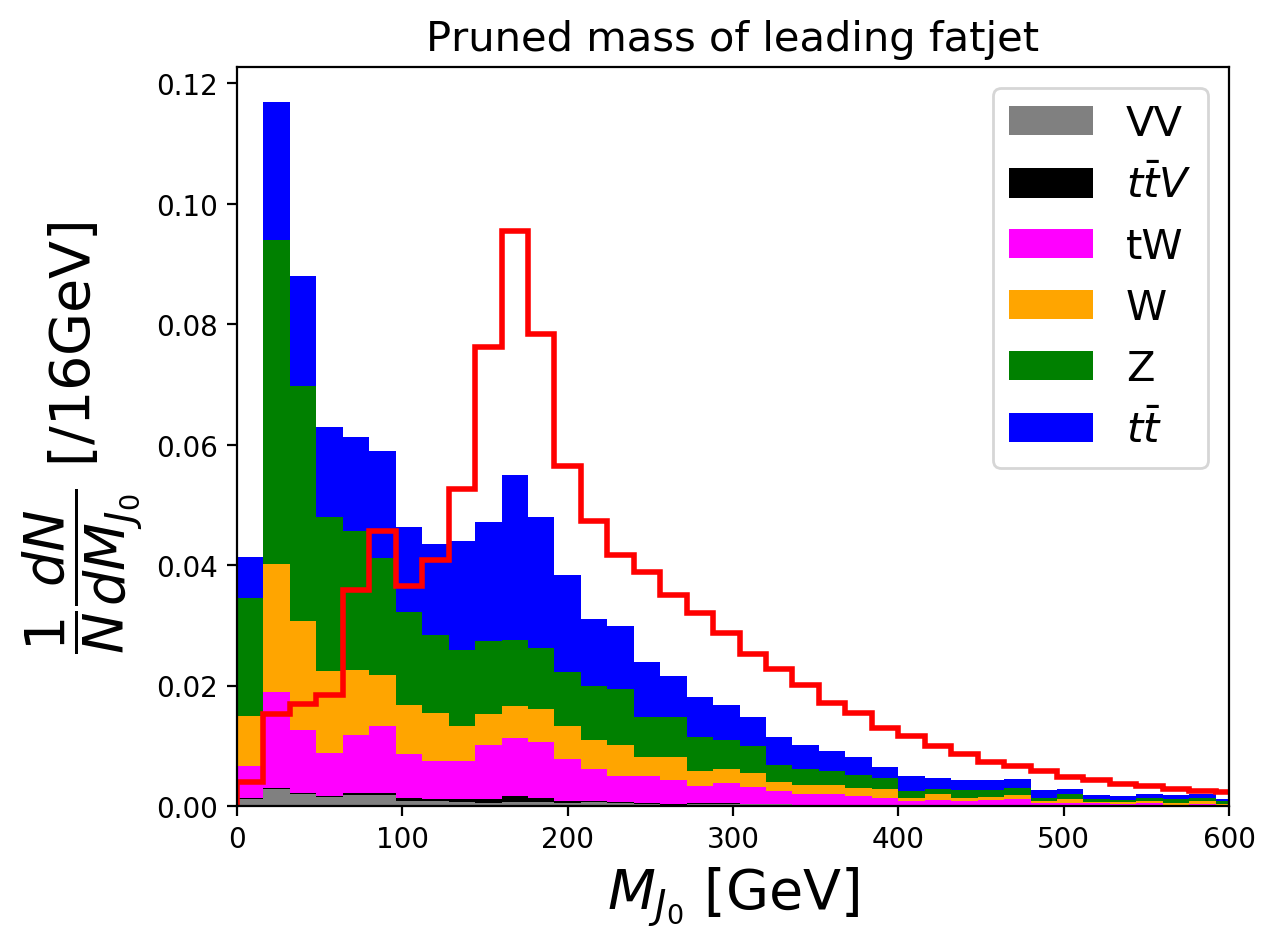}} 
\subfloat[] {\label{fig:FJ_MJ1} \includegraphics[width=0.33\textwidth]{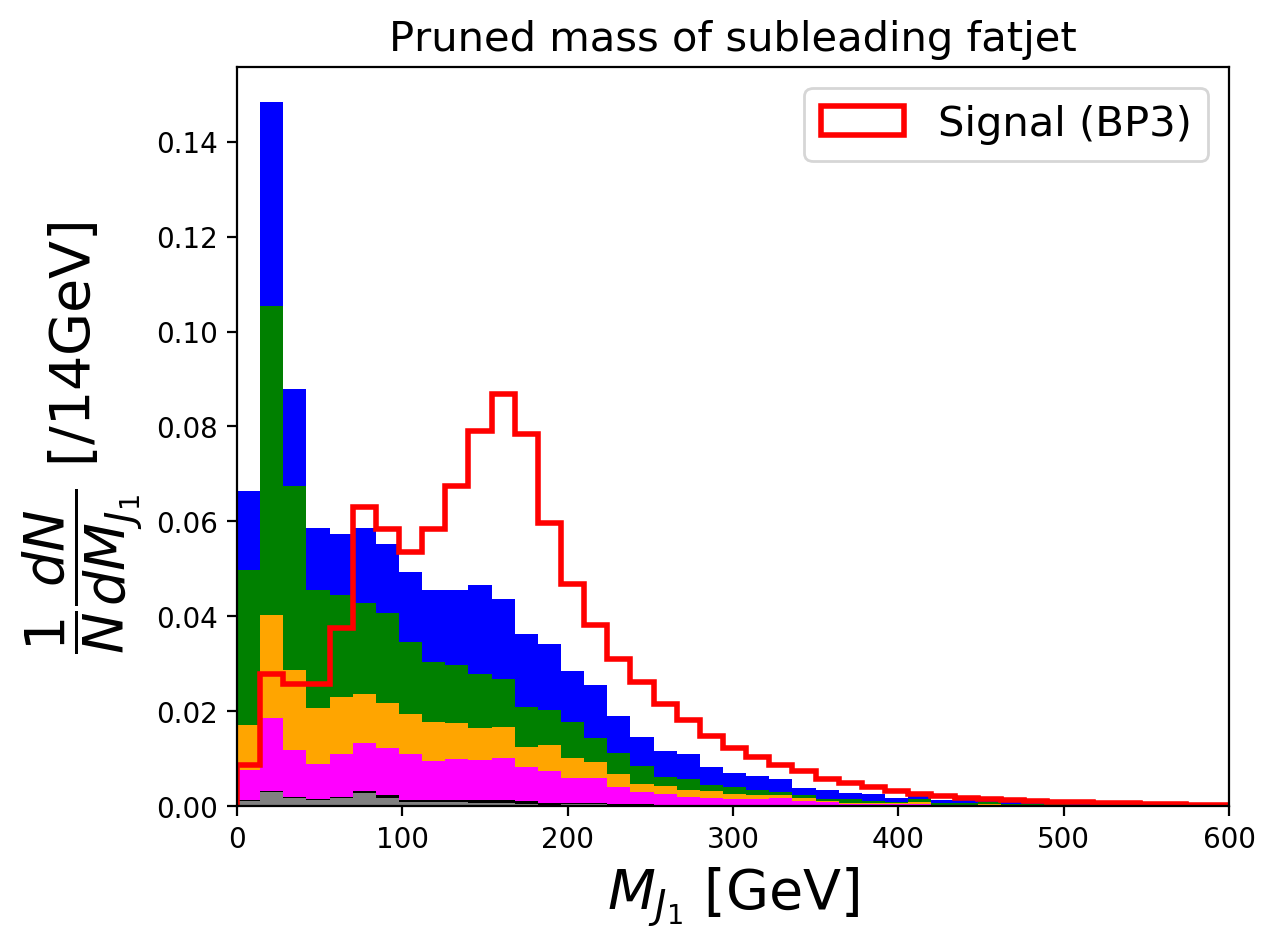}}
\subfloat[] {\label{fig:FJ_RJ0J1} \includegraphics[width=0.32\textwidth]{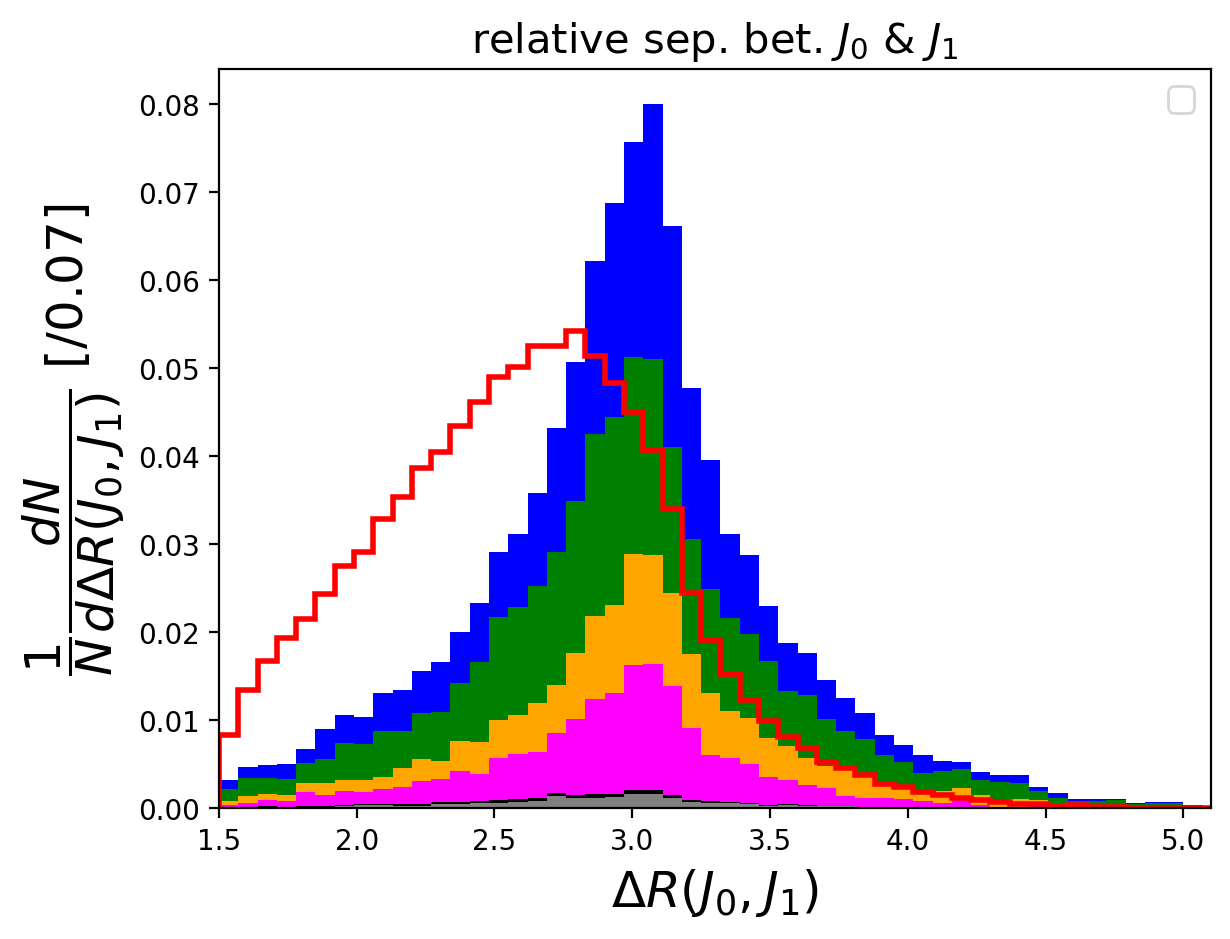}}\\
\subfloat[] {\label{fig:FJ_Phi_J0ET} \includegraphics[width=0.33\textwidth]{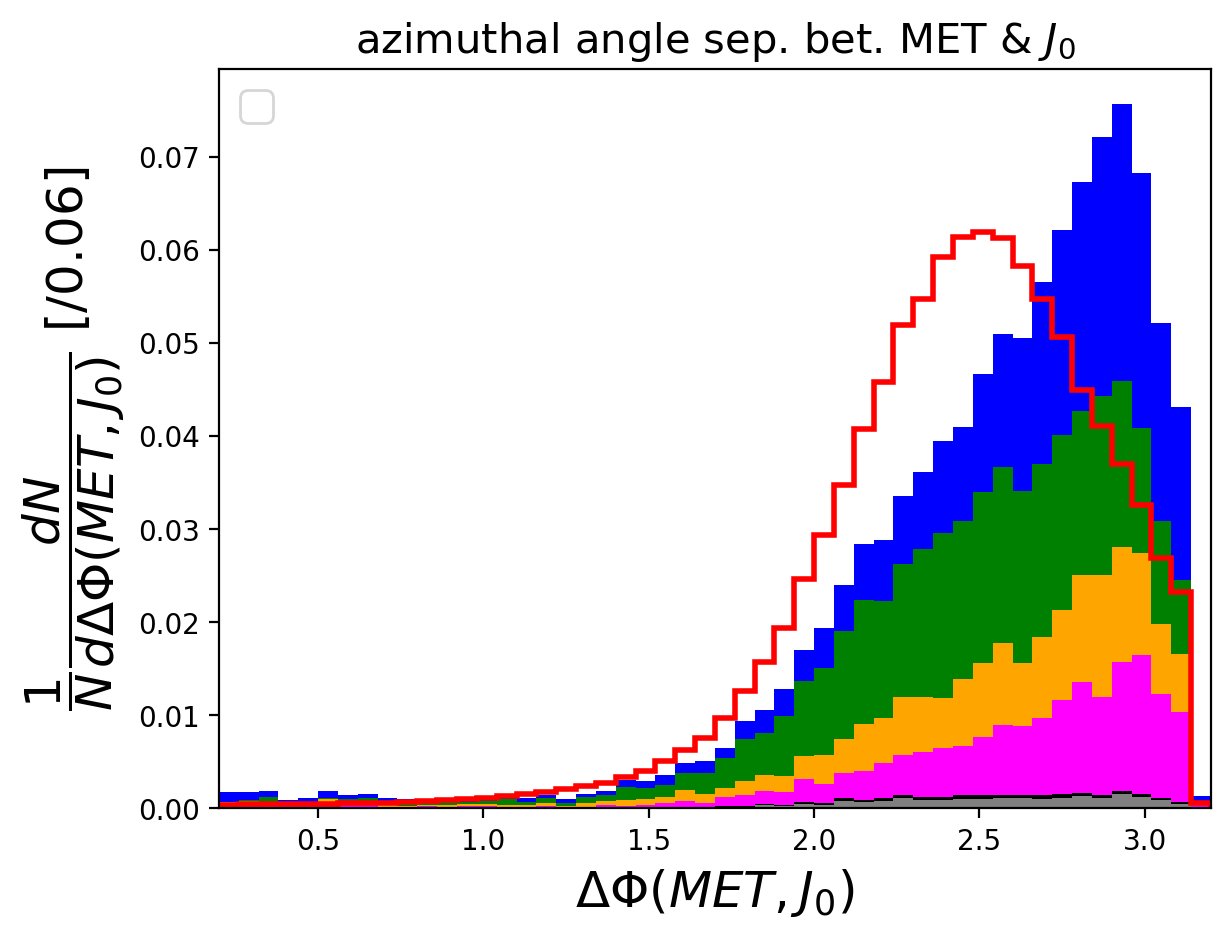}} 
\subfloat[] {\label{fig:FJ_Phi_J1ET} \includegraphics[width=0.33\textwidth]{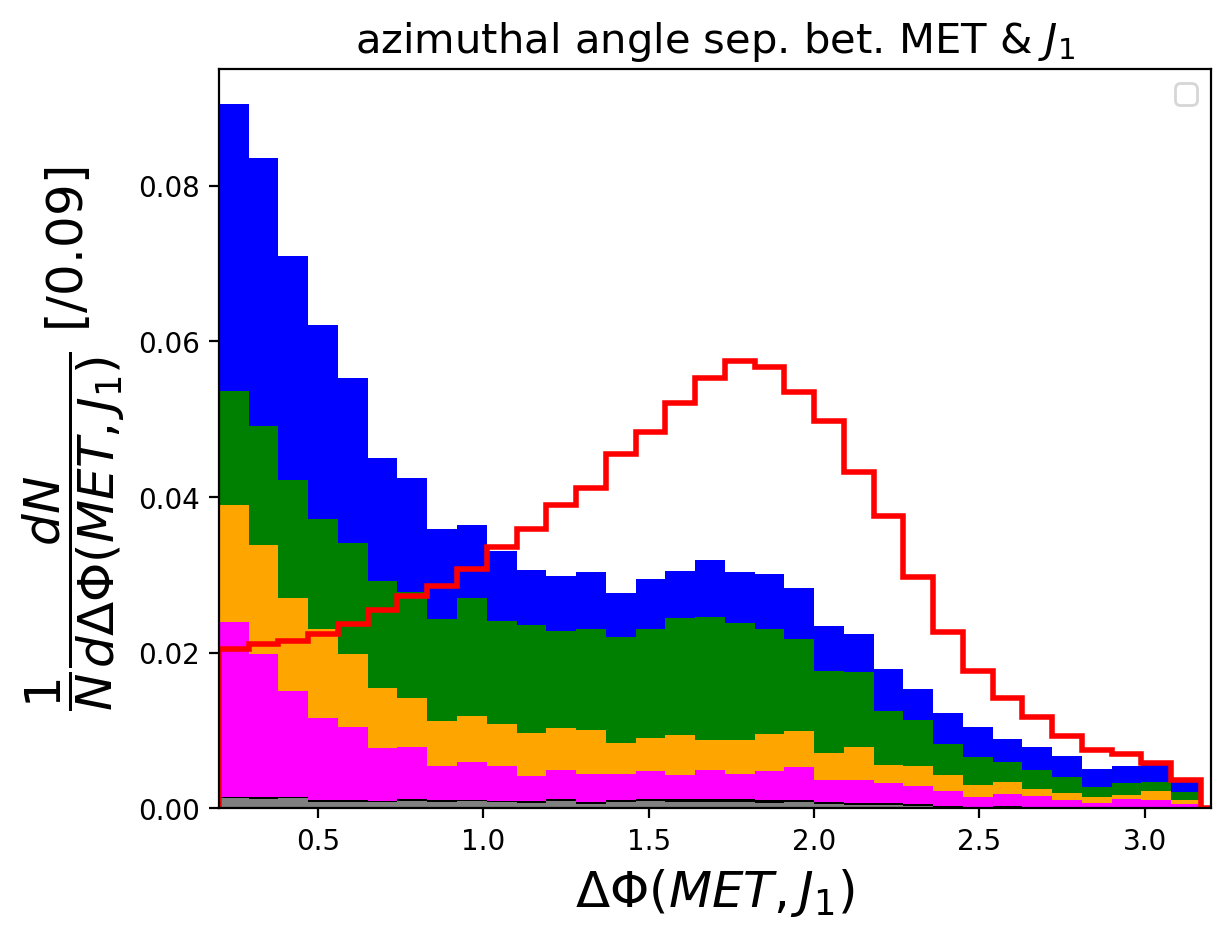}}
\subfloat[] {\label{fig:FJ_tau31_J0} \includegraphics[width=0.33\textwidth]{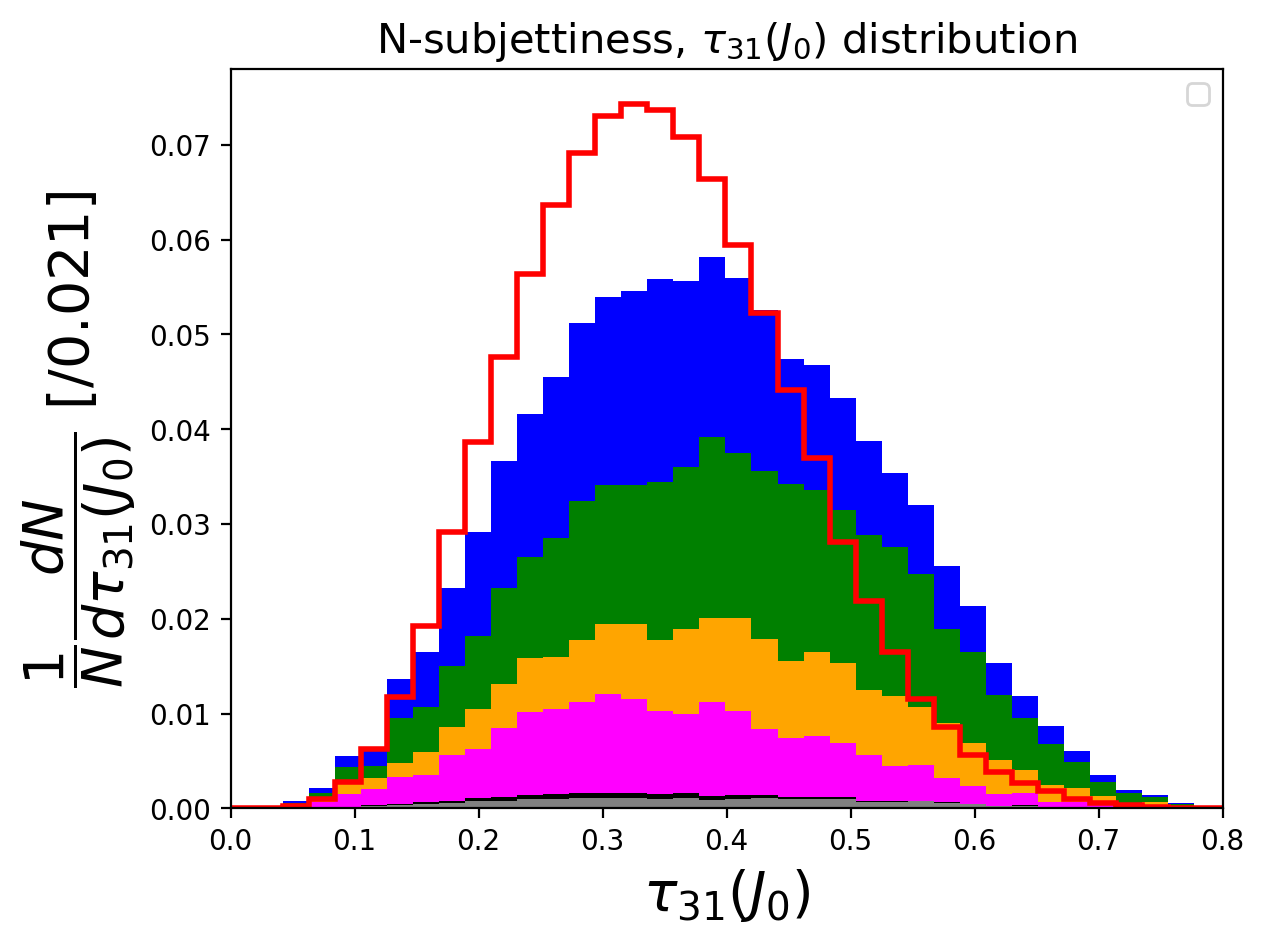}}\\ 
\subfloat[] {\label{fig:FJ_tau31_J1} \includegraphics[width=0.33\textwidth]{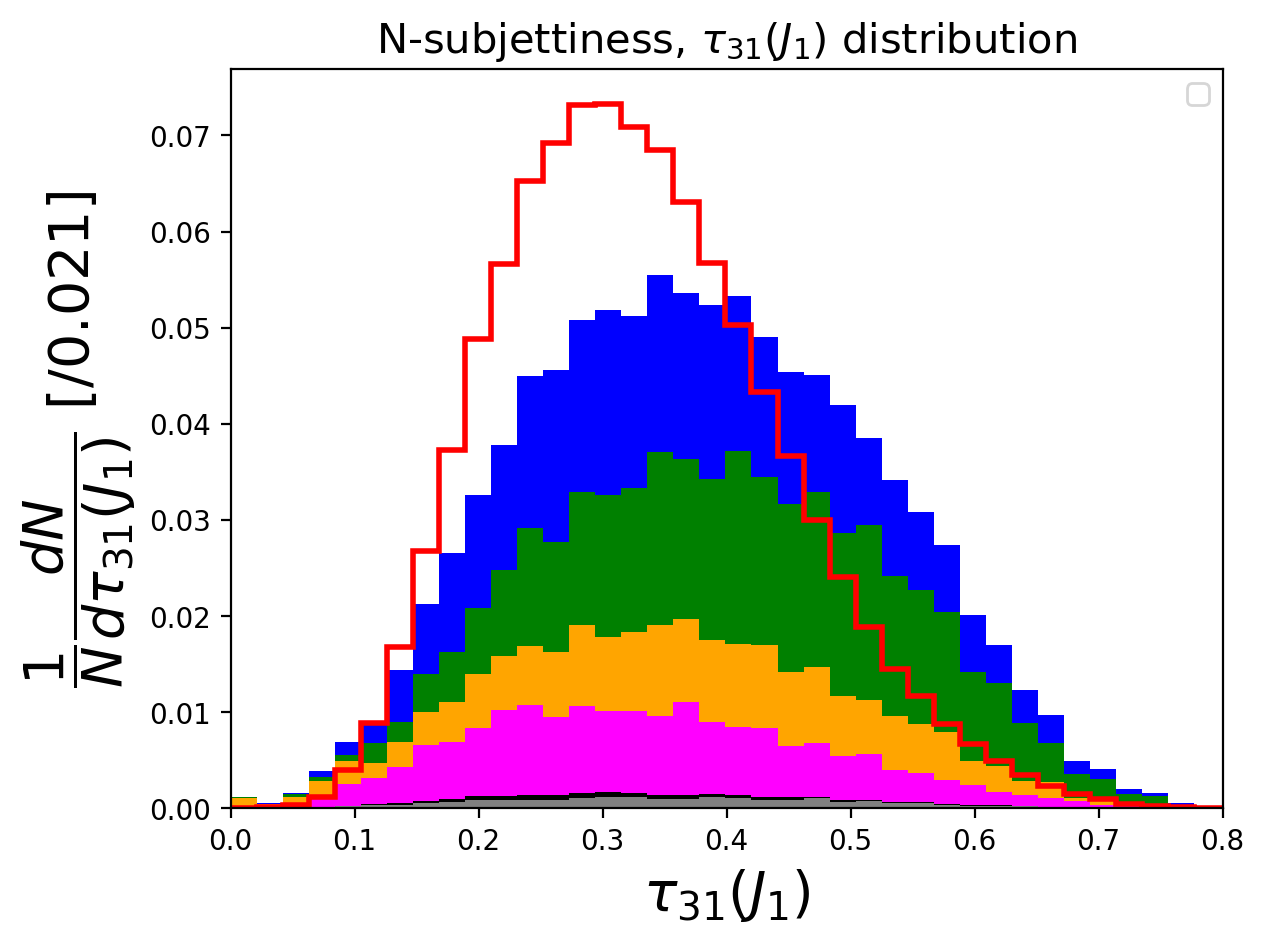}} 
\subfloat[] {\label{fig:FJ_tau32_J0} \includegraphics[width=0.33\textwidth]{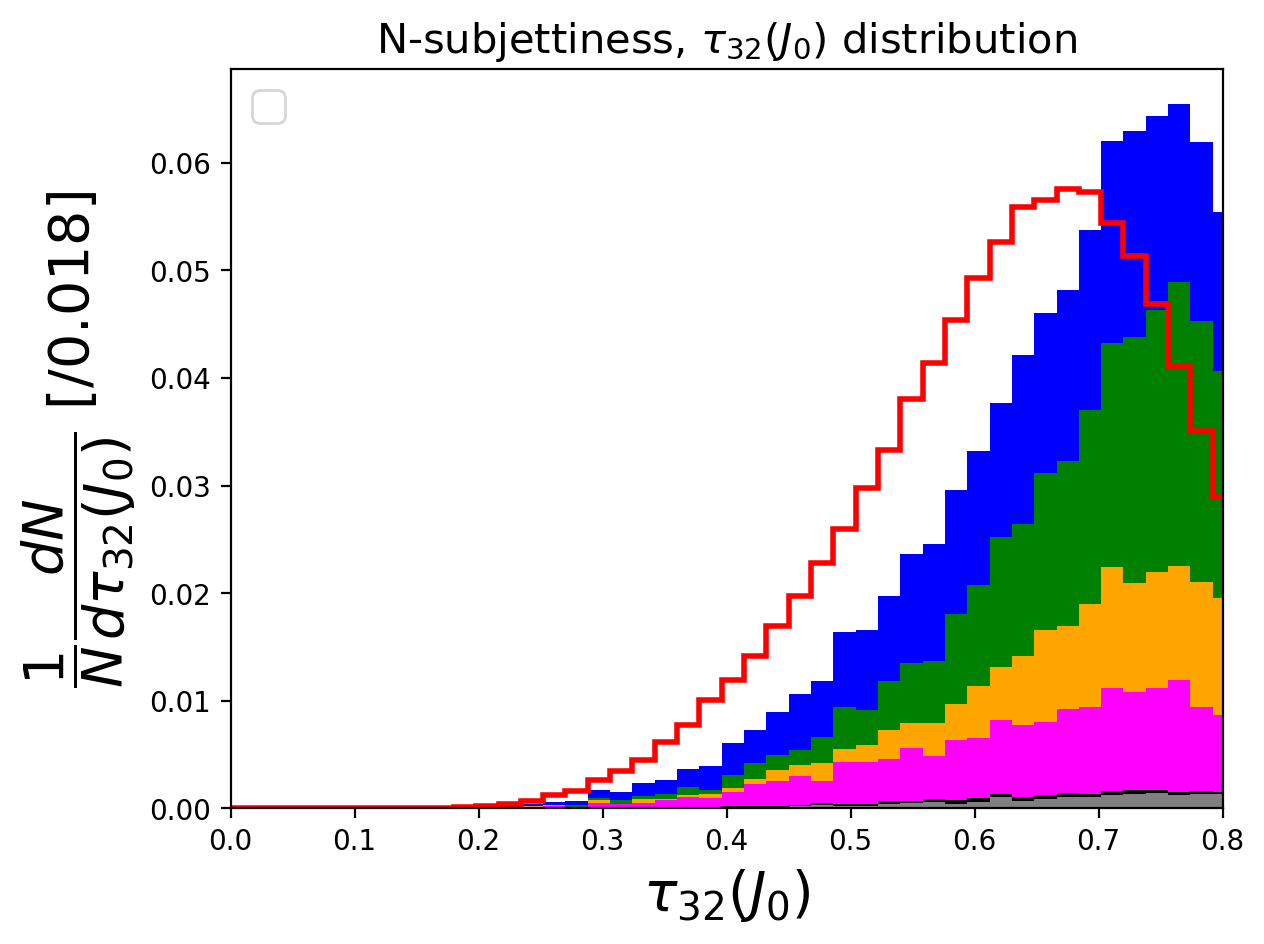}}
\subfloat[] {\label{fig:FJ_tau32_J1} \includegraphics[width=0.33\textwidth]{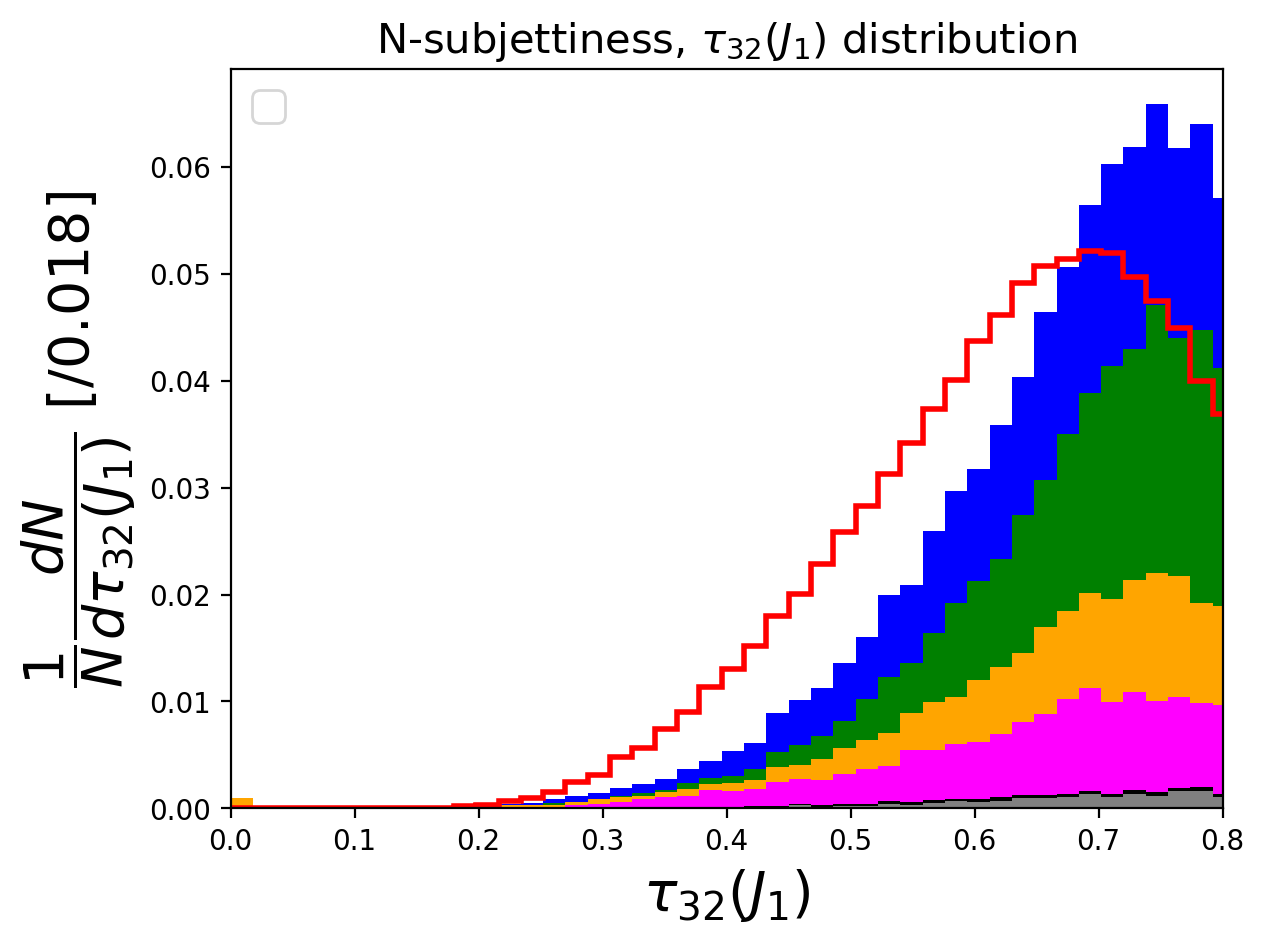}}\\
\subfloat[] {\label{fig:FJ_ET} \includegraphics[width=0.33\textwidth]{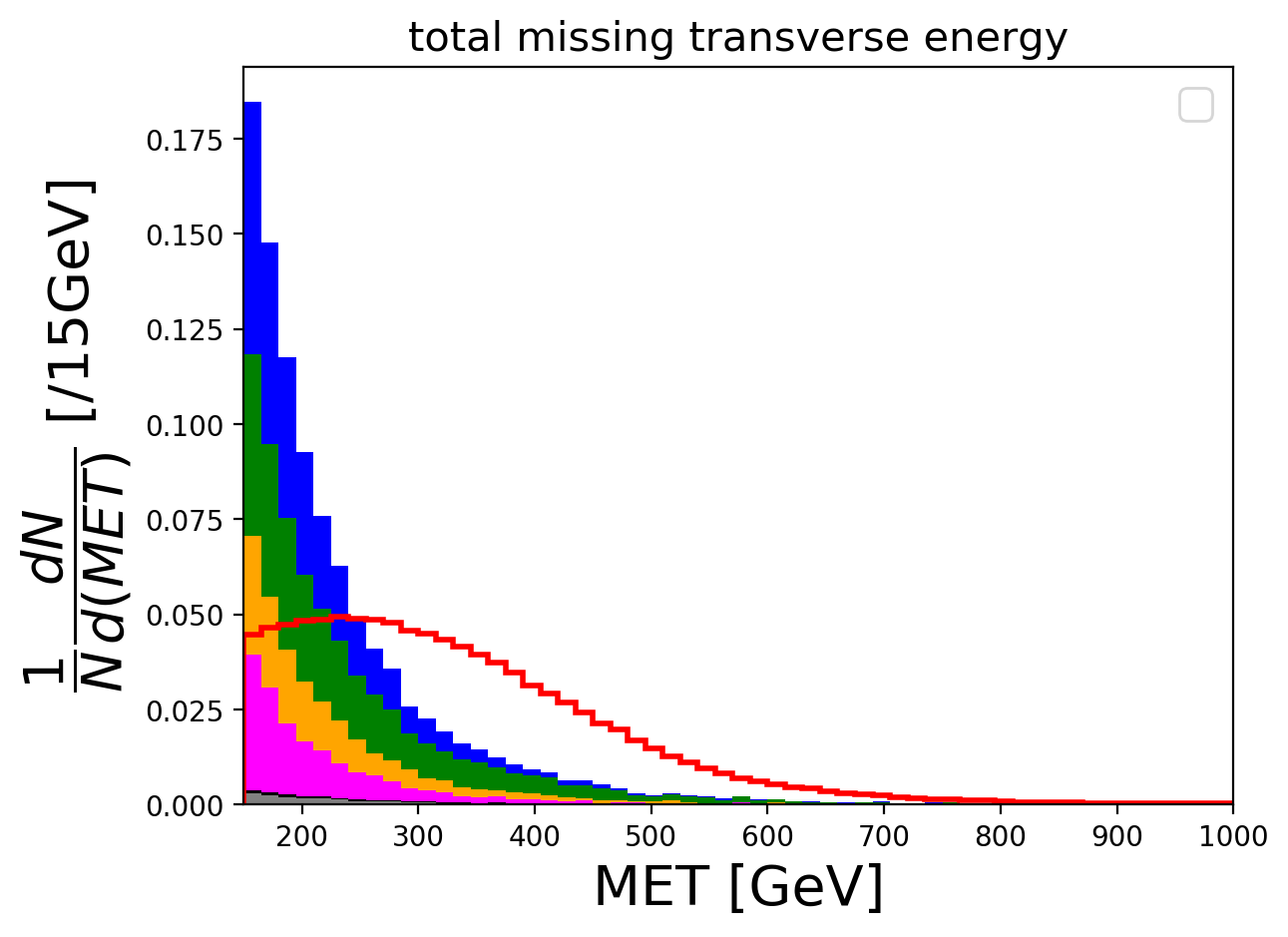}}
\subfloat[] {\label{fig:FJ_MT2} \includegraphics[width=0.33\textwidth]{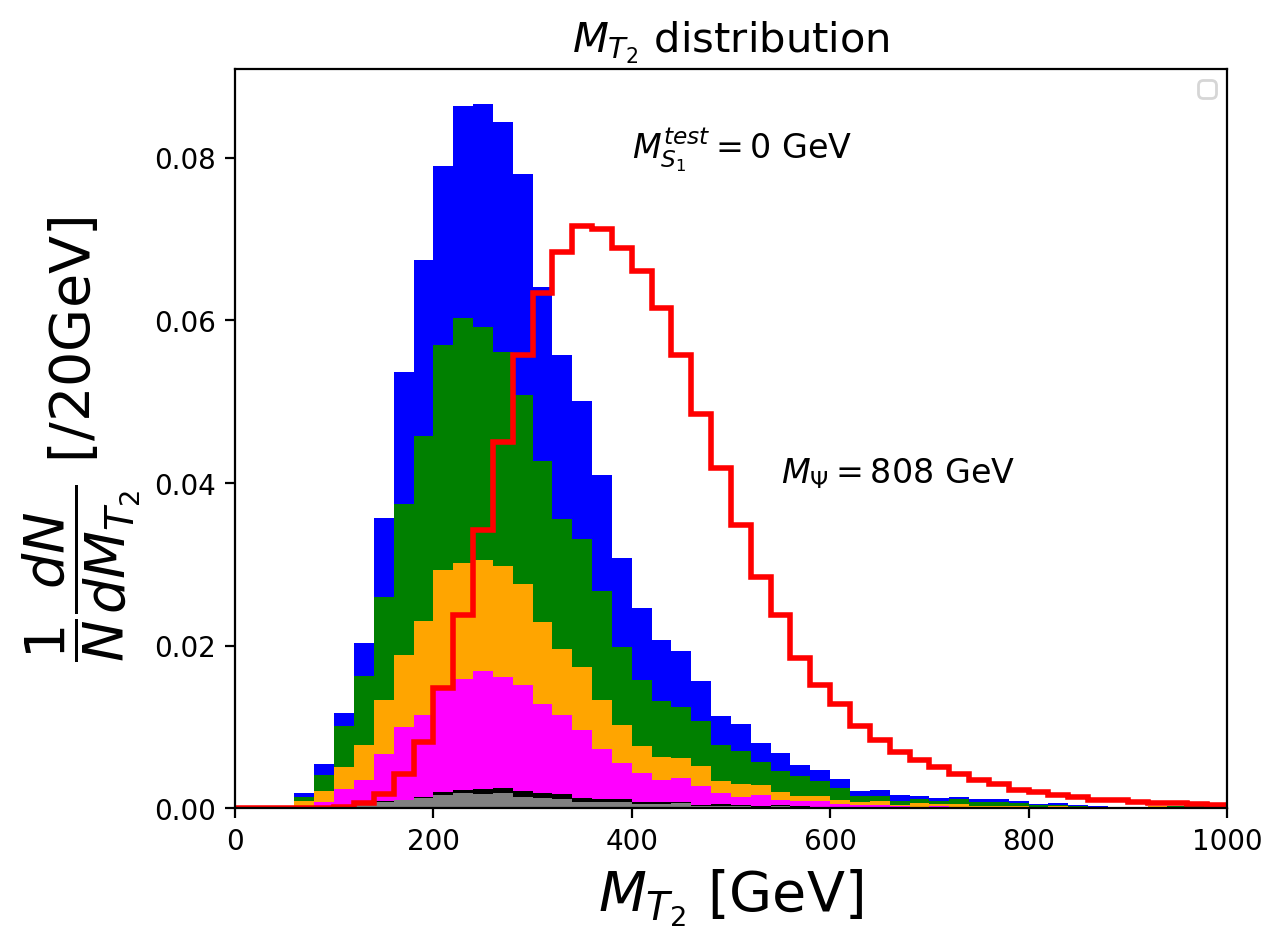}}
\subfloat[] {\label{fig:FJ_Meff} \includegraphics[width=0.32\textwidth]{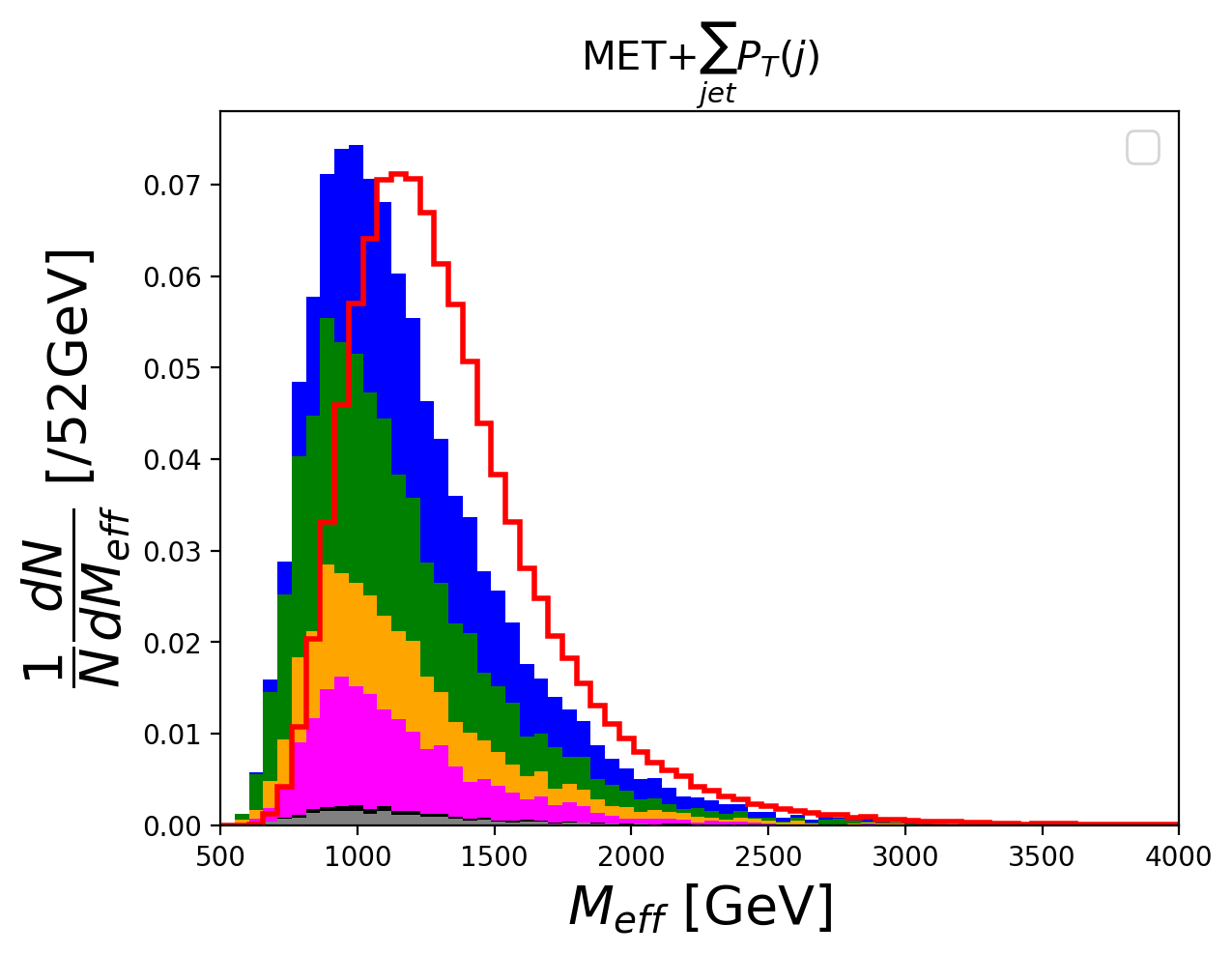}}
\caption{Distributions of different kinematical variables for the signal (BP3) and all the backgrounds contributing to the fatjets $+\slashed{E}_T$ final state after imposing the b tag within leading or subleading fatjet, $\slashed{E}_T>150$ GeV, and the primary event selection criteria (described in the text) for 14 TeV LHC. The normalized distribution for the signal is given by the solid red line. The events of each background process have been weighted by their cross-section and the cut efficiency after applying the previously mentioned cuts. Each background process is then normalized to the sum of individual cross-section times cut efficiency. Colors show the contribution of the individual background process.}
\label{fig:sig_bg_1}
\end{figure}

\noindent
\textbf{Primary event selection criteria :}
Based on our previous discussion and construction of high-level variables, we identify two large-radius jets, leptons, and missing transverse energy as per the following event selection criteria both for the signal and background events alike:
\begin{enumerate}
\item Each event should contain at least two fatjets constructed by CA algorithm with radius parameter $R=1.5$, and each of them has transverse momentum, $P_T(J_0), P_T(J_1)> 200$ GeV. Here, $J_0$ and $J_1$ represent the leading and subleading fatjet.
\item Each event is selected with a minimum missing transverse energy $\slashed{E}_T>100$ GeV.
\item Since our signal does not contain lepton, we veto any event if it contains any lepton with transverse momentum, $P_T(l)>10$ GeV within pseudorapidity $|\eta(l)|<2.4$.
\item To minimize jet mismeasurement contribution to $\slashed{E}_T$, we keep an azimuthal separation between each fatjet and $\slashed{E}_T$, $|\Delta \Phi(J_{0,1},\slashed{E}_T)|>0.2$.  
\end{enumerate}

The normalized distribution of different observables of a sample signal benchmark point, BP3, and bin-wise stacked histogram of all the backgrounds are shown in Fig. \ref{fig:sig_bg_1}. These plots are shown after demanding at least one the b tag within leading or subleading fatjet, enhanced $\slashed{E}_T>150$ GeV, over the preselection cuts already described for 14 TeV LHC.

The prime background $t \bar{t}+$ jets, where one of the top decay hadronically and the other decays leptonically, is shown by the top most blue shade, while the solid red line indicates the sample signal, BP3. The distributions of the pruned jet mass of the leading ($M_{J_0}$) and subleading ($M_{J_1}$) fatjets are given in Fig. \ref{fig:FJ_MJ0} and Fig. \ref{fig:FJ_MJ1}, respectively. At LO, $\Psi$ and $\bar{\Psi}$ produce back to back, each one followed by decay into an (anti)top quark and $S_{1,2}$. In most events in this benchmark point, these tops are boosted as they are produced from the decay of heavy particles. When the top is sufficiently boosted, all three constituents of the top quark fall within a single large-radius jet, giving a three-prong jet substructure and pruned jet mass very close to the top quark mass. For the signal, we get a sharp peak around the top quark mass for both the leading and subleading fatjet. These large radius jets sometimes misses some of the constituent sub-jets, especially when the boost of the top quark is relatively low, causing a secondary peak near the W/Z boson mass for both the fatjets of the signal. For semi-leptonic $t \bar{t}+$ jets background, the top which decays hadronically gives the leading fatjet for a significant number of events and causes a sharp peak near top mass in the leading fatjet mass distribution. From the demand for a very high missing transverse momentum, $t \bar{t}+$  background contributes to a phase space region where the b-jet from the leptonically decaying top quark generates the subleading fatjet predominantly. Consequently, subleading fatjet mass generates its peak near 20 GeV from QCD radiation.
 
\begin{table}[tb!]
\begin{center}
 \scriptsize
 \setlength\tabcolsep{2.7pt} 
 \begin{tabular}{|c||c||c|c|c|c|c|c|c|c|c|}
\hline
& \textbf{BP3} & \textbf{$t \bar{t}$+jets} & \textbf{$tW$+jets} & \textbf{$ttZ$} & \textbf{$ttW$} & \textbf{$Z$+jets} &\textbf{$W$+jets} & \textbf{$WZ$+j} & \textbf{$ZZ$+j} & \textbf{$WW$+j} \\
\hline \hline
\multirow{2}{*}{C1} & 5969 & $9.6\times 10^4$ & $5.1\times 10^4$  & 1048 & 111 
& $3.5\times 10^5$ & $1.9\times 10^5$ & $1.3\times 10^4$  & $1.6\times 10^3$ & $3.6\times 10^3$\\
& [$100\%$] & [$100\%$] & [$100\%$] & [$100\%$] & [$100\%$] 
& [$100\%$] & [$100\%$] & [$100\%$] & [$100\%$] & [$100\%$] \\
\hline
\multirow{2}{*}{C2} & 5296 & $4.2 \times 10^4$ & $2.12\times 10^4$ & 793 & 64 
& $2.28\times 10^5$ & $1.06\times 10^5$ & $8.11\times 10^3$ & 969.2 & $1.6\times 10^3$  \\
& [$88.73\%$] & [$43.89\%$] & [$41.96\%$] & [$75.71\%$] & [$57.53\%$] 
& [$65.06\%$] & [$53.78\%$] & [$64.34\%$] & [$61.73\%$] & [$43.97\%$] \\
\hline
\multirow{2}{*}{C3} & 4424 & $3.21\times 10^4$ & $1.59\times 10^4$ & 656 & 54.1 
& $3.36\times 10^4$ & $1.64\times 10^4$ & $1.5\times 10^3$ & 267 & 341.1 \\
& [$74.11\%$] & [$33.60\%$] & [$31.42\%$] & [$62.63\%$] & [$48.73\%$] 
& [$9.57\%$] & [$8.32\%$] & [$11.89\%$] & [$17.0\%$] & [$9.37\%$] \\
\hline
\multirow{2}{*}{C4} & 1005  &$4.02\times 10^3$  & $1.72\times 10^3$ & 185 & 16.7 
& $1.54\times 10^3$  & 926  &  72 & 10.4 & 26 \\
& [$16.85\%$] & [$4.20\%$] & [$3.39\%$] & [$17.66\%$] & [$15.07\%$] 
& [$0.44\%$] & [$0.47\%$] & [$0.57\%$] & [$0.66\%$] & [$0.71\%$] \\
\hline
\multirow{2}{*}{C5} & 666  & $2.46\times 10^3$  & $1.07\times 10^3$ & 132.5 & 12 
&842  & 493  &  42.5 & 7.1 & 15.7 \\
& [$11.16\%$] & [$2.57\%$] & [$2.12\%$] & [$12.64\%$] & [$10.84\%$] 
& [$0.24\%$] & [$0.25\%$] & [$0.337\%$] & [$0.45\%$] & [$0.43\%$] \\
\hline
\multirow{2}{*}{C6} & 432  & 411  & 197 & 54 & 3.1 
& 260  & 132  &  17.5 & 4.3 & 1.7 \\
& [$7.24\%$] & [$0.43\%$] & [$0.39\%$] & [$5.12\%$] & [$2.78\%$] 
& [$0.074\%$] & [$0.067\%$] & [$0.139\%$] & [$0.272\%$] & [$0.047\%$] \\
\hline
 \end{tabular}     
\caption{The cut efficiency and expected number of events after the corresponding cuts for the signal and all the backgrounds contribute to the fatjets $+\slashed{E}_T$ final state at the 14 TeV LHC and 139 $\text{fb}^{-1}$  integrated luminosity. 
The effectiveness of different selection cuts can be followed in the form a cut flow from top to bottom after applying (C1) Preselection cuts, (C2) $\slashed{E}_T>150$ GeV, (C3) requiring at least one b-tag within $J_0$ or $J_1$, (C4) 120 GeV$< M_{J_0}, M_{J_1}<230$ GeV, (C5) $\tau_{31}(J_0),\tau_{31}(J_1)<0.4$ and finally, (C6) $M_{T_2}>320$ GeV.
A sample benchmark point, BP3, is presented in this table.}
\label{tab:fatjet_cutflow}
\end{center}
\end{table}
 
The total missing transverse energy distribution is shown as another interesting variable in Fig. \ref{fig:FJ_ET}. In the case of signal, we have two missing DM particles coming from the decay of $\Psi$ pair, where they primarily produce back to back, so the $\slashed{E}_T$ has uniform distribution as two missing particles can avail entire phase space. In contrast, the background drops sharply for large $\slashed{E}_T$. Distributions of the azimuthal separation of the leading and subleading fatjets from the $\slashed{E}_T$ are presented in Fig. \ref{fig:FJ_Phi_J0ET} and Fig. \ref{fig:FJ_Phi_J1ET}, respectively. As stated earlier, two missing particles can avail the entire phase space for the signal, so both $\Delta \Phi(MET, J_{0,1})$ have a uniform distribution. For a significant amount of events of the $t \bar{t}+$ jets background, the b-jet from the leptonically decaying top quark behaves as a subleading fatjet ($J_1$), and the neutrino gives the $\slashed{E}_T$, where we select the events that have large $\slashed{E}_T$. Hence, the azimuthal separation of $J_1$ from $\slashed{E}_T$ gets a maximum at a lower value. In contrast, the azimuthal separation of the leading fatjet ($J_0$) from $\slashed{E}_T$ peaks near $\sim \pi \, \text{rad}$. The distribution of $\Delta R (J_0, J_1)$, angular distance between $J_0$ and $J_1$ in the transverse plane is given in Fig. \ref{fig:FJ_RJ0J1}. 

The distribution of the kinematic variable $\tau_{31}$ for both leading and subleading fatjet are shown in Fig. \ref{fig:FJ_tau31_J0}, Fig. \ref{fig:FJ_tau31_J1}, respectively. In both distributions, as expected, the signal has a peak for a smaller value of $\tau_{31}$ representing that signal fatjets have a three-prong structure. Similarly, the distribution of the kinematic variable $\tau_{32}$, which separates the three-prong fatjet from the two-prong fatjet, are presented in Fig. \ref{fig:FJ_tau32_J0}, Fig. \ref{fig:FJ_tau32_J1}. $\tau_{32}(J_0)$ has a peak near 0.6 and 0.75 for signal and background, respectively.
Note that we do not apply any mass window in generating these distributions, but if we do, the peaks of $\tau_{32}$ move towards a lower value. So, in the final event selection in the cut-based analysis, we apply a mass window to discriminate the signal from the background better. 

The distribution of kinematic stransverse mass variable $M_{T2}$ \cite{Lester:1999tx,Barr:2011xt,Konar:2009wn} is given in Fig. \ref{fig:FJ_MT2}. Assuming DM mass is unknown to us, we construct $M_{T2}$ after setting trial DM mass as zero in this construction. SM particles have a smaller mass compared to the mass of $\Psi$, so the $M_{T2}$ distribution of signal and background are well separated. Since we do not want to find the correct mass of the mother particle ($\Psi$), this variable is used to discriminate the signal from the background efficiently. The distribution of $M_{\text{eff}}$ is given in Fig. \ref{fig:FJ_Meff}. Effective mass is defined as
\begin{equation}
M_{\text{eff}} = \slashed{E}_T + H_T,
\label{Eq.Meff}
\end{equation}
where $H_T \equiv \sum \limits_{i=1}^{N_J} P_{iT} $ ($N_J$ is the number of visible jets) is the scalar sum of the transverse momentum of the jets. The above distributions show that all the variables are very good at distinguishing the signal from the background.
 
We apply the following selection cuts to demonstrate a cut-based analysis (CBA) over the preselection cuts (described before) to increase the signal-to-background ratio. Note that our final results are based on sophisticated multivariate analysis with improved statistics. So, the next part is for demonstration purposes without putting much effort into optimizing all the selection criteria. Here, we offer a cut-flow in cut-based analysis to better understand the signal and background differential distributions.


\begin{table}[tb!]
\begin{center}
 \begin{tabular}{|c||c|c|c|c|c|c|c|c|c|c|c|}
\hline
BP & BP1 & BP2 & BP3 & BP4 & BP5 & BP6 & BP7 &  BP8 & BP9 & BP10  & BP11  \\
\hline\hline
$\sigma$ & 11.4 & 12.8 & 11.1 & 9.9 & 6.9 & 5.3 & 4.0 & 3.6 & 2.5 & 1.2  & 2.8 \\
\hline
$\frac{S}{B}$ & 0.41 & 0.47 & 0.40 & 0.35 & 0.23 & 0.17 & 0.13 & 0.12 & 0.08 & 0.04  & 0.09\\
\hline
 \end{tabular} 
\caption{Statistical significance ($\sigma$) and the signal-to-background ratio ($\frac{S}{B}$) are shown for the signal corresponding to different benchmark points contributing to the fatjets $+\slashed{E}_T$ final state at 14 TeV LHC and 139 $\text{fb}^{-1}$  integrated luminosity.}
\label{tab:fatjet_sig}
\end{center}
\end{table}

\textbf{Final selection cuts:}
\begin{enumerate}
\item[5.] We increase $\slashed{E}_T$ from 100 GeV to 150 GeV since it reduces the background sharply than the signal.
\item[6.] Demand an additional b-tag withing either leading or subleading fatjet is applied. The b-tag efficiency for the signal within leading or subleading fatjet is $84\%$. This requirement reduces $Z+$ jets and $W+$ jets backgrounds substantially below $t \bar{t}+$ jets background. 
\item[7.] We select the events for which the pruned mass of the leading and subleading fatjets falls within 120 GeV$< M(J_0), M(J_1)<230$ GeV~\footnote{Note that, in the MVA next section, we retain only the lower mass threshold and let the framework select the non-linear cuts to get the optimal signal-to-background ratio.}. The lower threshold helps us reduce different backgrounds where one or both the fatjets originated from QCD radiation or $W/Z$ boson.
\item[8.] To discriminate further the fatjets from QCD jets, we use N-subjettines and collect the events that satisfy $\tau_{31}(J_0) \hspace{1mm}\text{and}\hspace{1mm} \tau_{31}(J_1)<0.4$ \footnote{One may use the N-subjettiness variables $\tau_{32}(J_0)\hspace{1mm}\text{and}\hspace{1mm} \tau_{32}(J_1)$ to discriminate the fatjets from two-prong fatjets originated from boosted $W/Z$ bosons. Since we analyze the same signal using MVA in the next section, we do not check $\tau_{32}$ variables in CBA.}.
\item[9.] We impose $M_{T_2}>320$ GeV. This requirement increases the signal-to-background ratio ($\frac{S}{B}$). For example, in the case of BP3, $\frac{S}{B}$ changes from 0.13 to 0.4 (Tab. \ref{tab:fatjet_cutflow}). 
\end{enumerate}


The expected number of signal (for a sample benchmark point, BP3) and background events and cut efficiency after imposing the preselection cuts and final selection cuts at 14 TeV LHC for 139 $\text{fb}^{-1}$ integrated luminosity are shown in Tab. \ref{tab:fatjet_cutflow}. Statistical significance and the signal-to-background ratio for different benchmark points are shown in Tab. \ref{tab:fatjet_sig}. $\sigma=\frac{N_S}{\sqrt{N_S+N_B}}$ defines the statistical significance, where $N_S$ and $N_B$ are the expected signal and background events after the cuts, respectively. The statistical significance for the signal of different benchmark points is above the discovery potential for an integrated luminosity of 139 $\text{fb}^{-1}$. We also have good statistics indicating that extracting the VLQ pair from the Standard Model background is not tough.

\subsection{Analysis based on the Multivariate Gradient Boosting Technique}
\label{MVA}


\begin{table}[tb!]
\begin{center}
\scriptsize
 \begin{tabular}{@{}*{12}{|p{.065\textwidth}@{}}|}
\hline
\multirow{2}{*}{\bf Signal} & BP1 & BP2 & BP3 & BP4 & BP5 & BP6 & BP7 & BP8 & BP9 & BP10 & BP11 \\
\cline{2-12}
 & 6625 & 3525 & 2341 & 2711 & 2176 & 1924 & 1424 & 1081 & 915 & 552  & 385 \\
\hline  
\end{tabular} 
 \begin{tabular}{@{}*{10}{|p{.0869\textwidth}@{}}|}
 \hline
\multirow{2}{*}{\bf SM BG} & tt+jets & tW+jets & ttZ & ttW & Z+jets & W+jets & WZ+jets & ZZ+jets & WW+jets \\
\cline{2-10}
 & 8928.06 & 3815.42 & 294.35 & 25.93 & 3527.96 & 2408.37 & 172.35 & 27.54 & 46.49   \\
\hline
\end{tabular} 
\caption{The expected number of signal and SM background events after applying $M_{J_0}>120$ GeV, $M_{J_1}>120$ GeV and b-tag (within leading or subleading fatjet), $\slashed{E}_T>150$ GeV in addition to preselection cuts at 14 TeV LHC for 139 $\text{fb}^{-1}$ integrated luminosity.}
\label{tab:MVA}
\end{center}
\end{table}

In the previous section, we constructed high-level variables and demonstrated their potential in a CBA. This section extends that idea to perform a more sophisticated MVA. In these analyses, MVA generally gives better sensitivity than CBA if appropriate kinematic variables are utilized, where we may get significance above the discovery limit for the benchmark points that is unable through CBA. The $M_{J_0}$ and $M_{J_1}$ distribution (Fig. \ref{fig:FJ_MJ0}, Fig. \ref{fig:FJ_MJ1}) have the largest peak around the top mass, and the signal is much harder than the background for $M_{J_{0,1}}>120$ GeV. Instead of both lower and upper mass thresholds to set an allowed window, we retain only a lower mass threshold of 120 GeV for both the fatjets for event selection in MVA for a higher number of events. We expect the MVA framework to select nonlinear variable space to get the optimal signal-to-background ratio. The 120 GeV cut on both fatjets reduces the backgrounds drastically compared to the signal for which fatjets arrises from the QCD jet (one-prong) or boosted W/Z boson (two-prong). We also demand at least one b tag within the leading or subleading fatjet for event selection in MVA, reducing the background much more than the signal. From the missing energy distribution (Fig. \ref{fig:FJ_ET}), we see most background events exist in low missing energy, so after demanding large missing energy, we reduce the background significantly compared to the signal. So we apply $\slashed{E}_T>150$ GeV for the event selection in MVA.

\begin{figure}[tb!]
\centering
 \includegraphics[width=0.48\textwidth]{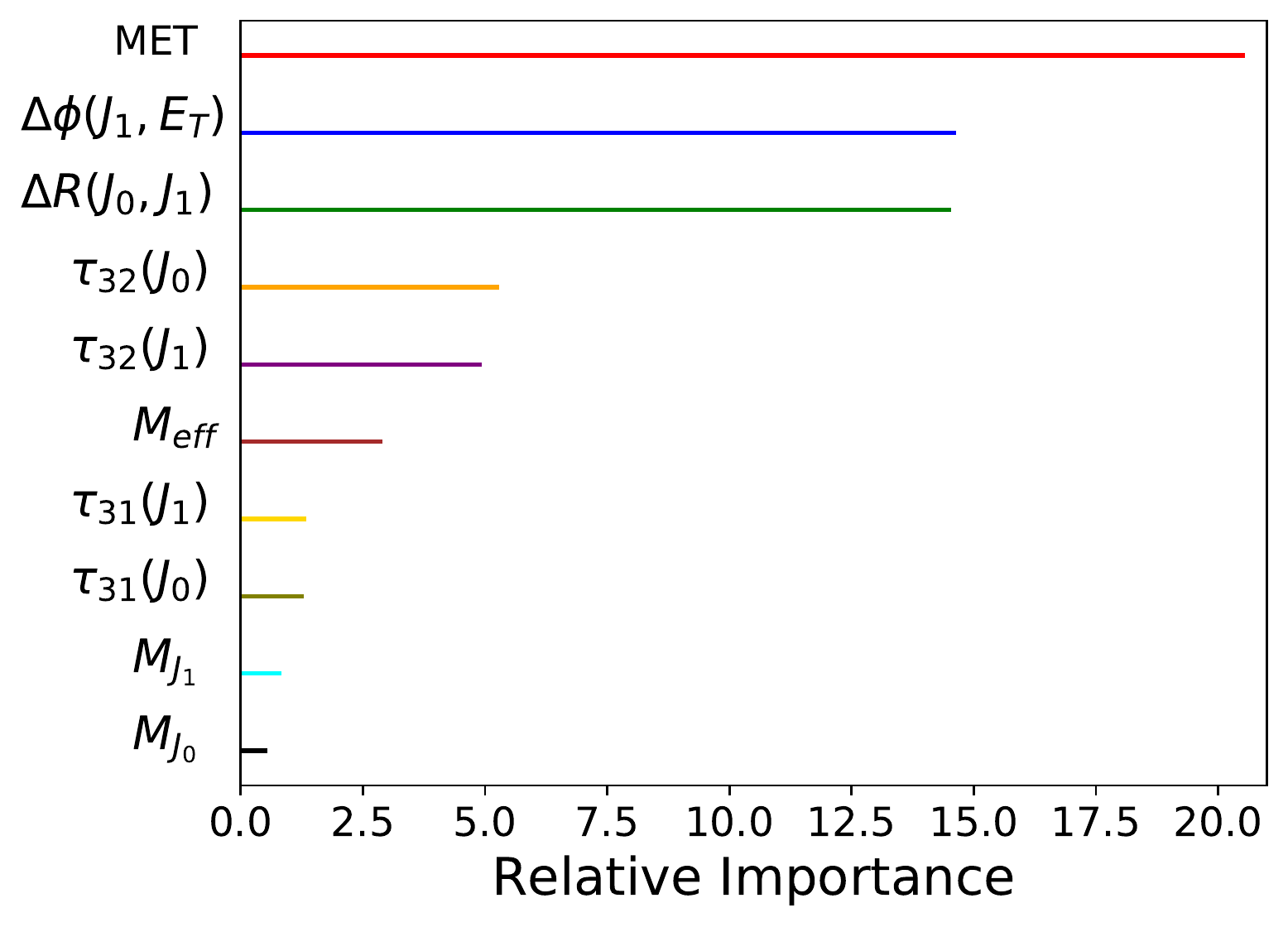}
  \caption{ Relative importance (Method unspecific) of the different kinematic variables used in MVA. We get those numbers for BP3 from the TMVA package. Those numbers can change a little bit if one chooses a different algorithm.}\label{MVA:relative_importance}
\end{figure}

\begin{figure}[tb!]
\centering
  \subfloat {\label{MVA:correlationS}\includegraphics[width=0.495\textwidth]{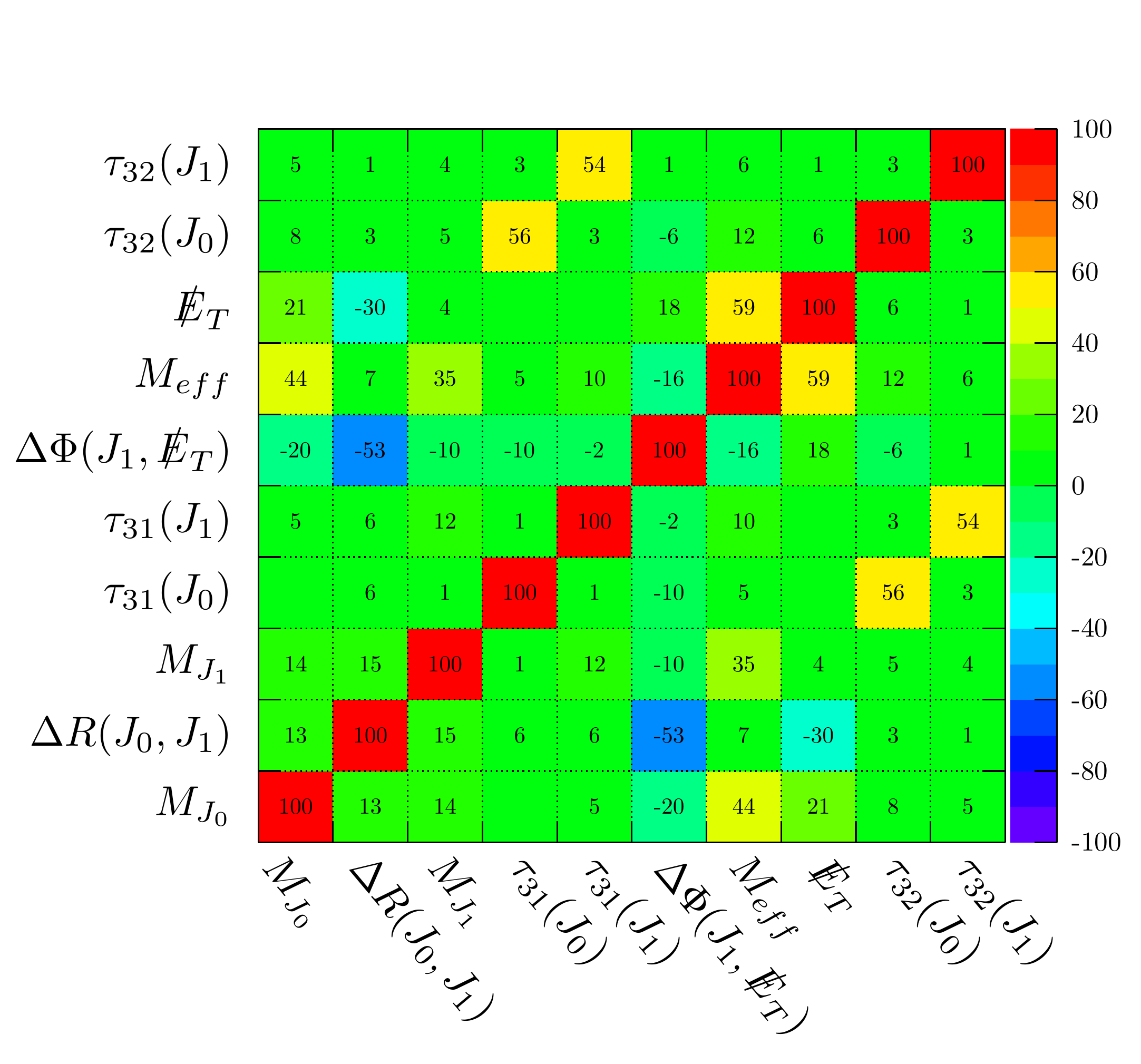}} 
  \subfloat {\label{MVA:correlationB}\includegraphics[width=0.495\textwidth]{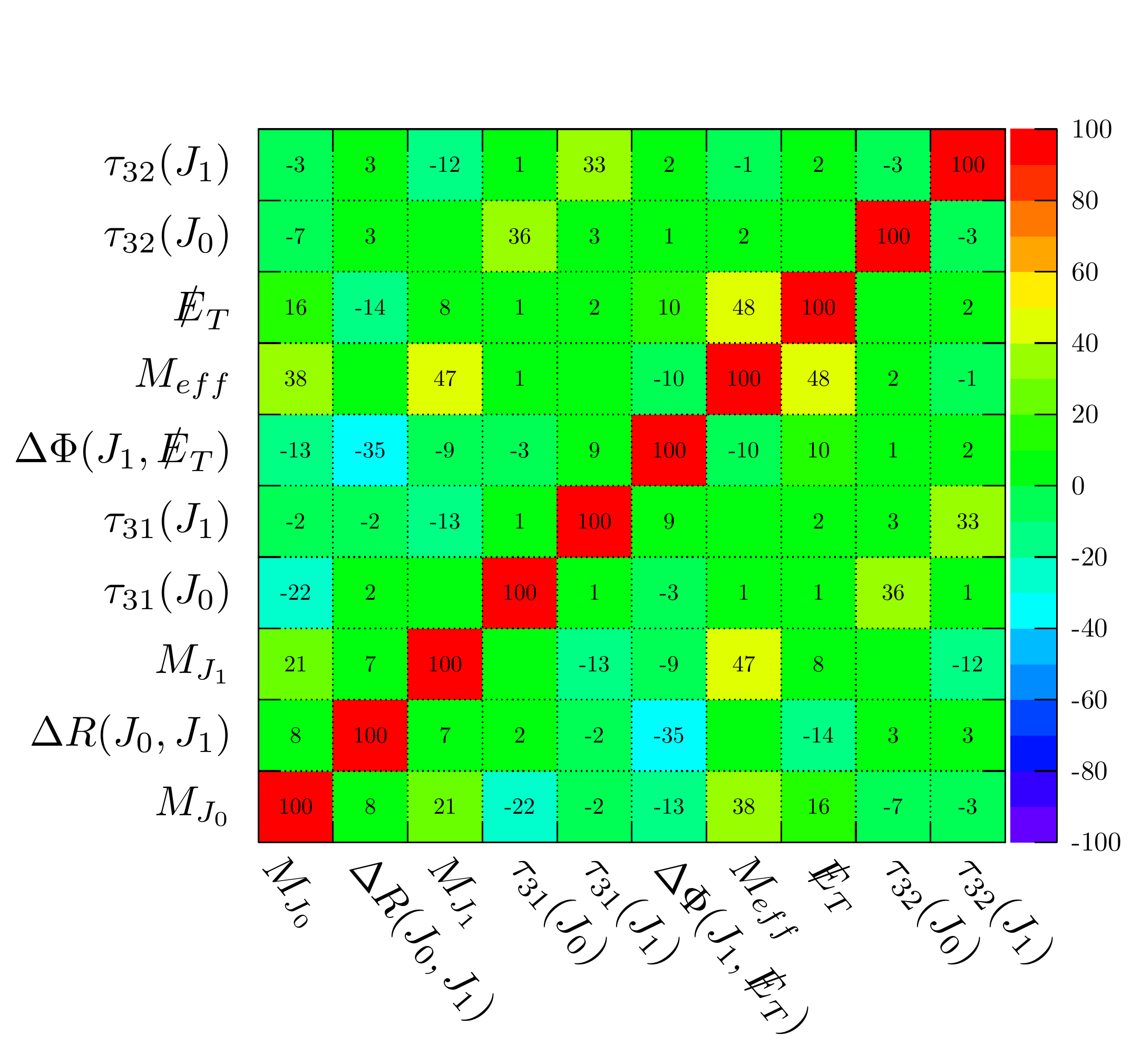}}
  \caption{Linear correlation coefficients (in percentage) between different kinematical variables for the signal (BP3, left panel) and background (right panel). Missing entries correspond to a negligible correlation smaller than one. Positive and negative coefficients indicate that two variables are correlated or anti-correlated, respectively. 
  }\label{correlation}
\end{figure}

\begin{figure}[tb]
\centering
  \subfloat {\label{MVA:sig_bg}\includegraphics[width=0.495\textwidth]{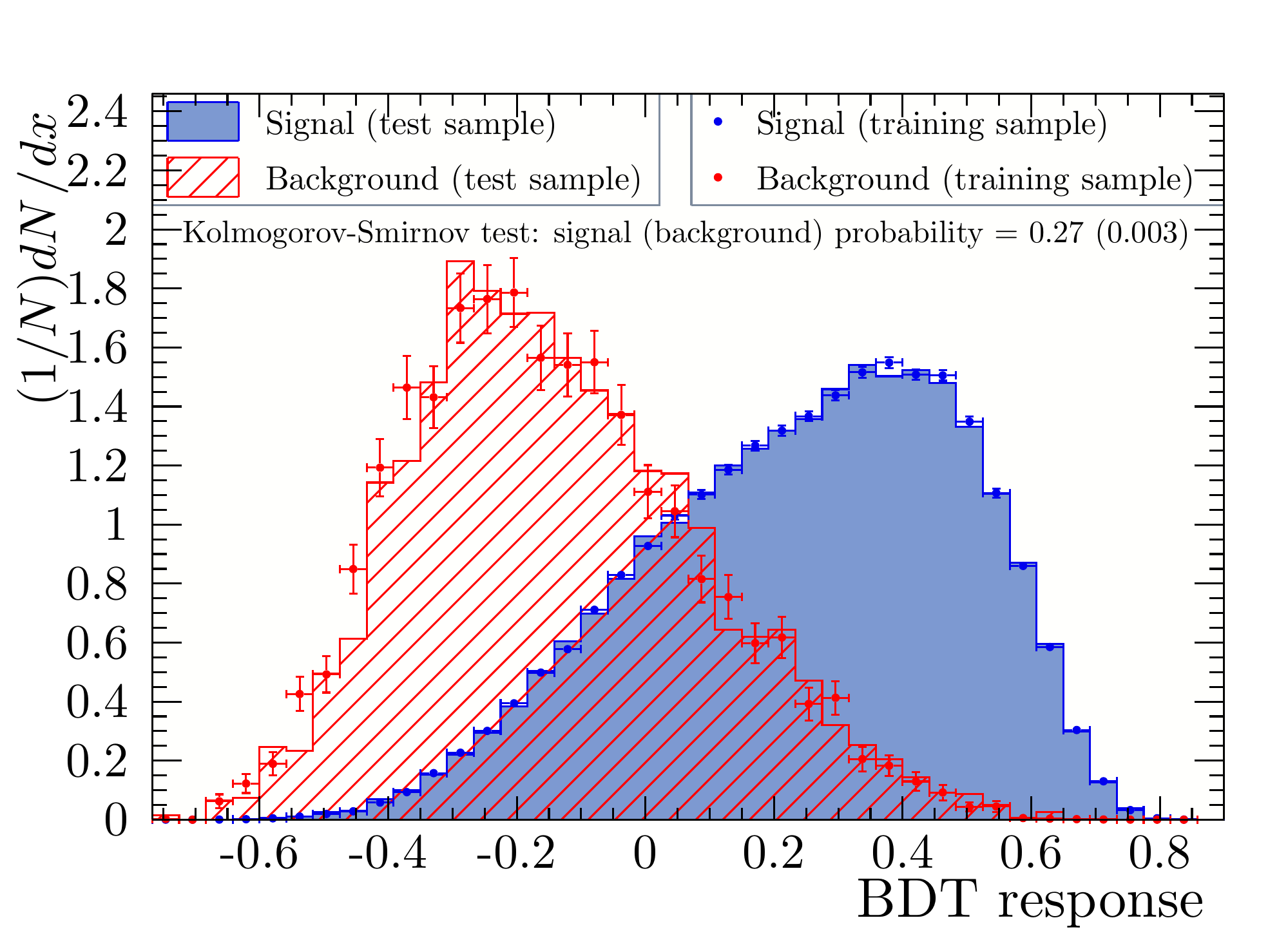}} 
  \subfloat {\label{MVA:significance}\includegraphics[width=0.495\textwidth]{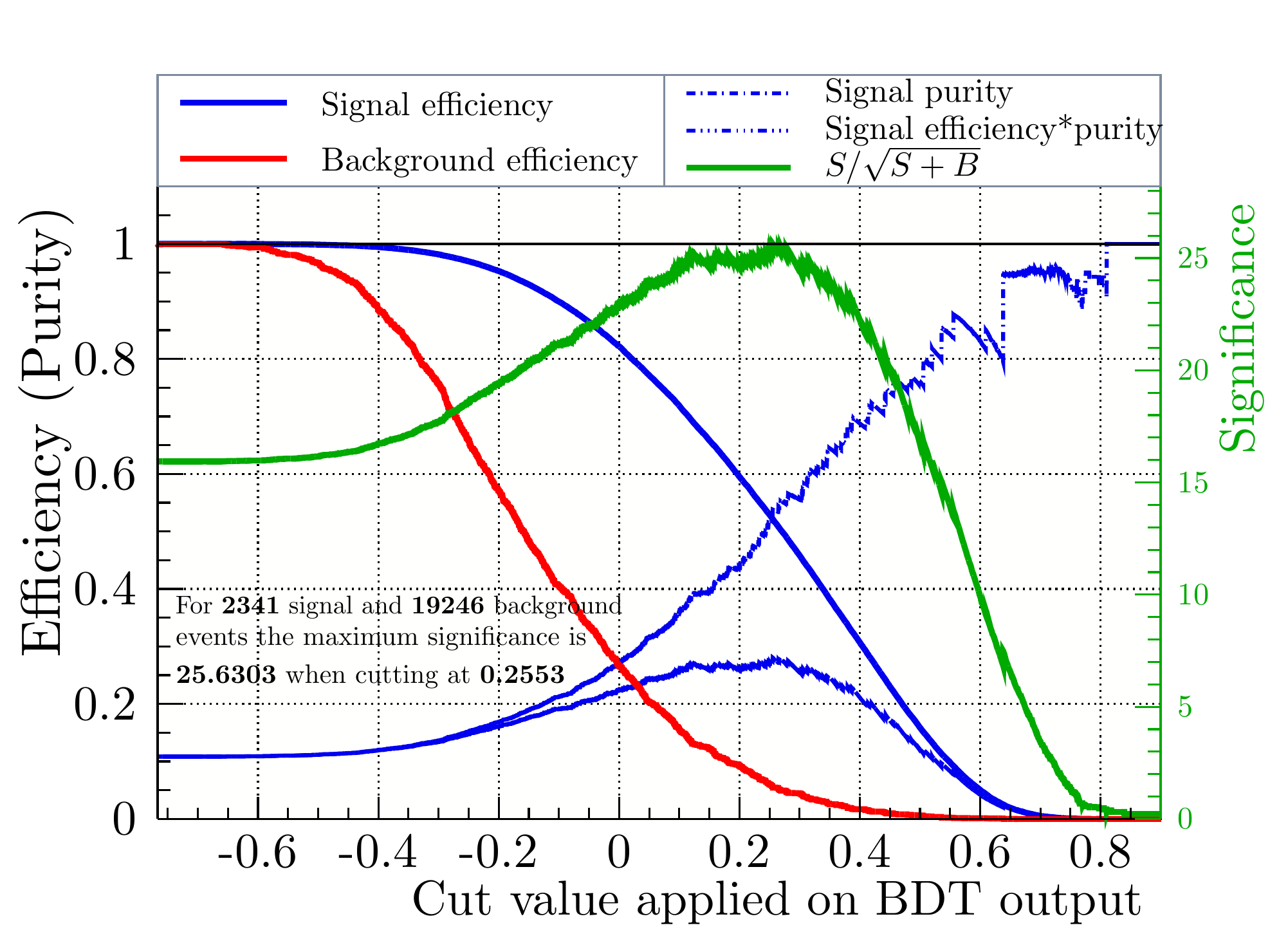}}
  \caption{Left panel: normalized distribution of the  BDT response for both signal (blue, BP3) and background (red) classes (both training and testing samples of both classes). Right panel: signal (blue) and background (red) efficiencies and the statistical significance of the signal (green) as a function of cut applied on BDT output.}\label{sig_bg_significance}
\end{figure}

With these selection criteria, we keep all other variables unrestrained, giving enough scope to the multivariate analysis to find an optimal nonlinear cut based on the suitable variables. The expected number of signal and background events after applying MVA selection cuts at 14 TeV LHC for 139 $\text{fb}^{-1}$ integrated luminosity is given in Tab. \ref{tab:MVA}. We apply the adaptive Boosted Decision Tree (BDT) algorithm in our analysis and construct statistically independent signal and background event samples. Each event sample is split randomly for training and testing purposes. Since multiple processes contribute to the total background, we generate them with two to four extra jets MLM matching separately and combine them in proportion to their proper weight to get a combined background sample. For multivariate analysis, a final set of kinematic variables are accepted from a larger set, where we retain only those variables that are less (anti) correlated in both signal and background and have larger relative importance. Even before implementing any model, a variable can have more relative importance than another when it has larger discriminating power separating the signal class from the background class. We find $P_T(J_0)$, $P_T(J_1)$, and $\sqrt{\hat{S}_{min}}$ \cite{Konar:2008ei, Konar:2010ma} are highly correlated with $M_{\text{eff}}$ in both signal and background. However, we keep $M_{\text{eff}}$ as it has larger relative importance than other variables. $\sqrt{\hat{S}_{min}}$ is defined as
\begin{equation}
\sqrt{\hat{S}_{min}} = \sqrt{(\sum_j E_j)^2 - (\sum_j P_{z,j})^2} + \slashed{E}_T 
\label{Eq.shat}
\end{equation}
where summation runs over all the visible jets. From Eq. \ref{Eq.Meff} and Eq. \ref{Eq.shat}, the above correlations are expected. We also observe that $\Delta \Phi (J_0,\slashed{E}_T)$ and $\Delta \Phi (J_1,\slashed{E}_T)$ are moderately anti-correlated in signal but highly anti-correlated in background. The moderate anti-correlation of the signal is because of the total availability of phase space of the two missing particles. In the case of background, for example, the principle $t \bar{t}+$ jets background, the only allowed phase space is when both top and anti-top are highly boosted and move almost in the opposite direction, where one of the reconstructed tops gives the leading fatjet, and another one gives both subleading fatjet and large missing transverse momentum. As a result, these two variables are highly anti-correlated in the background. We keep $\Delta \Phi (J_1,\slashed{E}_T)$ as it has larger relative importance than $\Delta \Phi (J_0,\slashed{E}_T)$. We notice that $M_{T2}$ and $\slashed{E}_T$ are also highly correlated in signal and moderately in background.  
Since $\slashed{E}_T$ has the largest relative importance, we choose $\slashed{E}_T$ over $M_{T2}$ for MVA analysis. The relative importance of the different kinematic variables used in MVA is presented in Fig. \ref{MVA:relative_importance} for sample benchmark point BP3. From the normalized distributions in the previous section, we notice that all variables used in MVA are outstanding in distinguishing the signal from the background. However, the $\slashed{E}_T$, $\Delta R(J_0,J_1)$, and $\Delta \Phi (J_1,\slashed{E}_T)$ are the finest among all these useful variables. The linear correlation coefficients (in percentage) between different kinematical variables for the signal (BP3, left panel) and background (right panel) is presented in Fig. \ref{correlation}. BDT algorithm may lead to overtraining for wrong choices of different (BDT specific) parameters during training. Such overtraining can be avoided if one checks the Kolmogorov-Smirnov probability during training. We train the algorithm separately for every benchmark point and confirm that no overtraining exists in our analysis. 

\begin{table}[tb]
\begin{center}
 \scriptsize
%
 \begin{tabular}{|c|c|c|c|c|c|c|c|}
\hline
\multirow{3}{*}{BP} & \multirow{3}{*}{$N_S^{bc}$} & \multirow{3}{*}{BD$T_{opt}$} & \multirow{3}{*}{$N_S$} & \multirow{3}{*}{$N_B$} & \multirow{3}{*}{$\sigma$} & \multirow{3}{*}{$\frac{N_S}{N_B}$} &  \multirow{3}{*}{ $\frac{N_S}{N_B}$ } \\
 & & & & & & & \\
 & & & & & &  (MVA) & (CBA)\\
\hline\hline
BP1 & 6625 & -0.0259 & 5643 & 8584 & 47.3 & 0.66 & 0.41\\
\hline
BP2 & 3525 & 0.1584 & 2325 & 2047 & 35.1 & 1.14 & 0.47\\
\hline
BP3 & 2341 & 0.2553 & 1222 & 1048 & 25.6 & 1.17 & 0.40\\
\hline
BP4 & 2711 & 0.1975 & 1556 & 1277 & 29.2 & 1.22 & 0.35 \\
\hline
BP5 & 2176 & 0.1446 & 1366 & 2502 & 21.9 & 0.55 & 0.23 \\
\hline
BP6 & 1924 & 0.1325 & 1170 & 2727 & 18.7 & 0.43 & 0.17\\
\hline
BP7 & 1424 & 0.1398 & 821 & 2845 & 13.6 & 0.29 & 0.13\\
\hline
BP8 & 1081 & 0.0942 & 624 & 2951 & 10.4 & 0.21 & 0.12\\
\hline
BP9 & 915 & 0.0785 & 627 & 4820 & 8.5 & 0.13 & 0.08\\
\hline
BP10 & 552 & 0.0506 & 311 & 3469 & 5.0 & 0.09 & 0.04\\
\hline
 BP11 & 385 & 0.3875 & 134 & 346 & 6.1 & 0.39 & 0.09 \\
\hline
$N_{SM}$ & 19246 & \multicolumn{6}{| c |}{} \\
\hline
 \end{tabular} 
%
 \begin{tabular}{|c|c|c|c|c|c|}
\hline
\multirow{2}{*}{$\Delta M$ }  & \multirow{2}{*}{$N_S^{bc}$} & \multirow{2}{*}{BD$T_{opt}$} & \multirow{2}{*}{$N_S$} & \multirow{2}{*}{$N_B$} & $\sigma=$  \\
 (GeV)& & & & & $\frac{N_S}{\sqrt{N_S+N_B}}$  \\
\hline\hline
75.1   & 2315 & 0.2645 & 1252 & 1240 & 25.1 \\
\hline
100 (BP3)  & 2341 & 0.2553 & 1222 & 1048 & 25.6 \\
\hline
153.6  & 2258 & 0.2129 & 1246 & 1176 & 25.3 \\
\hline
200.4  & 2035 & 0.1899 & 1284 & 1384 & 24.9 \\
\hline
244.5  & 2019 & 0.2963 & 1159 & 948 & 25.2 \\
\hline
302.7  & 2036 & 0.208 & 1254 & 1240 & 25.1 \\
\hline
351  & 2034 & 0.2271 & 1261 & 1199 & 25.4 \\
\hline
402.7  & 2023 & 0.2778 & 1133 & 882 & 25.2 \\
\hline
$N_{SM}$ & 19246 & \multicolumn{4}{| c |}{} \\
\hline
 \end{tabular}     
%
\caption{ The upper table demonstrates the effectiveness of the current search in terms of statistical significance ($\sigma$) for different benchmark points conceived for this study. The lower table illustrates the variation of this potential for one benchmark point, changing the mass of the heavy scalar, $S_2$, and shows that this mass does not have much impact in exploring the parameter space.
$N_S^{bc}$ and $N_{SM}$ are the total number of events for different signal benchmark points and the combined background before applying any cut on BDT output (as shown in Tab. \ref{tab:MVA}). After using an optimal selection on the BDT response (BD$T_{opt}$) surviving number of signal and background events are given by $N_S$ and $N_B$, respectively for 14 TeV LHC for an integrated luminosity 139 $\text{fb}^{-1}$. Corresponding statistical significance and the signal-to-background ratio are also presented for ready reference. To better compare the sensitivities between the different analysis methods, we add the $\frac{N_S}{N_B}$ ratio of CBA from Tab. \ref{tab:fatjet_sig} in the last column of the upper table.
}
\label{tab:BDT}
\label{tab:varry_S2}
\end{center}
\end{table}

The normalized distribution of the BDT response of the signal (BP3) and the background classes (both training and testing samples of both classes) is shown in Fig. \ref{sig_bg_significance}. We notice both the classes are well separated. We present the variation of the signal and background efficiencies and the statistical significance of the signal (BP3) with the cut applied on the BDT response in the right panel of Fig. \ref{sig_bg_significance}. Statistical significance is defined as $\sigma=\frac{N_S}{\sqrt{N_S+N_B}}$. The number of events that survive after applying the $\text{BDT}_\text{res}>\text{BDT}_{opt}$ cut for signal and background is $N_S$ and $N_B$, respectively. $\text{BDT}_{opt}$ is the optimal cut for which the significance is maximum. In Tab. \ref{tab:BDT} (upper) $N_S$, $N_B$, $\sigma$, and $\dfrac{N_S}{N_B}$ are presented for different benchmark points at 14 TeV LHC with integrated luminosity 139 $\text{fb}^{-1}$. We find that for a few of the chosen BPs, the number of signal events is larger than the background events after the $\text{BDT}_\text{res}>\text{BDT}_{opt}$ cut, and for all eleven benchmark points, we reach the discovery potential with integrated luminosity 139 $\text{fb}^{-1}$.

Our next interest would be to verify how significance varies with the mass of the scalar $S_2$. For that purpose, we generated the event samples separately for different masses of the $S_2$ with the same DM mass $M_{S_1}=403$ GeV, VLQ mass $M_{\Psi}=808$ GeV, and the coupling constant $f_t=1$. We train the algorithm separately for different samples of different $S_2$ masses and confirm that no overtraining exists in our analysis and perform the MVA. 

There are two possible hierarchies possible: $M_{S_2}>M_{\Psi}>M_{S_1}$ and $M_{\Psi}>M_{S_2}>M_{S_1}$. 
We consider two boosted tops associated with missing transverse momentum as our signal. It is interesting to note that the significance of the former hierarchy is always greater or equal to the second. In the case of the former, $S_2$ can decay into $\Psi (\rightarrow S_1 j) j$ or $S_1 j j$ (through off-shell $\Psi$), where $j$ is the up-type SM quark. If at least one of these jets is the top quark, then the signal efficiency increases and hence the significance. So the hierarchy $M_{\Psi}>M_{S_2}>M_{S_1}$ gives a lower statistical significance, and we consider this scenario throughout our result for a conservative estimation.
The total number of events coming from the signal topology for different masses of $S_2$ and background events after applying an optimal cut (BD$T_{opt}$) is given in Tab. \ref{tab:varry_S2} (lower) for the hierarchy $M_{\Psi}>M_{S_2}>M_{S_1}$. The statistical significance variation with the $S_2$ mass is also shown here for a given mass of $\Psi$ and couplings. The mass of $S_2$ has no effect on the statistical significance of the boosted top fatjets plus a large missing momentum signature. However, if one of the decay products of $S_2$ ($S_2 \rightarrow S_1 j j$) is at least a top quark, then it can increase the significance.

\begin{figure}[tb!]
\centering
\includegraphics[width=0.495\textwidth]{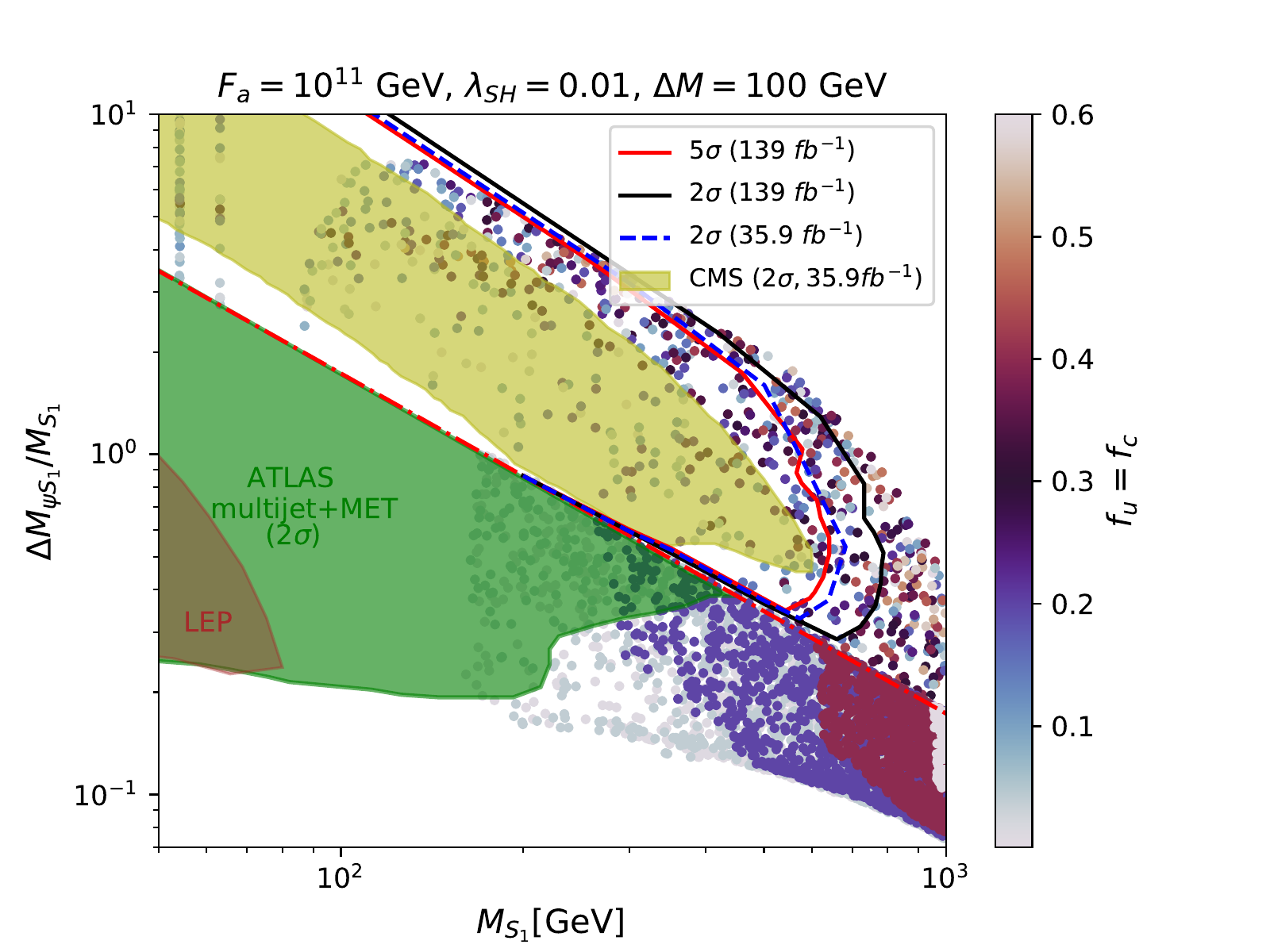}
\includegraphics[width=0.495\textwidth]{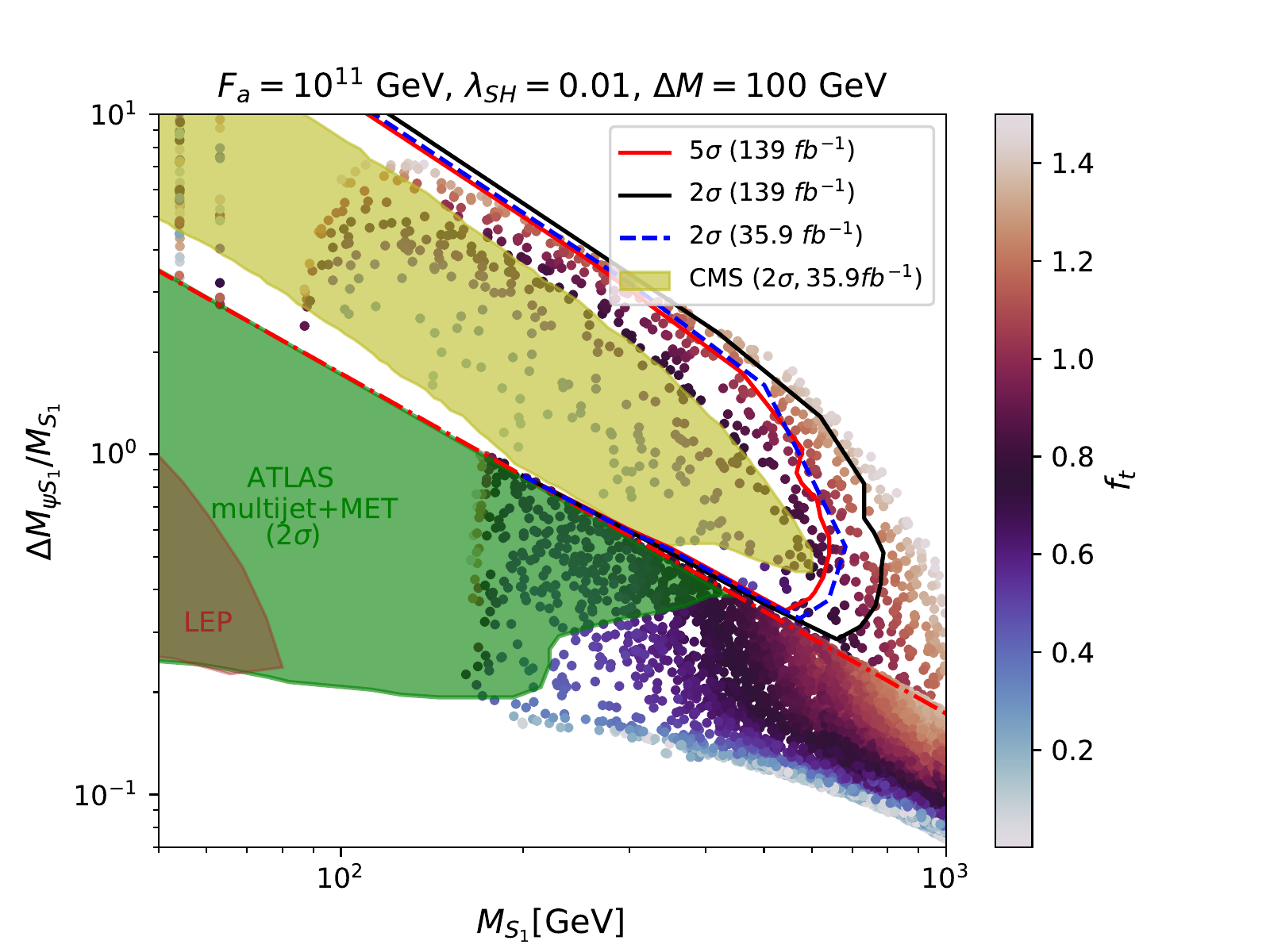}
\caption{The solid red line is the $5\sigma$ discovery contour, and any point inside the red line has a statistical significance $>5\sigma$ at 14 TeV LHC for an integrated luminosity 139 $\text{fb}^{-1}$. The dashed red line corresponds to $\Delta M_{\Psi S_1}=M_{\Psi}-M_{S_1}=M_{\text{top}}$. Below the dashed red line, we can not probe with the boosted tops plus missing energy signal, as we can not get any on-shell top from the decay of $\Psi$. The solid black and blue dashed lines are the exclusion contour ($2\sigma$) of our analysis for an integrated luminosity of 139 $\text{fb}^{-1}$ and 35.9 $\text{fb}^{-1}$, respectively. The exclusion region ($2\sigma$) from the existing LEP, ATLAS (multijet $+$ MET), and CMS ($t \bar{t}$ $+$ MET) analysis are shown by brown, green, and olive color, respectively. }
\label{result_final}
\end{figure}
 
Finally, we present the discovery ($5 \sigma$) and exclusion ($2 \sigma$) contours from our analysis at 14 TeV LHC for an integrated luminosity 139 $\text{fb}^{-1}$ in the bi-dimensional plane of $\dfrac{\Delta M_{\Psi S_1}}{M_{S_1}}-M_{S_1}$ in Fig. \ref{result_final} by solid red and solid black lines, respectively. Our analysis is effective when the on-shell top is produced from $\Psi$ decay. Hence the region below the dashed red line can not be probed in the present channel. Considering the $100\%$ branching fraction of the decay of $\Psi$ into the top quark associated with the scalar, the existing search~\cite{Colucci:2018vxz} can exclude vector-like quark masses up to 1 TeV. We find the masses of the vector-like quark ranging up to 1.41 TeV can be excluded, while the masses extent to 1.28 TeV can be discovered at 14 TeV LHC with an integrated luminosity of 139 $\text{fb}^{-1}$. In the region below the dashed red line, the mass difference between a vector-like quark and the scalar DM is less than the top quark's mass, and the vector-like quark fully decays into a light quark associated with a scalar when kinematically allowed. So, one can probe those regions using multi jets plus missing transverse momentum signature, which is beyond the scope of our present analysis.

\section{Summary and Conclusion}
\label{conclusions}

In this work, we analyze a hybrid KSVZ setup, where the model is extended by an extra complex scalar singlet whose lightest component plays the role of dark matter. We highlight the fact that the presence of a colored vector-like quark that occurs naturally in the KSVZ model plays a crucial role both in the dark sector and collider phenomenology of the setup. Being charged under $U(1)_{PQ}$ allows the VLQ to couple with all up-type quarks and the DM through the Yukawa interactions. When appropriately tuned, this coupling can enhance the DM parameter space in comparison to what is observed in a pure scalar singlet DM scenario. In this work, we demonstrate that the Yukawa couplings play a non-trivial role in obtaining the observed relic density. Moreover, the same couplings also allow the parameter space from the direct search bounds by entering into extra Feynman diagrams that contribute toward the direct detection cross-section of the dark matter.

A search of vector-like quarks in events with two boosted top fatjets with large missing transverse momentum is presented. The analysis is done for 139 $\text{fb}^{-1}$ integrated luminosity at 14 TeV LHC. We discuss all the significant backgrounds that can potentially mimic the signal. Jet substructure variables and various other variables are used in our analysis. Sophisticated multivariate analysis is performed to increase the sensitivity over cut-based one. Different jet substructure variables, $\Delta R (J_0,J_1)$, N-subjettiness ratios, and $M_{eff}$ are outstanding in distinguishing the signal from the background and take a central role in getting very high significance. However, the missing transverse momentum distribution and the azimuthal separation between the subleading fatjet and the missing transverse energy direction have the uppermost importance in separating the signal from the background. With a conservative estimation, we give discovery and exclusion contours in Fig. \ref{result_final} in the region where the mass difference between the vector-like quark and the scalar dark matter is larger than the top quark mass.

\acknowledgments
 This work is supported by the Physical Research Laboratory (PRL), Department of Space, Government of India. Computational work was performed using the HPC resources (Vikram-100 HPC) and TDP project at PRL. R.R. also acknowledges the National Research Foundation of Korea (NRF) grant funded by the Korean government (NRF-2020R1C1C1012452).

\appendix 
\section{Feynman Diagrams}
\label{appendixFR}

\begin{figure}[H]
\centering
\includegraphics[width=0.32\textwidth]{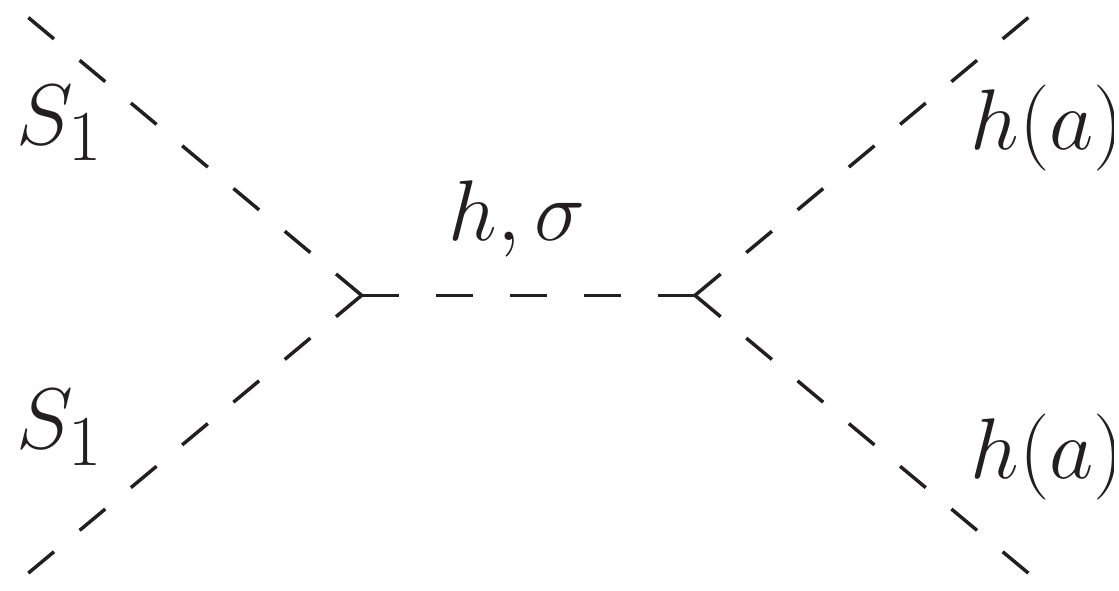}
\includegraphics[width=0.32\textwidth]{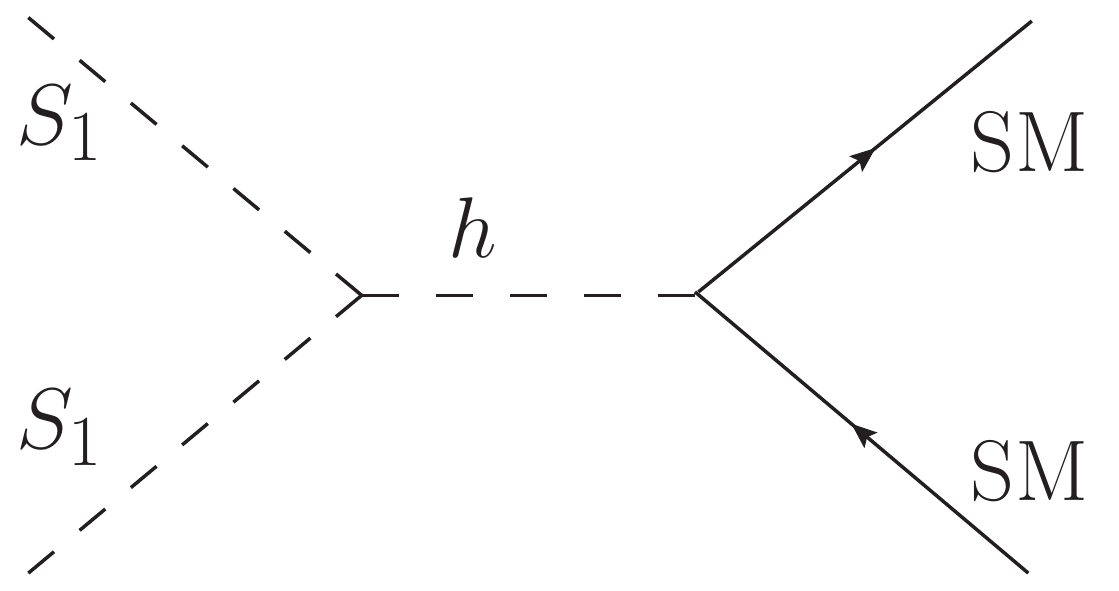}
\includegraphics[width=0.32\textwidth]{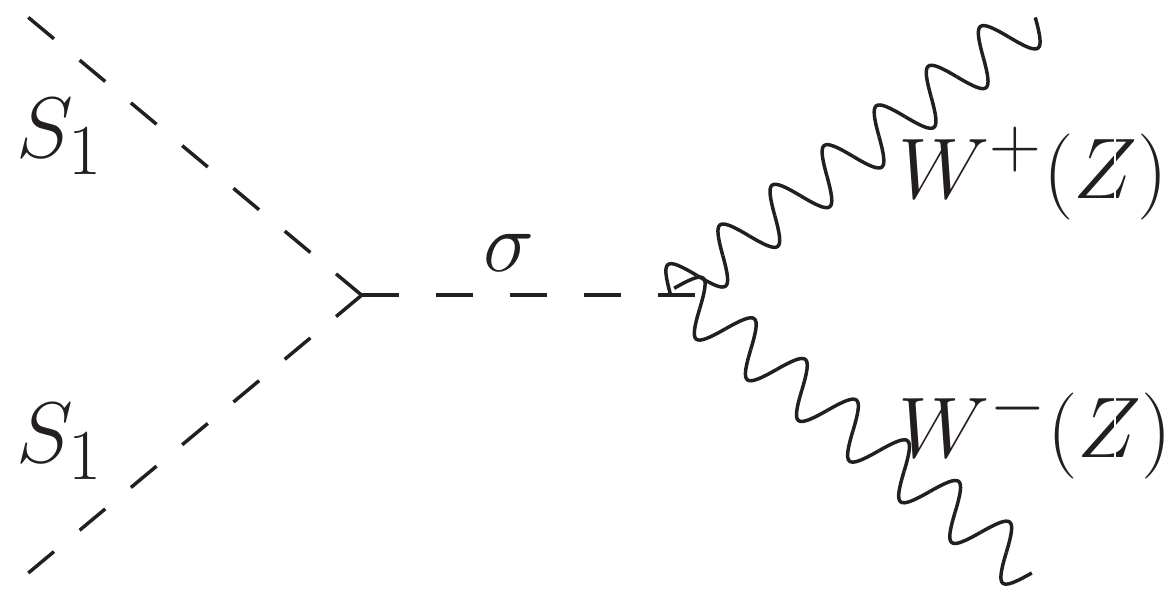}
\includegraphics[width=0.32\textwidth]{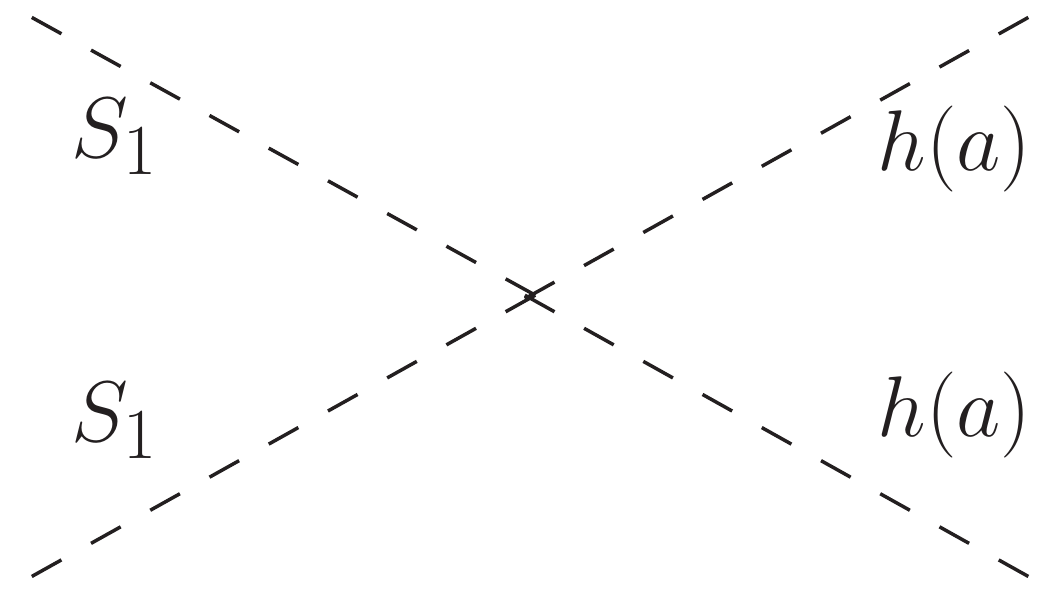}
\includegraphics[width=0.32\textwidth]{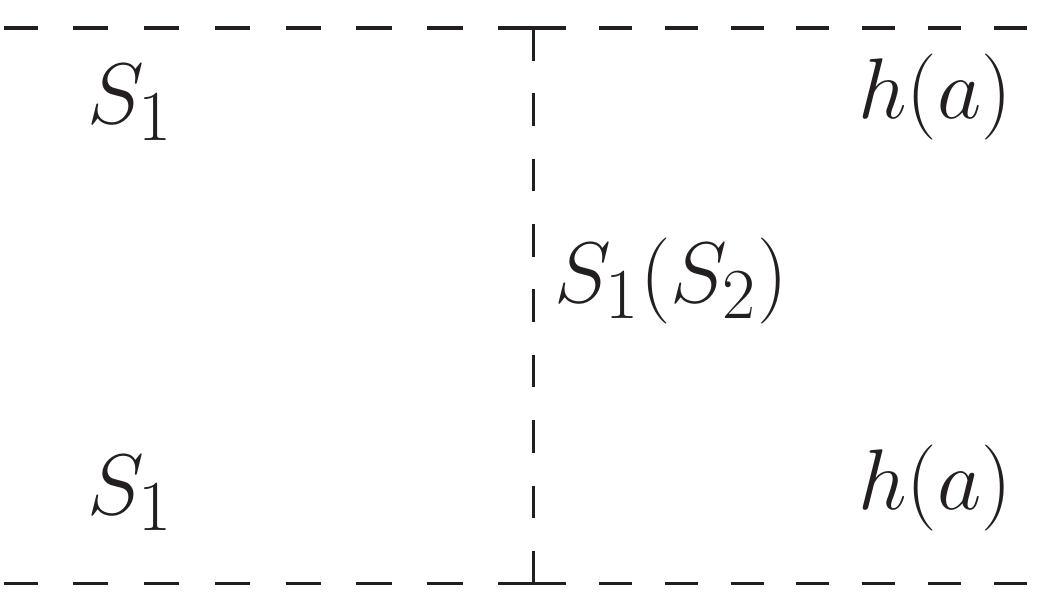}
\includegraphics[width=0.325\textwidth]{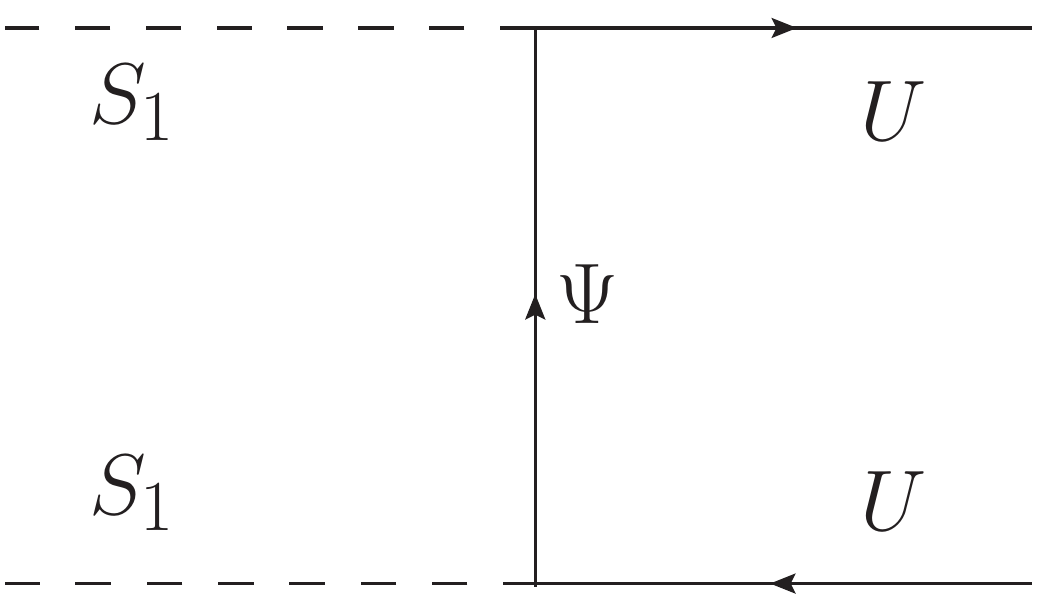}
\caption{Annihilation channels of scalar dark matter $S_1$. U denotes the SM up-type quark ($U \equiv u,c,t, \bar{u}, \bar{c}, \bar{t}$)}
\label{annihilation}
\end{figure}

\begin{figure}[H]
\centering
\includegraphics[width=0.32\textwidth]{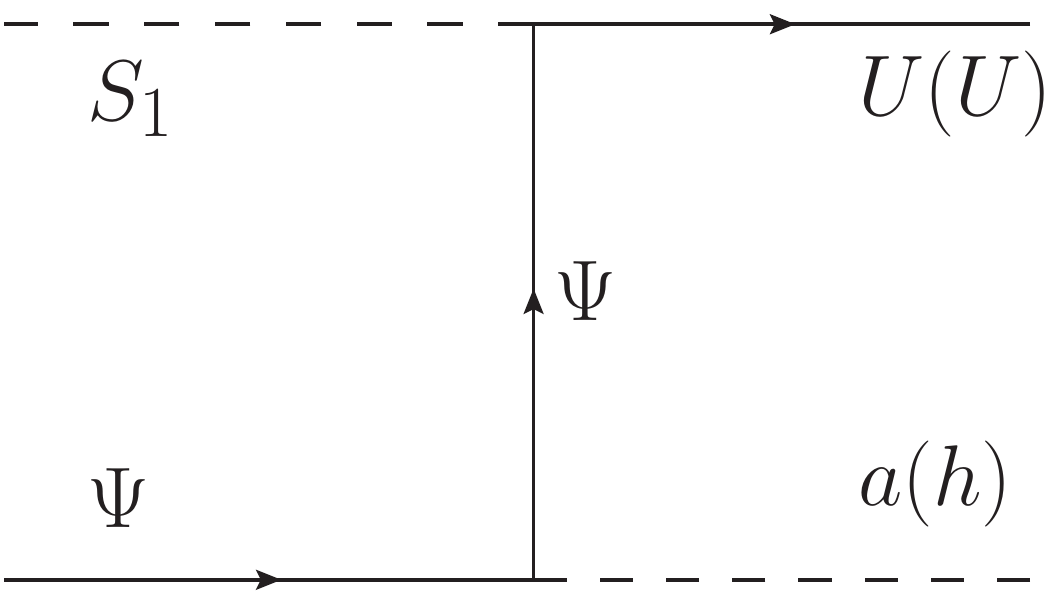}
\includegraphics[width=0.32\textwidth]{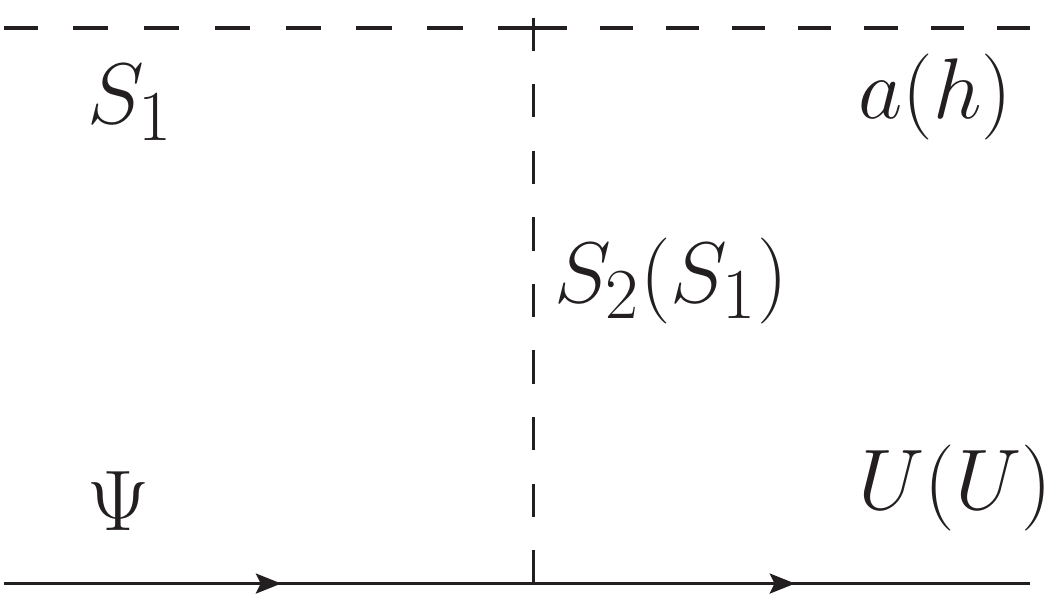}
\includegraphics[width=0.32\textwidth]{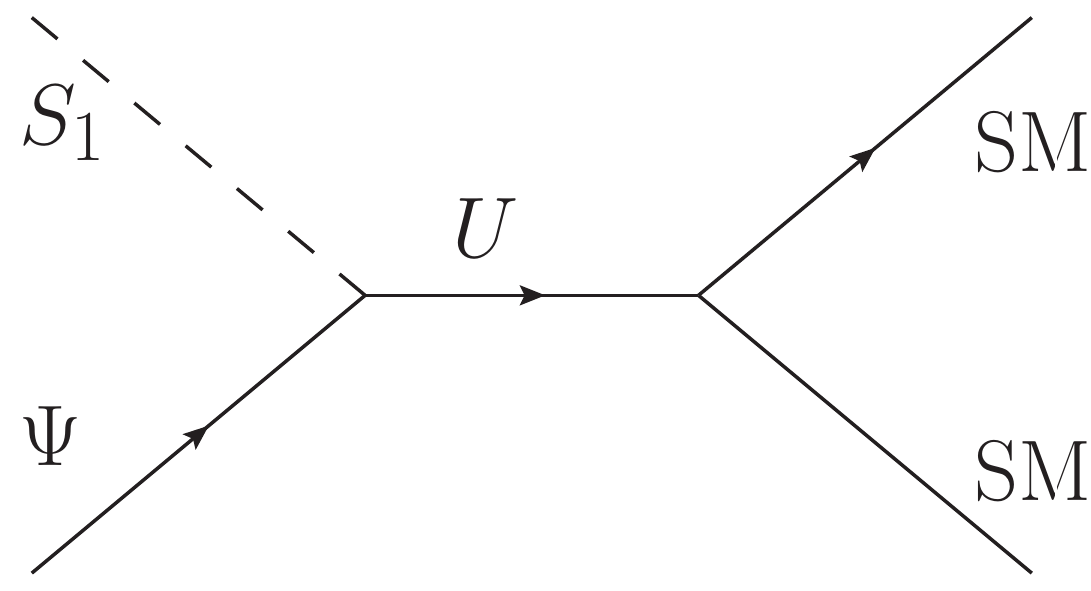}
\includegraphics[width=0.32\textwidth]{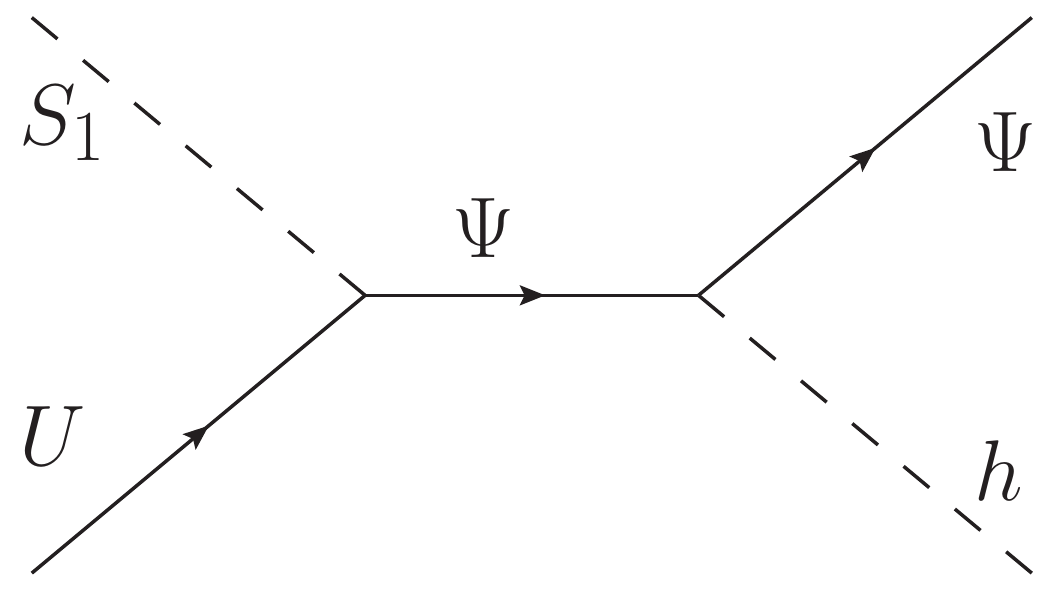}
\includegraphics[width=0.32\textwidth]{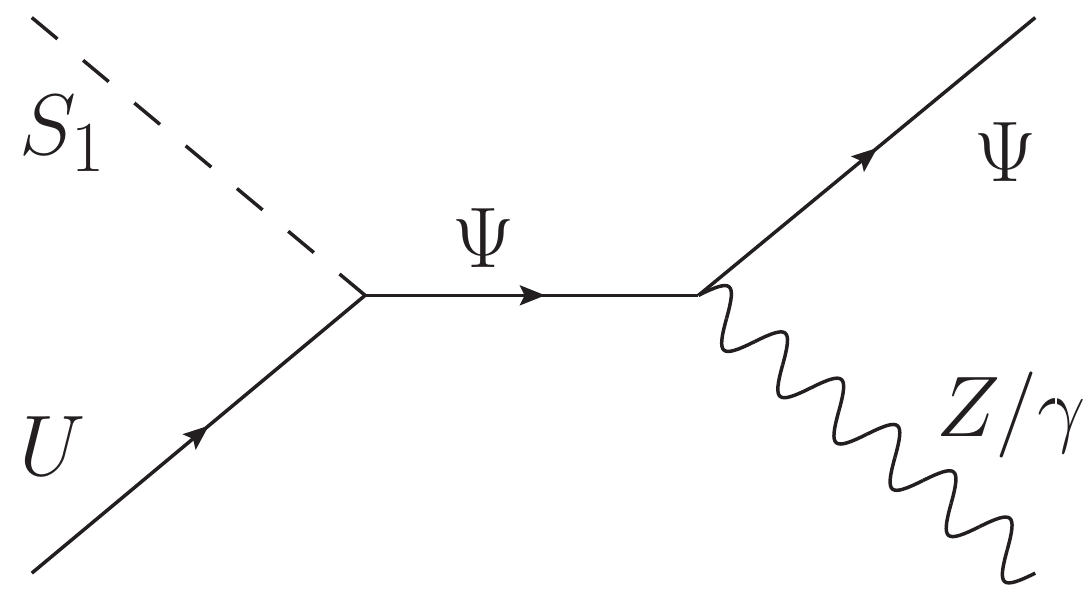}
\includegraphics[width=0.32\textwidth]{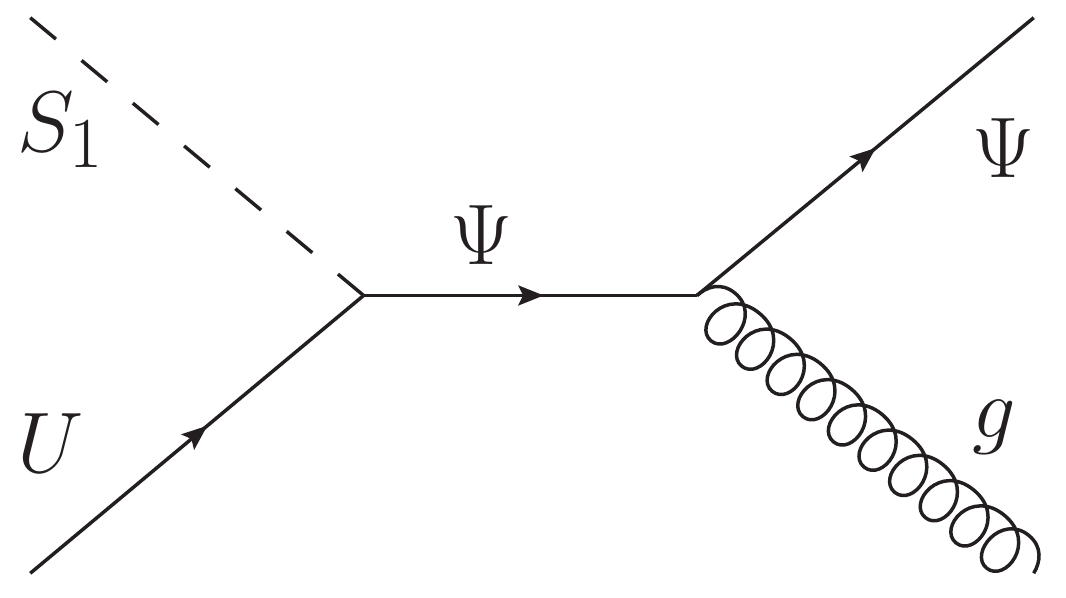}
\includegraphics[width=0.32\textwidth]{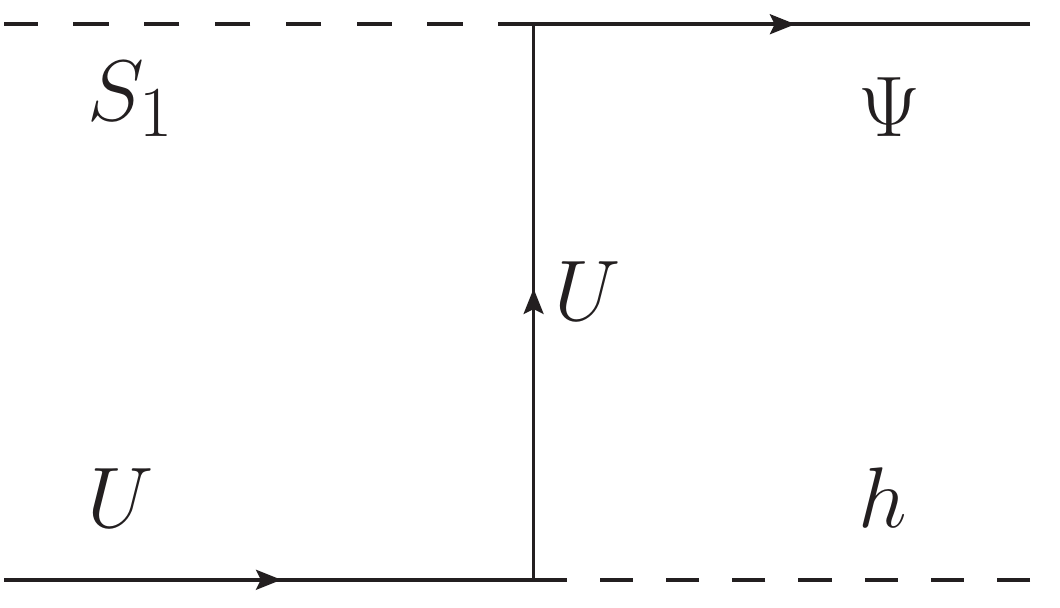}
\includegraphics[width=0.32\textwidth]{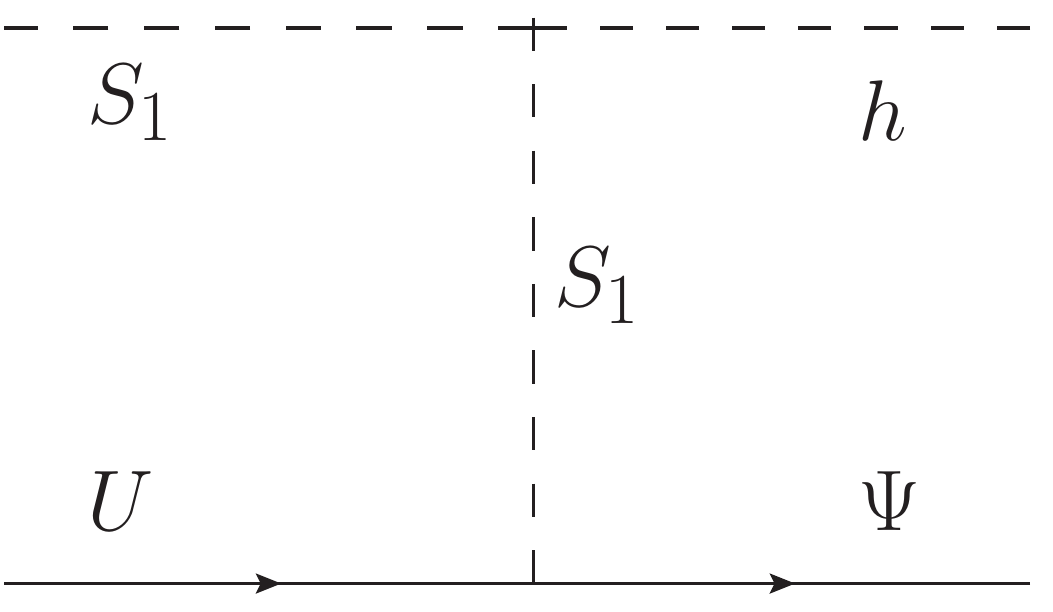}
\includegraphics[width=0.32\textwidth]{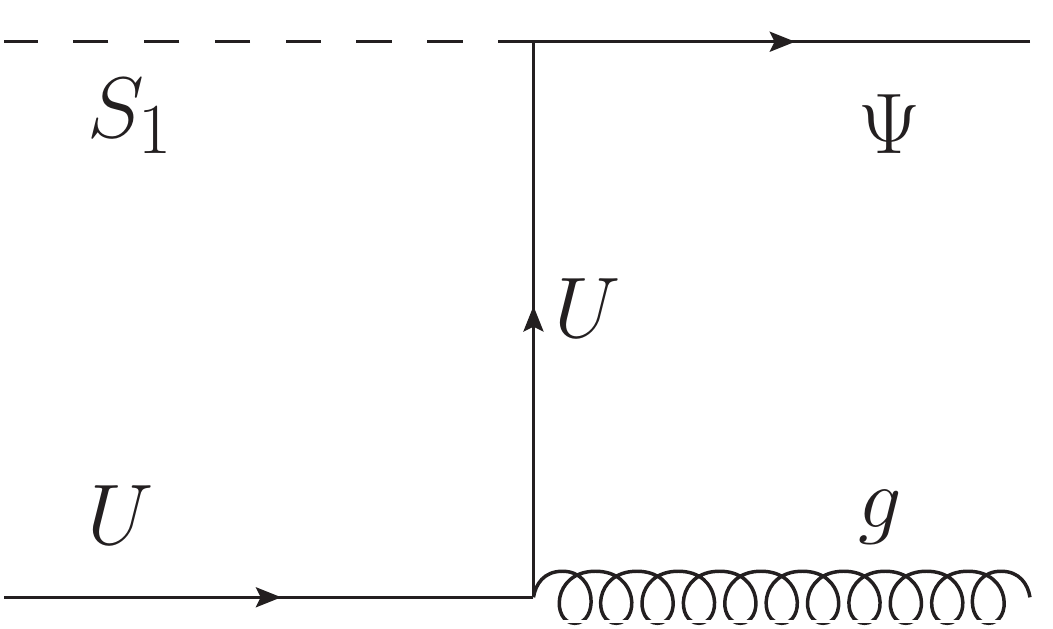}
\includegraphics[width=0.32\textwidth]{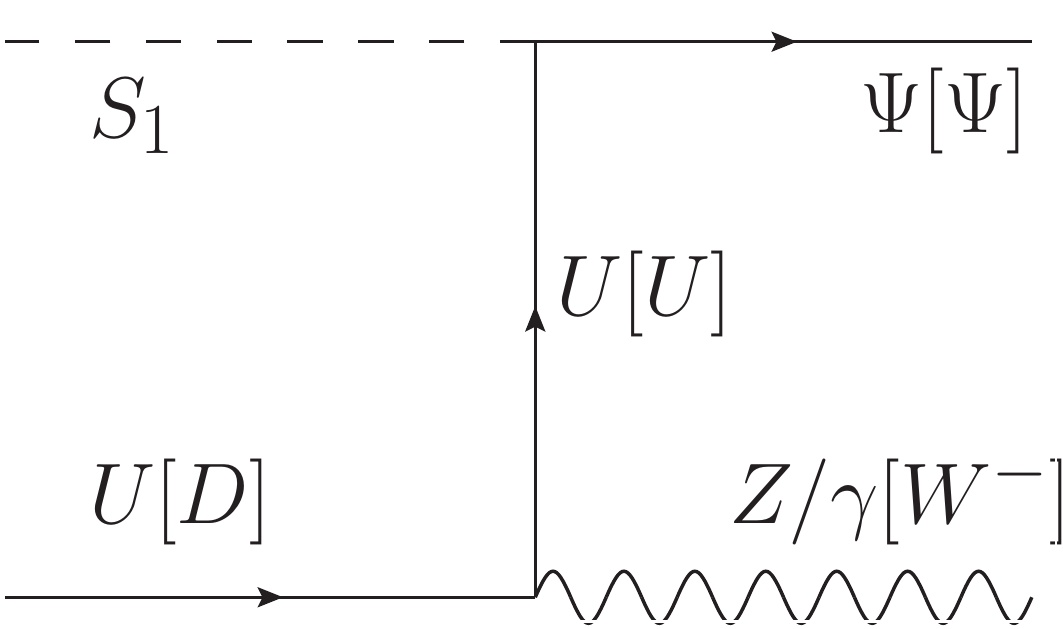}
\includegraphics[width=0.32\textwidth]{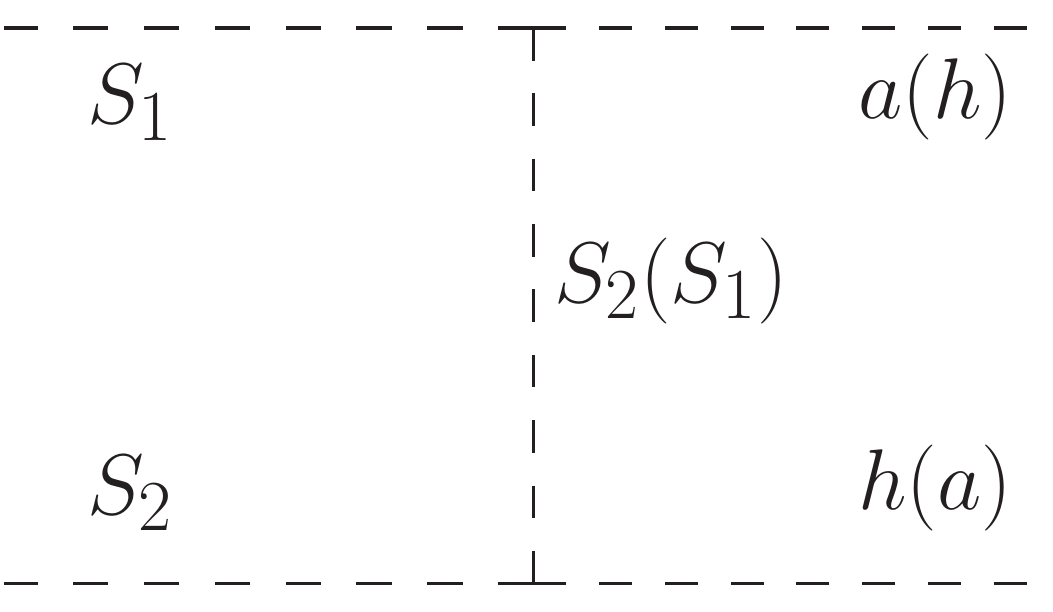}
\includegraphics[width=0.32\textwidth]{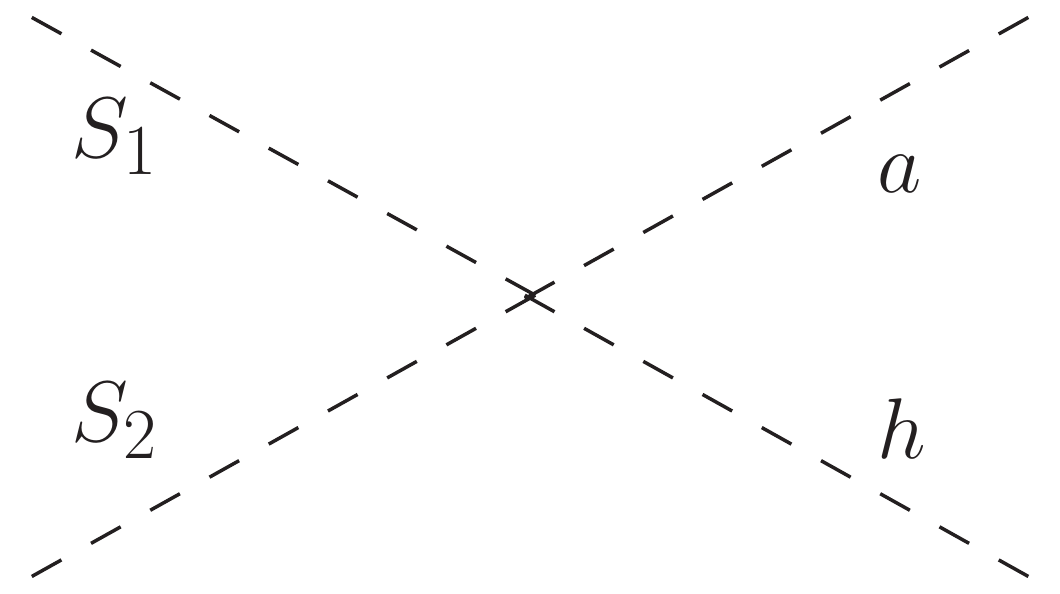}
\includegraphics[width=0.32\textwidth]{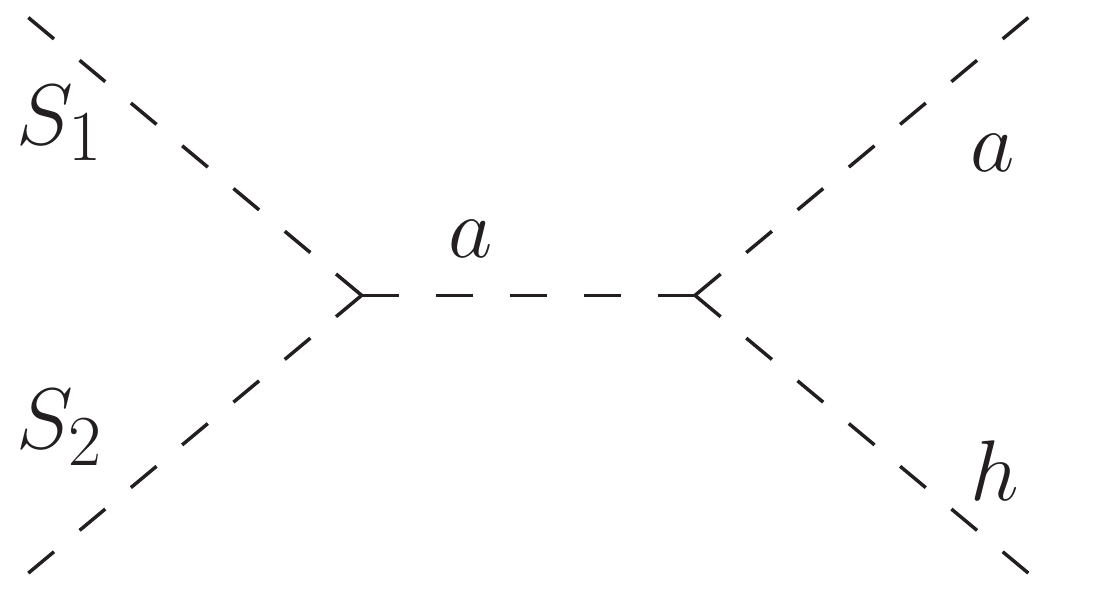}
\includegraphics[width=0.32\textwidth]{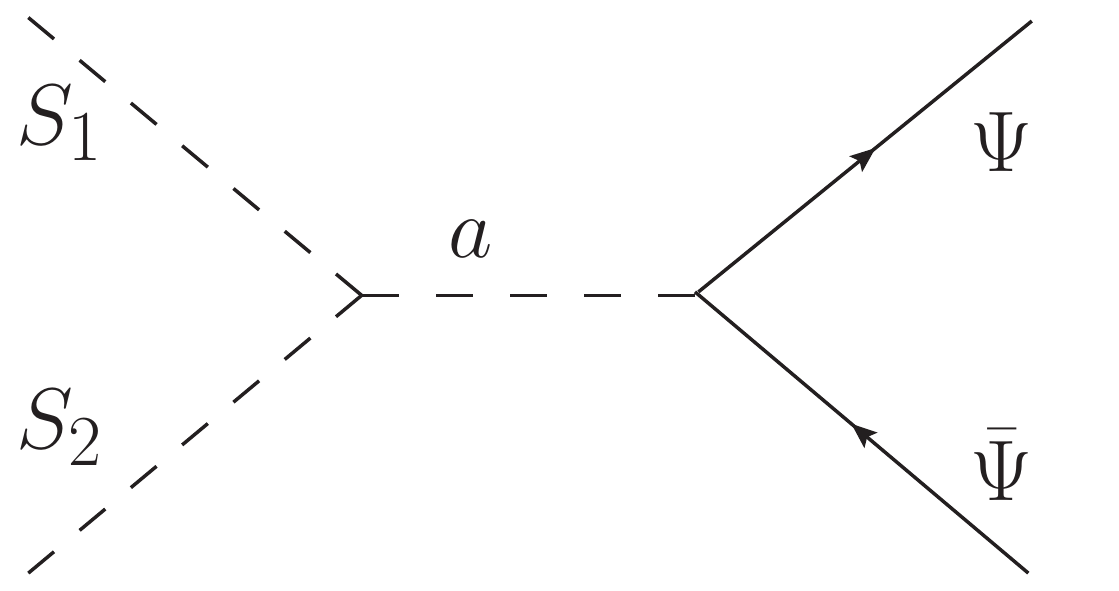}
\includegraphics[width=0.32\textwidth]{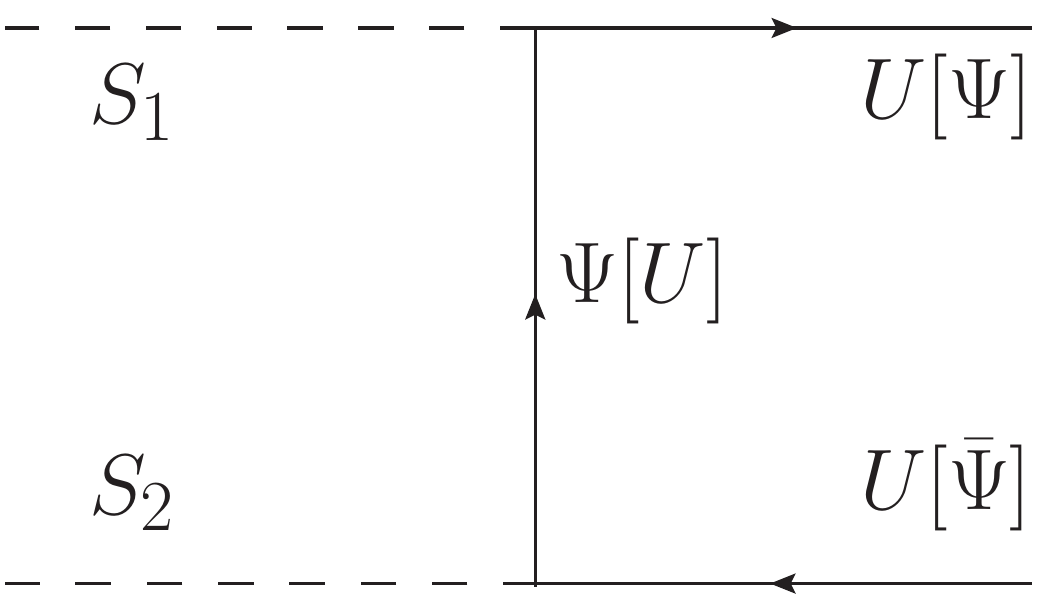}
\caption{Co-annihilation channels of scalar dark matter $S_1$. U and D denote the SM up-type and down-type quark, respectively; $U \equiv u,c,t, \bar{u}, \bar{c}, \bar{t}$, $D \equiv d,s,b, \bar{d}, \bar{s}, \bar{b}$}
\label{co-annihilation}
\end{figure}

\begin{figure}[H]
\centering
\includegraphics[width=0.32\textwidth]{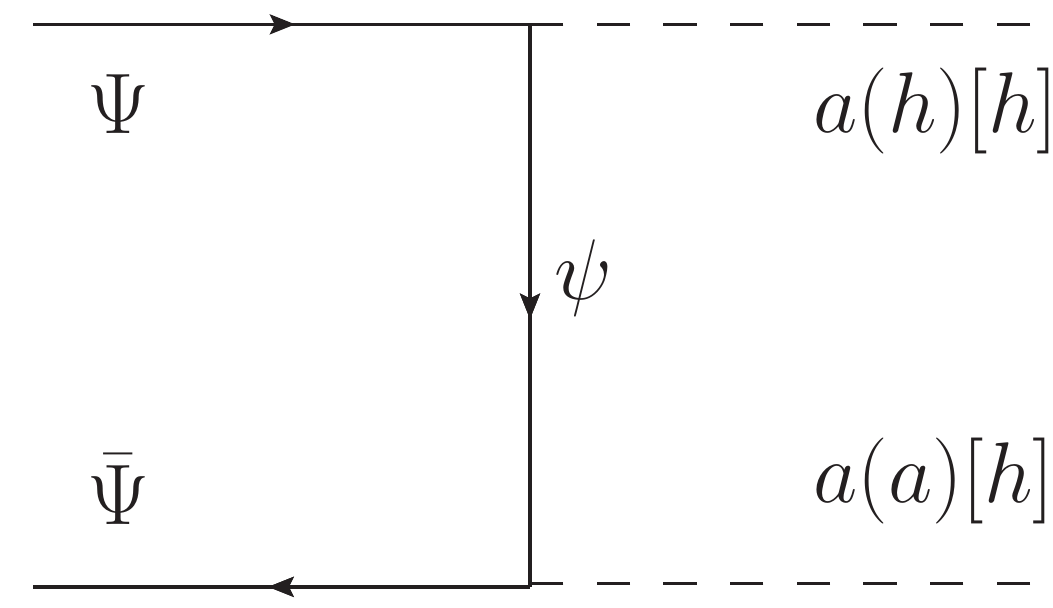}
\includegraphics[width=0.32\textwidth]{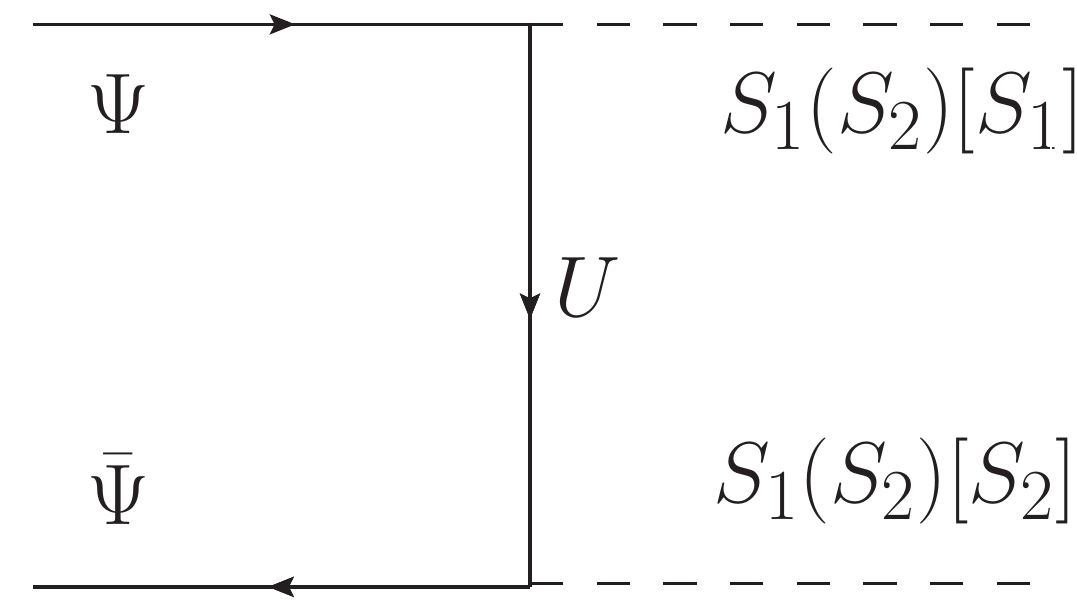}
\includegraphics[width=0.32\textwidth]{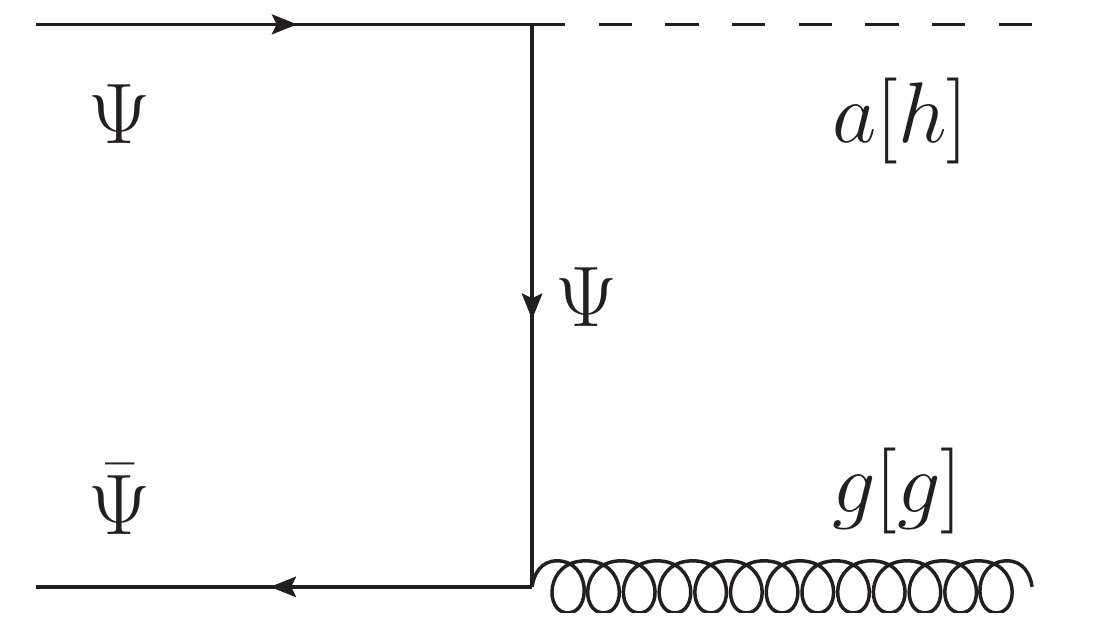}
\includegraphics[width=0.32\textwidth]{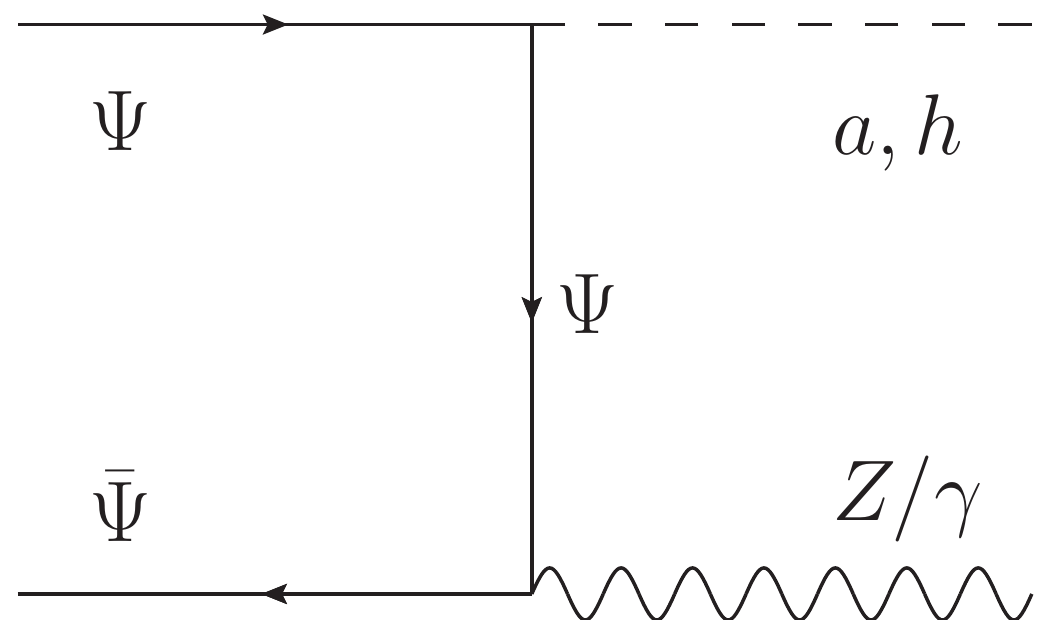}
\includegraphics[width=0.32\textwidth]{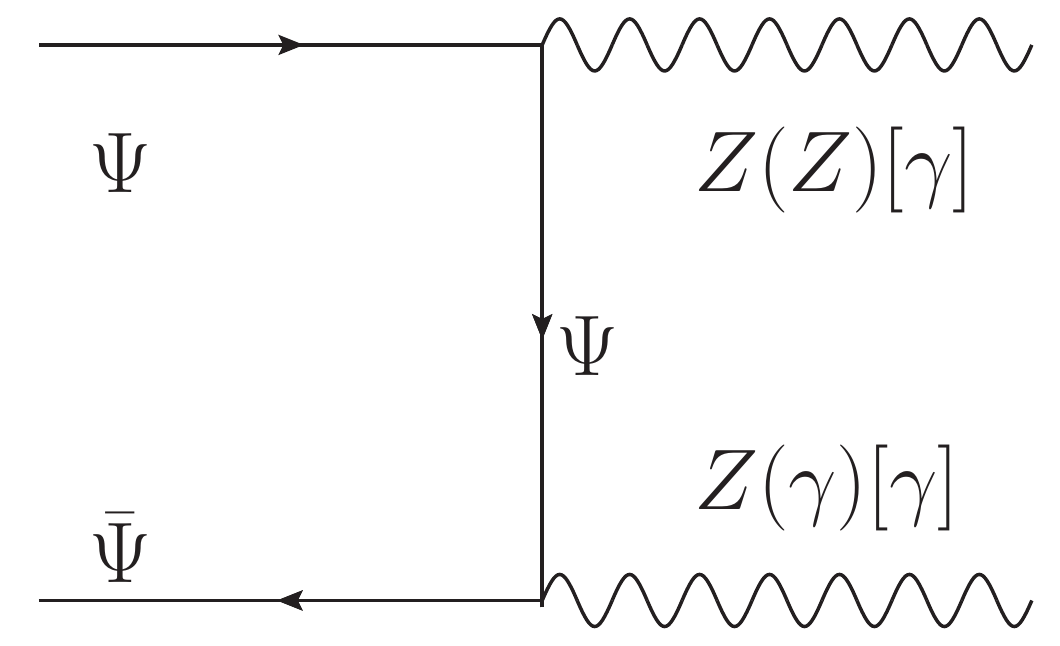}
\includegraphics[width=0.32\textwidth]{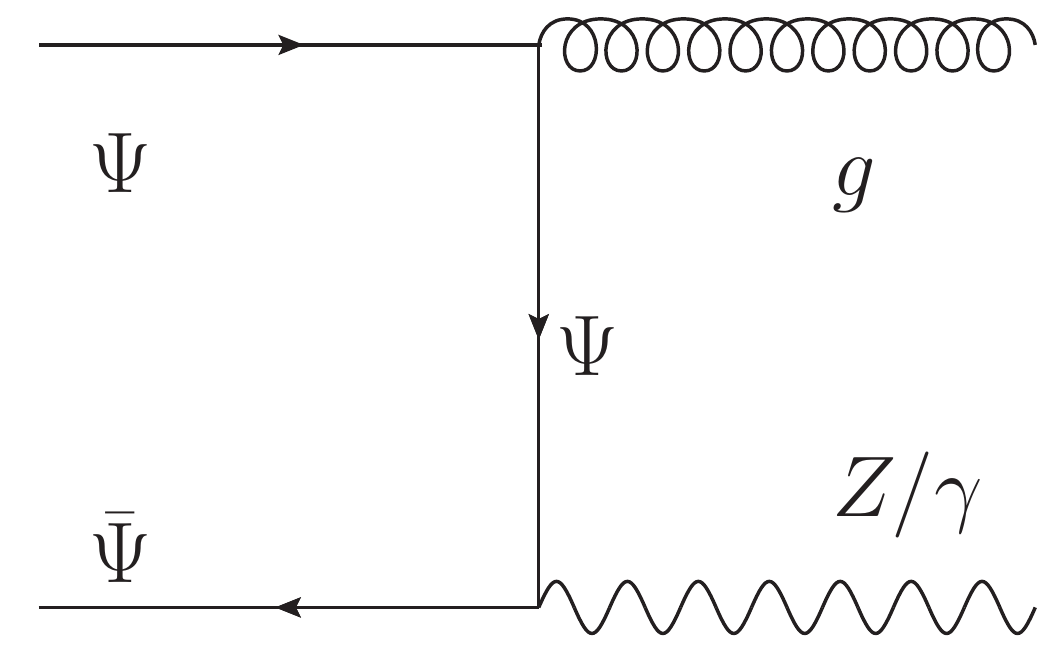}
\includegraphics[width=0.32\textwidth]{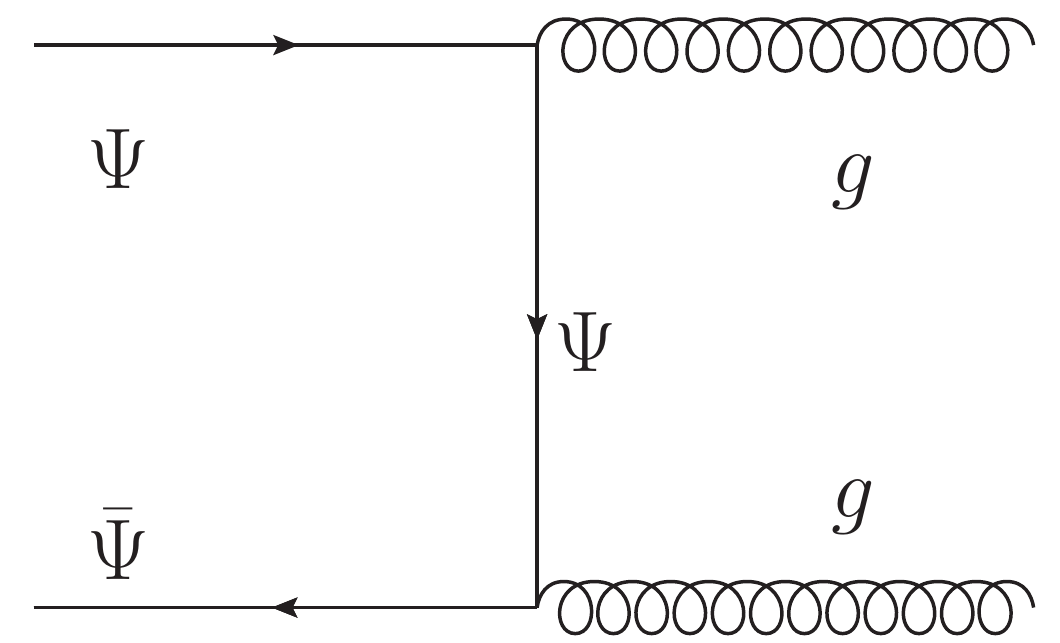}
\includegraphics[width=0.32\textwidth]{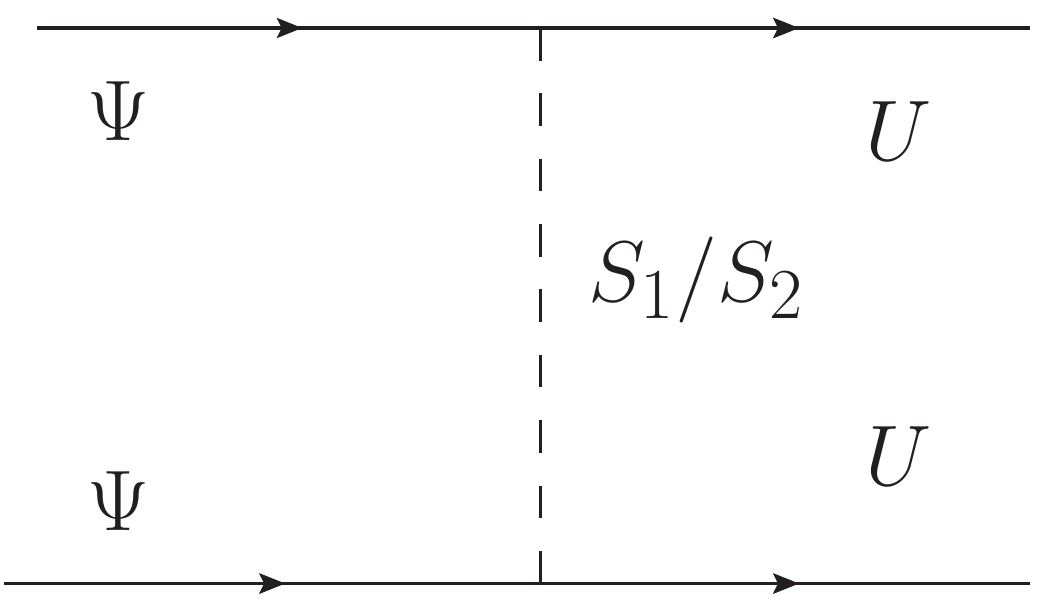}
\includegraphics[width=0.32\textwidth]{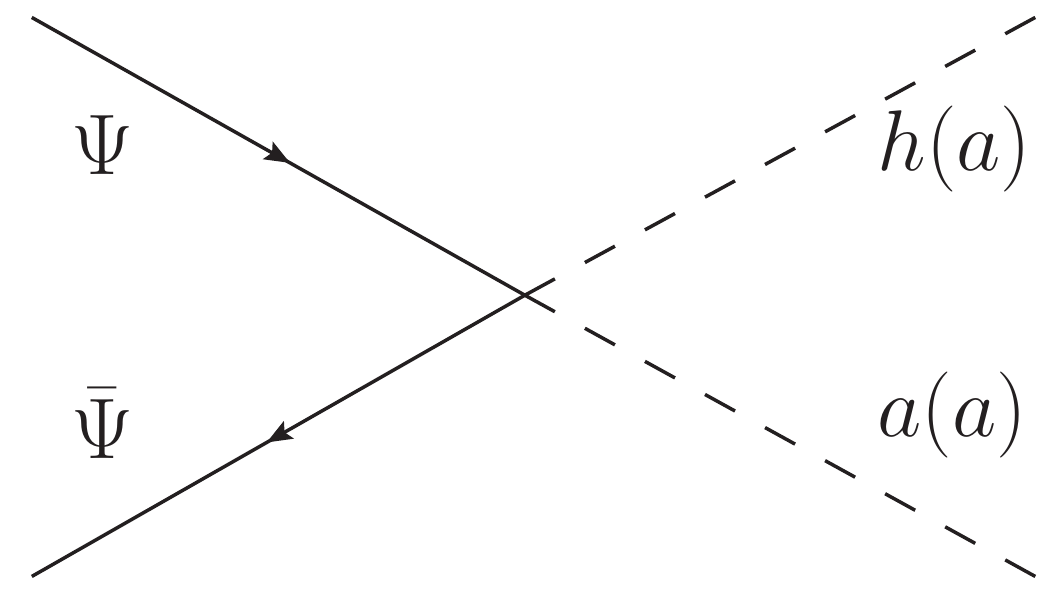}
\includegraphics[width=0.32\textwidth]{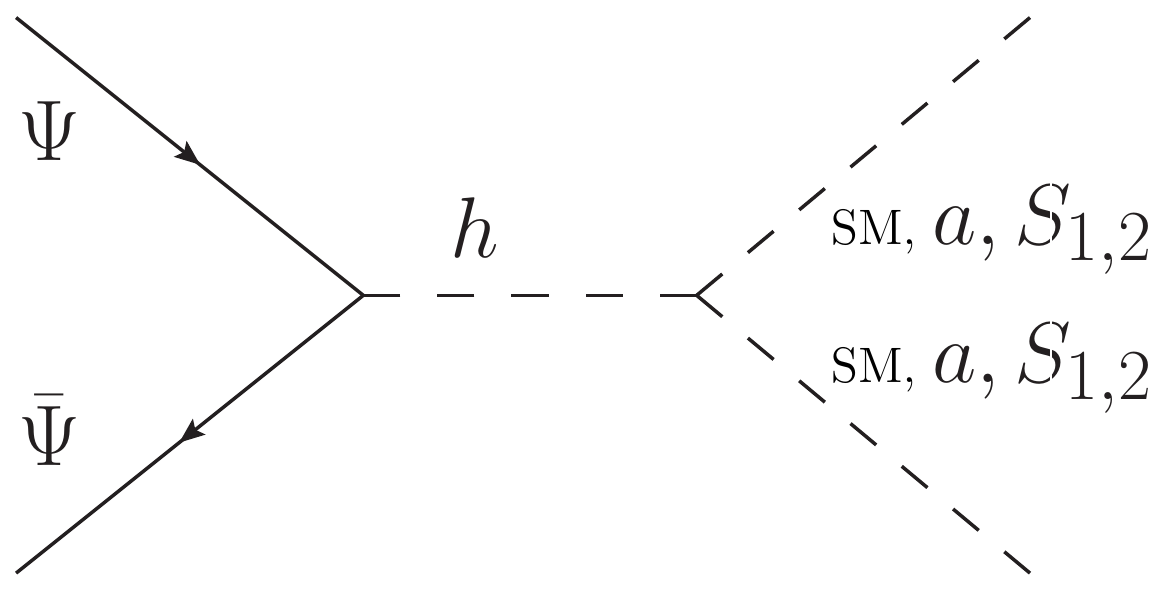}
\includegraphics[width=0.32\textwidth]{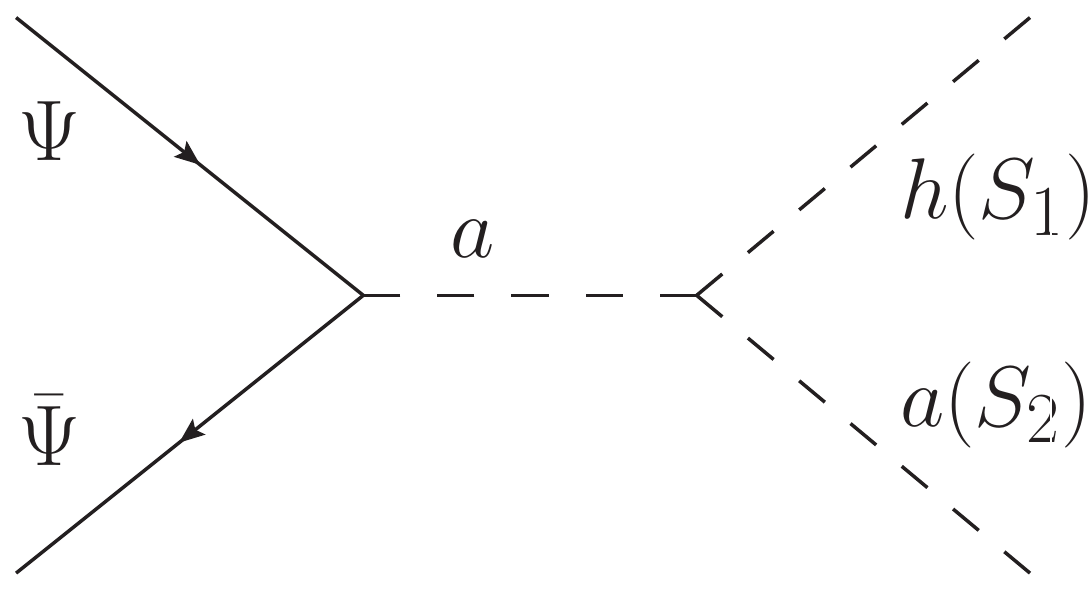}
\includegraphics[width=0.32\textwidth]{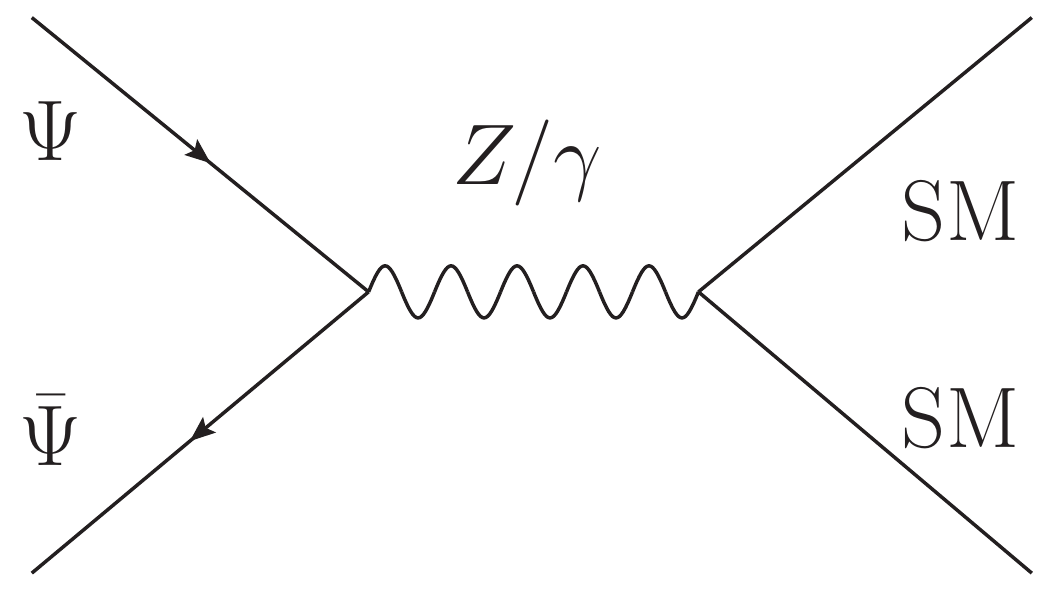}
\includegraphics[width=0.32\textwidth]{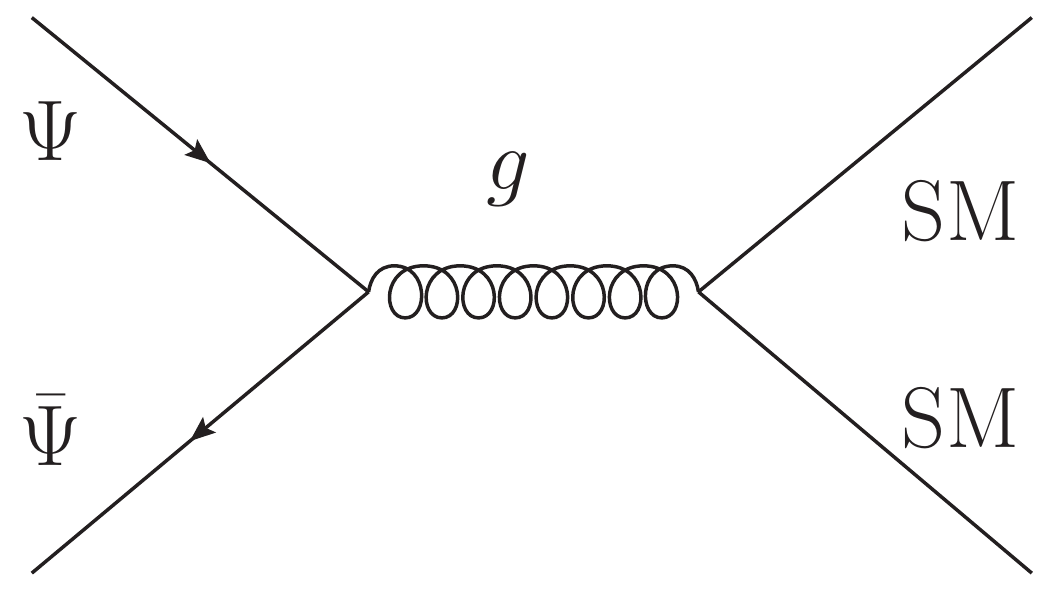}
\caption{Annihilation channels of vectorlike quark $\Psi$. U denotes the SM up-type quark ($U \equiv u,c,t, \bar{u}, \bar{c}, \bar{t}$)}
\label{VLQ-annihilation}
\end{figure}

\begin{figure}[H]
\centering
\includegraphics[width=0.32\textwidth]{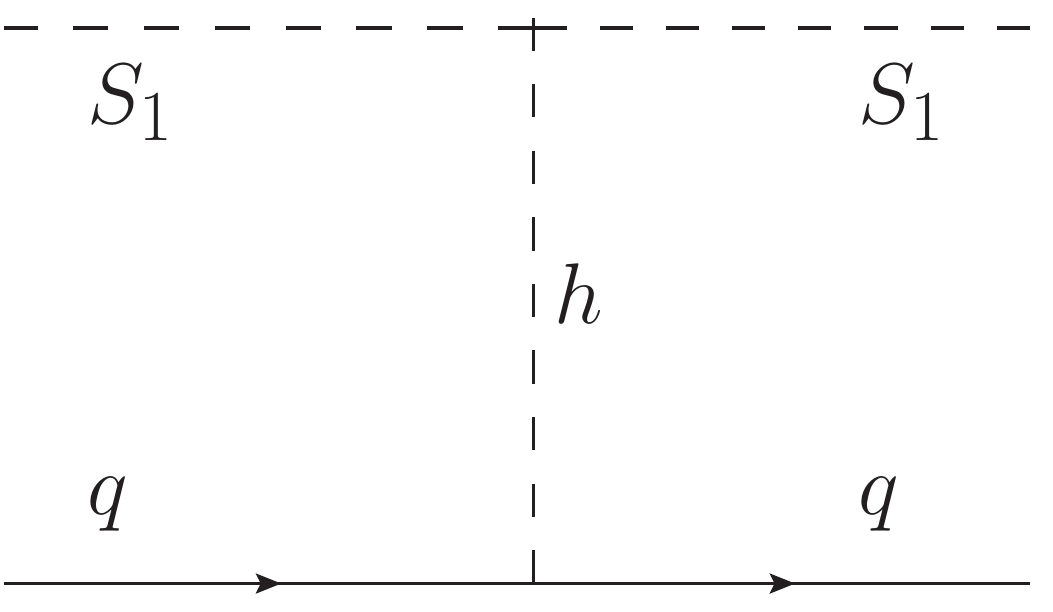}
\includegraphics[width=0.32\textwidth]{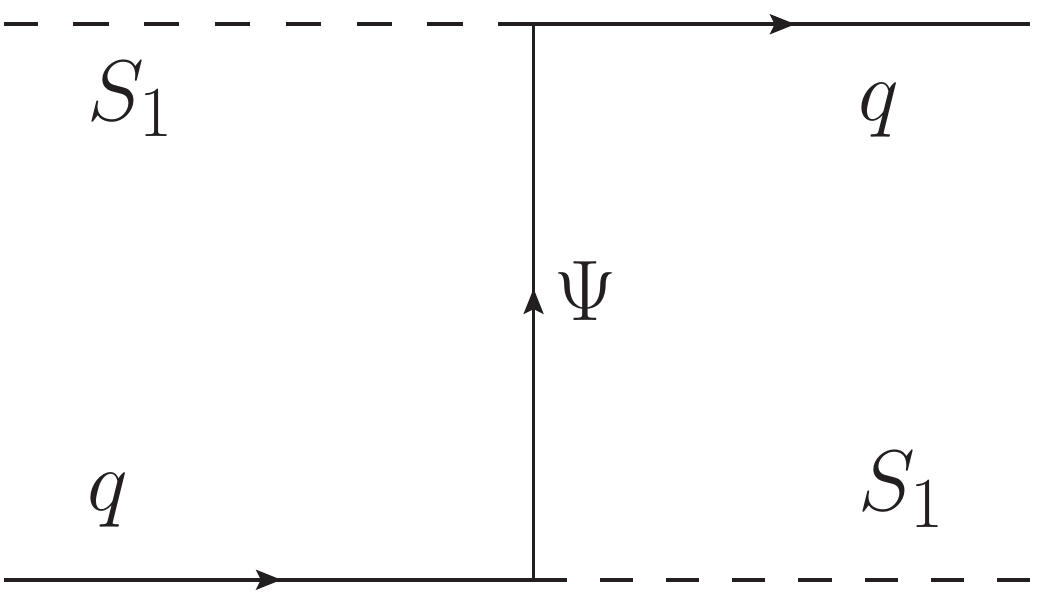}
\includegraphics[width=0.32\textwidth]{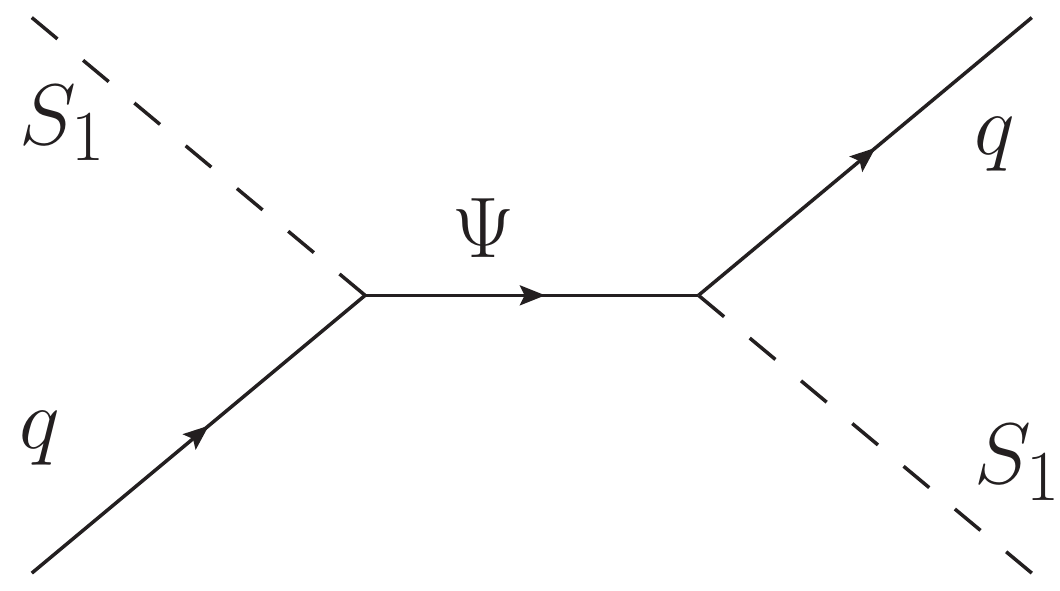}
\caption{Spin independent elastic scattering between dark matter ($S_1$) and nucleon}
\label{directDetection}
\end{figure}

\begin{figure}[H]
\centering
\includegraphics[width=0.24\textwidth]{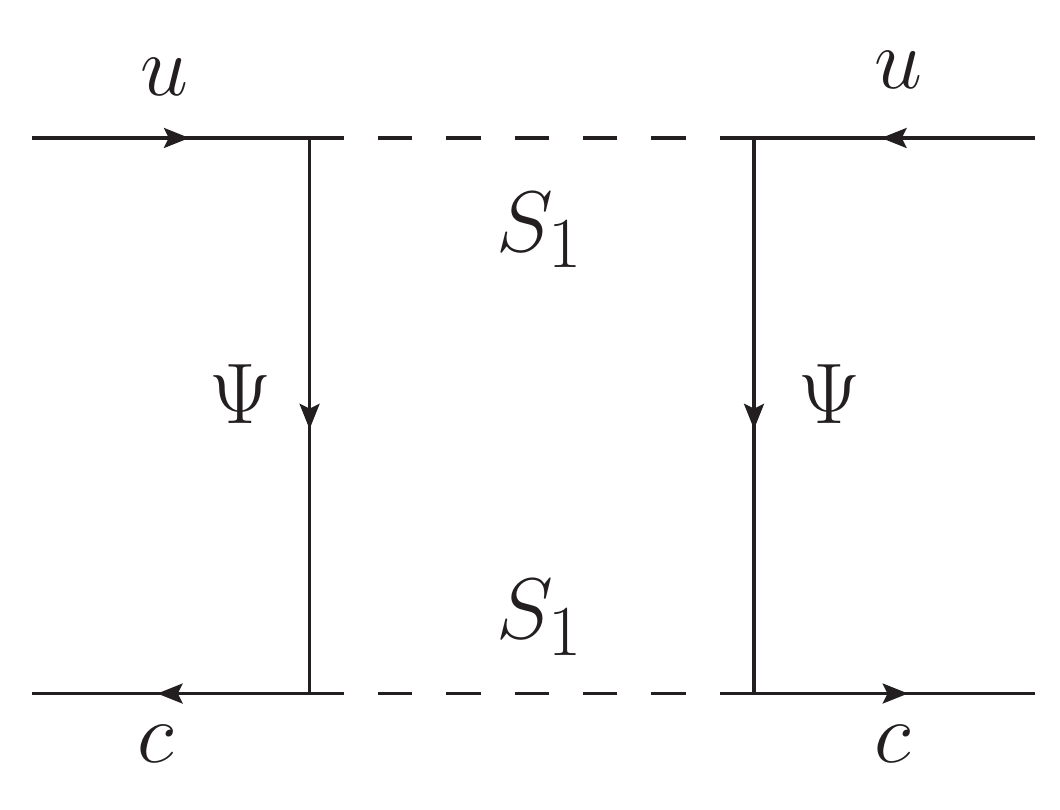}
\includegraphics[width=0.24\textwidth]{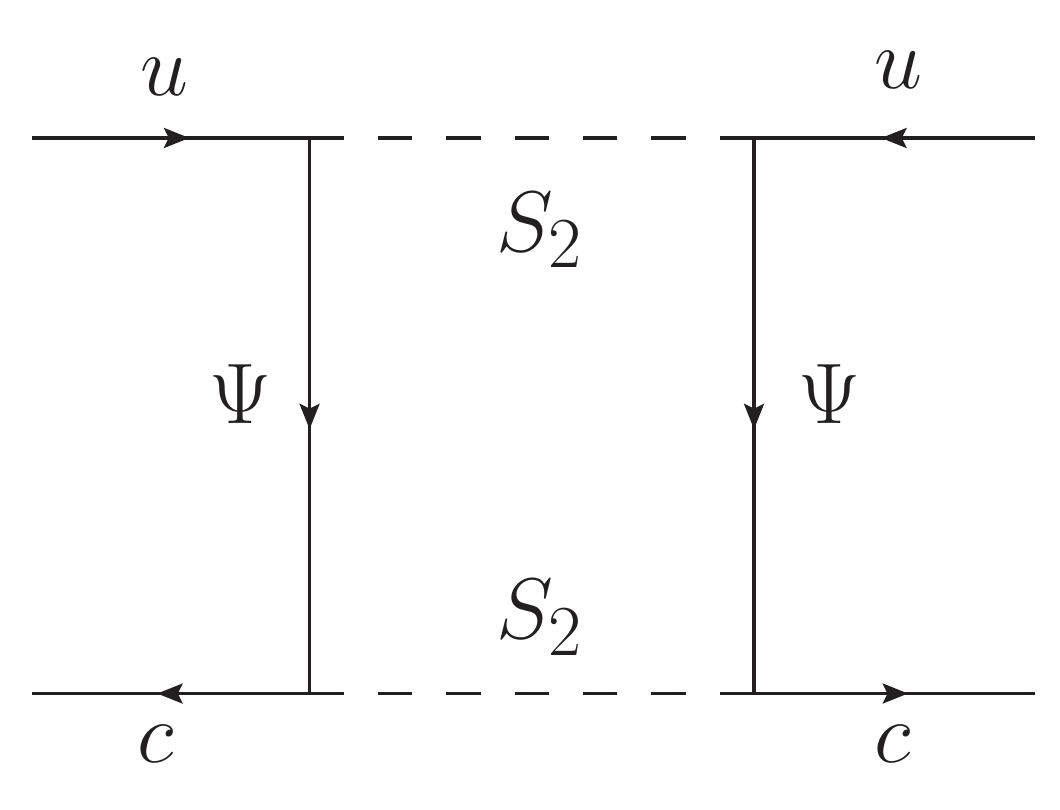}
\includegraphics[width=0.24\textwidth]{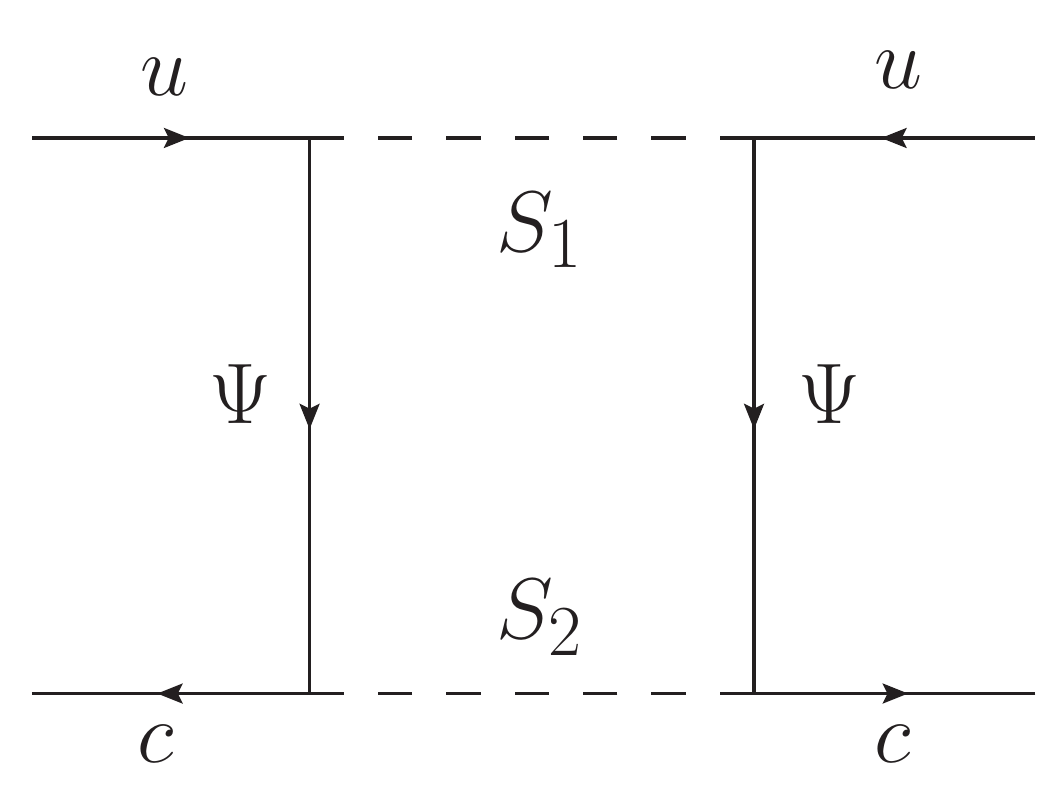}
\includegraphics[width=0.24\textwidth]{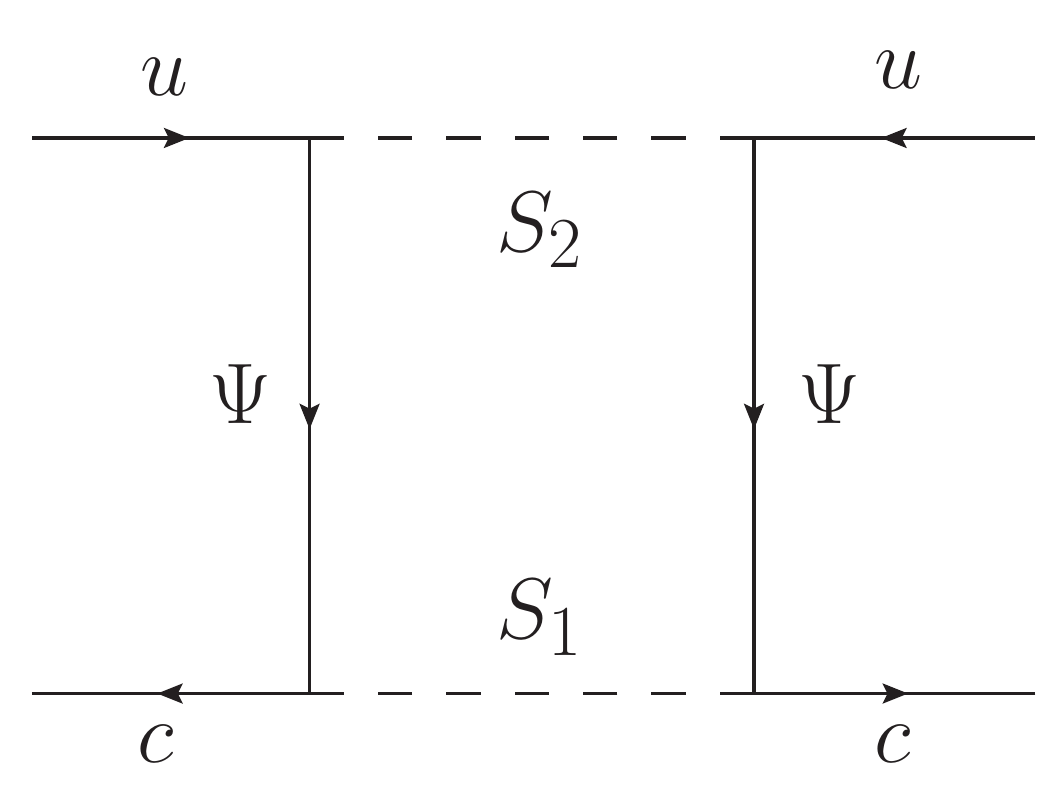}
\caption{Diagrams contributing to the $D^0-\bar{D}^0$ mixing.}
\label{DDbar}
\end{figure}

\begin{itemize}
\item Annihilation channels of scalar dark matter $S_1$ are shown in Fig. \ref{annihilation}.
\item Co-annihilation channels of scalar dark matter $S_1$ are shown in Fig. \ref{co-annihilation}.
\item Annihilation channels of vector-like quark $\Psi$ are shown in Fig. \ref{VLQ-annihilation}.
\item Spin independent elastic scattering between dark matter ($S_1$) and nucleon channels are shown in Fig. \ref{directDetection}.
\item Diagrams contributing to the $D^0-\bar{D}^0$ mixing are shown in Fig. \ref{DDbar}.
\end{itemize}


\bibliographystyle{JHEP}
\bibliography{ref.bib}
\end{document}